\documentclass[12pt]{article}

\usepackage{moreverb}

\usepackage[dvips,colorlinks,bookmarksopen,bookmarksnumbered,citecolor=red,urlcolor=red]{hyperref}

\usepackage{amssymb,bm,amsmath,psfrag,graphicx,times}

\usepackage[numbers,sort]{natbib}

\usepackage{correct}
\newcommand\BibTeX{{\rmfamily B\kern-.05em \textsc{i\kern-.025em b}\kern-.08em
T\kern-.1667em\lower.7ex\hbox{E}\kern-.125emX}}

\begin{document}

\title{Fujiwhara interaction of tropical cyclone scale vortices  using a weighted residual collocation method}%\footnotemark[2]}

\author{Raymond P Walsh, Jahrul~M~Alam\thanks{Corresponding author's email: alamj@mun.ca}\\
Department of Mathematics and Statistics\\ Memorial University, Canada, A1C 5S7\\
\date{}
}

\maketitle

%\corraddr{Department of Mathematics and Statistics, Memorial University, Canada, A1C 5S7}

\begin{abstract}
The fundamental interaction between tropical cyclones was investigated through a series of water tank experiements by~\citet{Fujiwhara21,Fujiwhara23,Fujiwhara31}. However, a complete understanding of tropical cyclones remains an open research challenge although there have been numerous investigations through measurments with aircrafts/satellites, as well as with numerical simulations.  
This article presents a computational model for simulating the interaction between cyclones. The proposed numerical method is presented briefly, where the time integration is performed by projecting the discrete system onto a Krylov subspace. The method filters the large scale fluid dynamics using a multiresolution approximation, and the unresolved dynamics is modeled with a Smagorinsky type subgrid scale parameterization scheme. Numerical experiments with Fujiwhara interactions are considered to verify modeling accuracy. An excellent agreement between the present simulation and a reference simulation at $\mathcal Re=5\,000$ has been demonstrated. At $\mathcal Re=37\,440$, the kinetic energy of cyclones is seen consolidated into larger scales with concurrent enstrophy cascade -- suggesting a steady increase of energy containing scales -- a phenomena that is typical in two-dimensional turbulence theory. The primary results of this article suggest a novel avenue for addressing some of the computational challenges of mesoscale atmospheric circulations.\\
\end{abstract}

% \keywords{
% cyclone interaction; interpolating wavelet; scaling function; dyadic refinement; numerical simulation; Navier-Stokes;}

\section{Introduction}
Numerical modeling of the interaction between two (or more) tropical cyclones is a challenging mesoscale atmospheric phenomena in the field of meteorology ({\em e.g.} see \cite{Brand70,Dong83,Lander93}). For example, the physical mechanisms that lead to a secondary eyewall formation in tropical cyclones is not fully clear~\cite{Moon2010}. Numerical prediction of the track of a cyclone that interacts with other cyclones and/or with other convection induced mesoscale vortices is much more challenging than that of a single cyclone~\cite{Prieto2003}. Notable disagreements between the observed track of hurricane Sandy (2012) and that predicted by state-of-the-art computer models were mainly due to its binary interaction with another winter system, resulting in an unexpected left turn onto the New Jersey coastline. Clearly, developments of numerical methods are useful in this field. Note, the study of the primary mechanism for the binary interaction of cyclones dates back to the experimental investigations of~\citet{Fujiwhara21,Fujiwhara23,Fujiwhara31}. This study showed that two tropical cyclones rotate cyclonically around each other while the separation distance between them decreases with time if they are initially separated by not more than a critical distance. Two cyclones may eventually merge completely or partially, continue to rotate around each other, or one of the cyclones may strain out fully or partially~\cite{Prieto2003,Dritschel92}. However, the interaction between a cyclone and a mesoscale vortex may exhibit phenomena that is completely different than what was seen from Fujiwhara experiments~\cite{Moon2010}. Consequently, a continuous stream of research on this topic has been well documented using observations~\cite{Brand70,Dong83,Lander93} and numerical simulations~\cite{Prieto2003,Moon2010,Nomura2007,Kuo2004}. The present article reports on the investigation of a numerical modeling approach for the interaction between two or three tropical cyclone like vortices and similar other phenomena. One objective is to demonstrate the application of wavelet filtering on modeling mesoscale atmospheric phenomena such that partial derivatives are discretized with a multiresolution/multiscale framework.

A primary motivation for the present research is \Add{to investigate the development of a numerical methodology that employs} \Del{to extend and examine }some of the multiscale computational methods for simulating geophysical fluid flow problems. This includes, for example, the principle of multiresolution approximation~\cite{Mallat89,Mallat2009}, the weighted residual collocation method~\cite{Bruce}, and the Newton-Krylov method~\cite{Edwards94,Knoll2004}. First, due to the wide range of scales in the geophysical flows, only the most significant large scales or eddies are resolved~(LES), and \Add{the effects of an unresolved cascade of smaller eddies are parameterized to complement the resolved large eddies}~\cite{dear70,dear72}. Typically, the numerical method is assumed to apply the LES filter indirectly; however, the cut-off scale $\Delta$ is often overshot when an upwind-biased higher order discretization is used~\cite{Piotr2014}. We have studied a multi-resolution approximation~(MRA) method~({\em e.g.}~\cite{Mallat89}), in order to filter a geophysical flow at a cut-off scale $\Delta\sim\mathcal O(10$~km) which is typical for simulating tropical cyclones~\cite{Prieto2003}. Second, the subdivision scheme of~\citet{Dubuc89} (DD subdivision) is investigated towards the development of a weighted residual collocation methodology~\cite{Bruce} for solving the equations. Third, a Krylov space projection method is developed for modeling the simultaneous dependence of multiphysics phenomena associated with cyclone interactions. Readers are cautioned that the present work is neither a hybrid scheme consisting of three or more methods stated here, nor a new numerical scheme for solving a partial differential equation. In contrast, \Add{it is a novel multiscale modeling idea} being developed~\cite{Alam2011,Alam2014,Alam2015}, is not fully understood, and the necessary ingredients behind the model stand on some powerful computational methods. 

To understand how effectively the large scale dynamics are filtered by the present method, a comparison with a reference direct numerical simulation~(DNS) at $\mathcal Re=5\,000$ is presented. The reference model is equivalent to applying a box filter so that an exact solution $u(\bm x)$ is approximated by $\mathcal N$ discrete values $\{u_k\}$, where the subgrid scale processes are directly represented by the viscous term. In contrast, the present method applies a wavelet filter so that an exact solution $u(\bm x)$ is approximated by a continuous function $u^{\mathcal N}(\bm x)$, along with a Smagorinsky type subgrid scale model. The accuracy of approximating $u(\bm x)$ by $u^{\mathcal N}(\bm x)$ is demonstrated by comparison with respect to a Fourier spectral method. In addition to simulating several cyclone interactions, we have also demonstrated how the equation for water vapor can be solved accurately. These numerical examples demonstrate how the quality of numerical simulations can be improved by the proposed numerical method. 

\subsection{Plan}
In section~\ref{sec:cm}, a brief outline of the governing equations and the numerical methodology is presented. We begin with a brief review of similar methods and introduce the governing equations along with subgrid scale parameterization. The numerical methodology is described step-by-step so that interested readers may want to implement the technique with other applications. In section~\ref{sec:bi}, results from numerical experiments are summarized, where numerical examples on the binary interaction between cyclones and comparison results with reference models are also summarized. Finally, concluding thoughts along with possible future research directions are summarized in section~\ref{sec:cr}.

\section{Computational methodology}\label{sec:cm}
\subsection{Existing similar methods}
At this point, it appears natural to review similar methods that are based on the DD subdivision. The adaptive wavelet collocation method~\cite{Oleg2000,Oleg2005,Alamphd2006,Alam2006} employs the DD subdivision to develop a lifted interpolating wavelet compression (for a comprehensive review, see~\cite{Vasilyev2010}). \citet{Kevlahan2014} used the DD subdivision for developing the spherical wavelet method. Briefly, the beauty of this wavelet method lies mainly in representing turbulent flows with a highly compressed grid. As explained in detail by~\citet{Oleg2000}, this wavelet method discretizes the partial derivatives with a finite difference approach~\cite{Vasilyev2010}. This wavelet method may also be extended to simulate geophysical flows~({\em e.g.}~\cite{Alam2011,Alam2015}). 

Meteorological models focus on achieving accurate explicit time integration of advection dominated flows, where the advection terms are typically treated with a upwind-biased higher order discretization~\cite{Pielke2002,Prieto2003,Piotr2014,Skamarock2008}. Such a scheme maintains numerical stability through numerical diffusion which controls spectral blocking associated with the enstrophy (or energy) cascade to high wavenumbers.  Thus, energy spectra often begin to decay at relatively shorter wavelengths when there is insufficient numerical diffusion or at relatively larger wavelengths due to excessive diffusion. To avoid this artifact, \citet{Dritschel92} approximated the advection terms on a Lagrangian frame of reference without requiring numerical diffusion for stability, and used this model to classify binary interaction of cyclones~\cite{Prieto2003,Kuo2004}. \citet{Piotr2014} explained in detail how the computational challenges of state-of-the-art meteorological models lean heavily on the lack of accurate numerical discretization. 

The present article demonstrates a numerical method that discretizes derivatives on a multiresolution approximation space through the DD subdivision. The method treats the time integration by projecting the divergence of stresses and fluxes on a Krylov subspace in order to maintain numerical stability. The multiresolution approximation filters more significant large scale flow from less significant cascade of subgrid scale background. If the resolution approaches to infinity, we recover the exact solution (Chapter 7,~\cite{Mallat2009}). Benefits of this model are illustrated by applying the method on the interactions between cyclones and mesoscale vortices.  A brief literature review indicates that the present approach for simulating cyclone interactions and similar phenomena is not fully understood.

\subsection{Governing equations}
Details of the two-dimensional atmospheric model used in this study was discussed by~\citet{Piotr2014} and \citet{Pielke2002}.
For brevity, the equations governing the atmospheric phenomena are presented briefly; {\em i.e.},
\begin{equation}
  \label{eq:pi}
  \frac{D\pi'}{Dt} = -\pi'\frac{\partial u_i}{\partial x_i}, 
\end{equation}
\begin{equation}
  \label{eq:me}
  \frac{Du_i}{Dt} = -c_p\theta_0\frac{\partial\pi'}{\partial x_i}  -\epsilon_{ijk}f_j(u_k-G_k) -\frac{\partial\tau_{ij}}{\partial x_j},
\end{equation}
\begin{equation}
  \label{eq:vpr}
  \frac{Dr_v}{Dt} = -\dot{r}_{\hbox{cond}} - \dot{r}_{\hbox{dep}} - \dot{r}_{\hbox{diff}}.
\end{equation}
The conservation of mass and the equation of state are combined into eq~(\ref{eq:pi}) (see~\cite{Pielke2002,Bryan2002,Piotr2014}) where the Exner function of pressure
$$\pi' = \left(\frac{R_d}{p_0}\rho\theta\right)^{R_d/c_v} $$
depends on the density ($\rho$) and the potential temperature
$$
\theta = T\left(\frac{p_0}{p}\right)^{R_d/c_p}.
$$
In~(\ref{eq:pi}-\ref{eq:vpr}), $\frac{D}{Dt}$ is the material derivative, $c_p$ and $c_v$ denote the specific heat at constant pressure and constant density, respectively, $R_d$ is the gas constant, $\theta_0$ and $p_0$ are reference potential temperature and pressure, respectively, $G_k$ represents the mean geostrophic wind, and $f_i$ denote the Coriolis parameter. The terms  $\dot{r}_{\hbox{cond}}$, $\dot{r}_{\hbox{dep}}$, and $\dot{r}_{\hbox{diff}}$ denote conduction, deposition, and diffusion of water vapor, respectively~\cite{Bryan2002,Bannon2002}. 

In the momentum equation~(\ref{eq:me}), a spatial filtering operator is already applied such that $u_i(\bm x) = u_i^{\mathcal N}(\bm x) + u'_i(\bm x)$. The velocity $u_i(\bm x)$ is decomposed into a resolved part $u_i^{\mathcal N}(\bm x)$ ($i=1,\,2$)  defined by a linear combination of $\mathcal N$ wavelet basis functions, which represents the large eddies of a desired length scale $\Delta$, and an unresolved part $u'_i(\bm x)$ associated to a cascade of subgrid scale eddies.   For simplicity, the symbol $u_i$ may be retained to denote the resolved component $u_i^{\mathcal N}(\bm x)$ ({\em i.e.} the superscript $\mathcal N$ may be dropped). When the nonlinear advection operator acts on such a decomposition of $u_i(\bm x)$, a subgrid scale stress $\tau_{ij}$ appears in eq~(\ref{eq:me}), which models the effect of the unresolved component $u'_i(\bm x)$. 
If $\mathcal N\rightarrow\infty$, the influence of $\tau_{ij}$ vanishes, and eq~(\ref{eq:me}) takes the form of the `all scale' compressible model ({\em e.g.}~their eq~13) that is presented by~\citet{Piotr2014}. Note that a detailed derivation of~(\ref{eq:me}) is also given by~\citet{Pielke2002} with respect to a grid volume filtering. To test the proposed numerical model for simulating the transport of moisture by tropical cyclones, we have also considered the equation for the water vapor mixing ratio, $r_v=\rho_v/\rho_a$ ({\em e.g.}~\cite{Bannon2002,Bryan2002}). In other words, this article demonstrates a numerical methodology for solving the coupled system of nonlinear PDEs~(\ref{eq:pi}-\ref{eq:vpr}).
\subsection{Subgrid scale processes}
In principle, the subgrid scale parameterization in the present model is similar to what was adopted by~\citet{Moon2010} and~\citet{Prieto2003}, except the present filtering approach. In addition, a \Add{generalized Ekman} balance~\cite{Kundu} among surface friction, geostrophic pressure gradients, Coriolis forces, and vertical mixing of momentum has been adopted, {\em i.e.}
\begin{equation}
  \label{eq:gb}
0 = - c_p\theta_0\frac{\partial\pi'}{\partial x_i} -\epsilon_{ijk}f_jG_k + \frac{\partial\tau_{i3}}{\partial x_3},
\end{equation}
where eq~(\ref{eq:gb}) is estimated through a Rayleigh friction model~\cite{Stevens2002} and a fixed value $10^{-5}~\hbox{s}^{-1}$ is used for the frictional coefficient~({\em e.g.} \cite{Pielke2002}).

The model of~\citet{smagorinsky} is adapted to represent the subgrid scale stress,
\begin{equation}
  \label{eq:tau}
\tau_{ij} = -\underbrace{2(C_s\Delta)^2|S|}_{\nu_{\tau}}S_{ij},
\end{equation}
where 
$$
|S| = \sqrt{2S_{ij}S_{ij}}, \quad 
S_{ij} = \frac{1}{2}\left(\frac{\partial u_i}{\partial x_j} + \frac{\partial u_j}{\partial x_i}\right),
$$
$C_s$ is the Smagorinsky constant.  For all reported simulations, we have used $C_s=0.18$ and $\Delta$ is taken equal to the smallest scale associated with the DD subdivision scheme. Instead of computing $|S|$ dynamically, we have estimated a value of $|S|$ {\em a priori} for all reported simulations. In other words, the coefficient of eddy viscosity $\nu_{\tau} = -2(C_s\Delta)^2|S|$ is adapted to the resolved scale $\Delta$ and the resolved rate of strain. 

\subsection{Multiscale filtering with Deslauriers-Dubuc subdivision of order $2p$}
Since the Deslauriers-Dubuc (DD) subdivision scheme is detailed in several other references~\cite{Dubuc89,Alam2014}, we present briefly how this method provides a multiscale filtering approach according to the principle of multiresolution approximation developed by~\citet{Mallat89}.

A given structured collection of rectangles (or rectangular prisms) constitute a $d$-dimensional mesh $\mathcal G^0$ that can be refined successively to obtain a finite collection of quad-tree (or octree in 3D) meshes ({\em e.g.} Chapter 7.8 of~\cite{Mallat2009}) 
$$
\mathcal G^0\subseteq\cdots\subseteq\mathcal G^{s-1}\subseteq\mathcal G^{s}\subseteq\mathcal G^{s+1}\cdots.
$$
At each refinement, $\mathcal G^s$ grows to $\mathcal G^{s+1}$ by inserting $2^d-1$ child nodes in the neighbourhood of a parent node $\bm x_k\in\mathcal G^s$~\cite{Alam2014}. 
Let $\varphi(\bm x)$ be the fundamental function developed by applying the DD subdivision on the mesh $\mathcal G^0$.
\citet{Mallat89} showed that a set of a scaling functions $\{\varphi_k(\bm x)\}$ can be associated with each node $\bm x_k\in\mathcal G^0$ with appropriate translations of $\varphi(\bm x)$ after setting $\varphi_0(\bm x) = \varphi(\bm x)$. If $\mathcal G^0$ is refined, $\varphi_k(\bm x)$ is first dilated and then translated in order to get $2^d-1$ additional functions associated with each $\bm x_k$, which represent ``details'' of small scale information in the neighbourhood of $\bm x_k$. Since each node $\bm x_k$ is associated to a scale or refinement level $s$, a set of such functions would resolve  scale-by-scale information contained in, {\em e.g.} $u(\bm x)$~\cite{Mallat89}. If the mesh $\mathcal G^s$ contains $\mathcal N$ nodes, which corresponds to a length scale $\Delta$, there exists a continuous function~({\em e.g.} theorem 7.3,~\cite{Mallat2009})
\begin{equation}
  \label{eq:mra}
  u^{\mathcal N}(\bm x) =\sum_{k=0}^{\mathcal N-1}c_k\varphi_k(\bm x) 
  %u^{\mathcal N}(\bm x) =\sum_{\bm x_k\in\mathcal G^0}^{}c_k\varphi_k(\bm x) + \sum_{l=0}^{s}\sum_{\bm x_k\in\mathcal G^{l+1}\backslash\mathcal G^l}^{}c_k^l\varphi_k^l(\bm x) = \sum_{k=0}^{\mathcal N-1}c_k\varphi_k(\bm x),
\end{equation}
such that for a given $\epsilon > 0$, one has $|u(\bm x)-u^{\mathcal N}(\bm x)|_{\infty} < \epsilon$.
Note that the complement $\displaystyle\sum_{\bm x_k\in\mathcal G^{1}\backslash\mathcal G^0}^{}c_k\varphi_k(\bm x)$ provides the ``details'' of $u(\bm x)$ that appear at the scale $2^{s-1}\Delta$ ({\em e.g.} $\mathcal G^1$) but disappear at the coarser scale $2^{s}\Delta$ ({\em e.g.} $\mathcal G^0$). Thus, starting with a large scale representation $\displaystyle\sum_{\bm x_k\in\mathcal G^0}^{}c_k\varphi_k(\bm x)$, additional detail of the flow is added at each refinement, and finally, the decomposition~(\ref{eq:mra}) filters $u(\bm x)$ such that $u^{\mathcal N}(\bm x)$ does not oscillate at a frequency~$>\frac{1}{2\Delta}$. %On the last equality of~(\ref{eq:mra}), the superscript $l$ is dropped because the subscript $k$ is sufficient to represent the physical position of a node and its associated scale.

\begin{figure}
  \centering
  \begin{tabular}{cccc}
    \includegraphics[trim=0cm 0cm 0cm 0cm,clip=true,height=3cm]{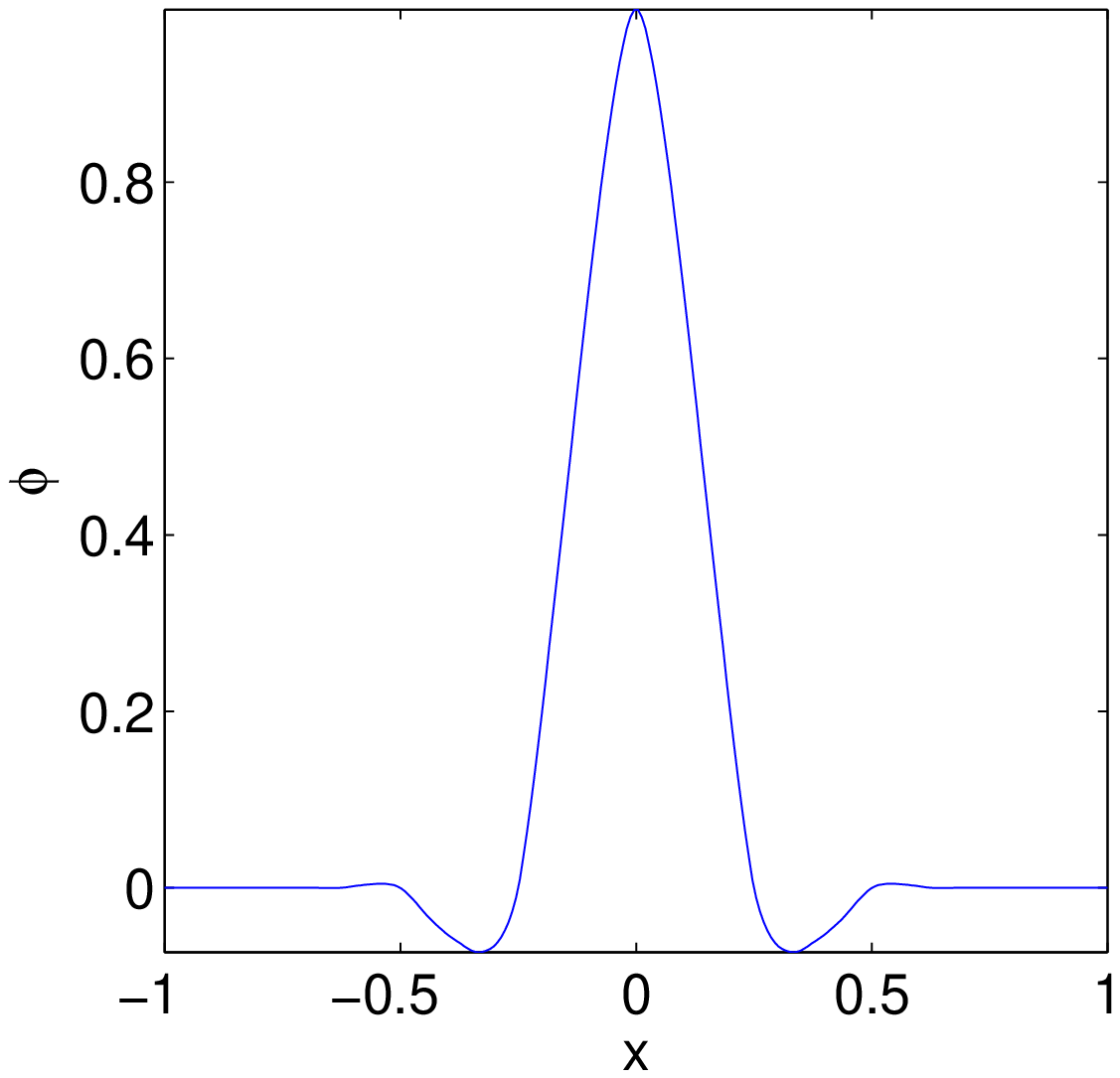}&
    \includegraphics[trim=0cm 0cm 0cm 0cm,clip=true,height=3cm]{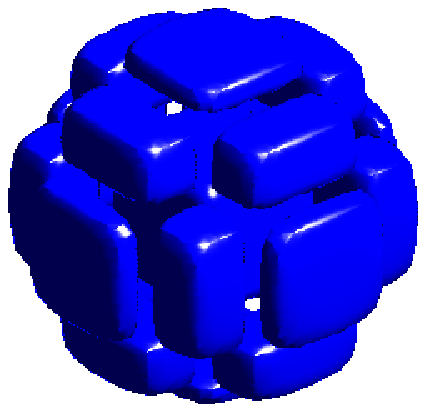}&
    \includegraphics[trim=0cm 0cm 1cm 0cm,clip=true,height=3cm]{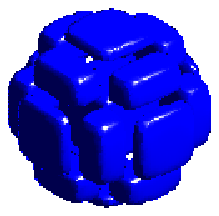}&
    \includegraphics[trim=1cm 0cm 0cm 0cm,clip=true,height=3cm]{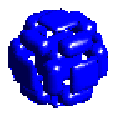}\\
    $(a)$ & $(b)$ & $(c)$ & $(d)$
  \end{tabular}
  \caption{The idea of capturing small scale features by the dilation of a fundamental scaling function $\varphi(\bm x)$ has been demonstrated where the scaling function is developed in $[-1,1]\times[-1,1]\times[-1,1]$ using the Deslauriers-Dubuc subdivision scheme of order $4$. $(a)$~The one-dimensional cross-section $\varphi(x,0,0)$ of the label surface plot of $\varphi(x,y,z)$ as shown in $(b)$, $(c)$~$\varphi(2x, 2y, 2z)$ is the factor of $1/2$ dialation of $\varphi(x,y,z)$, and $(d)$~$\varphi(4x,4y,4z)$ is the factor of $1/4$ dialation of $\varphi(x,y,z)$.}
  \label{fig:phi3d}
\end{figure}

In $1$D, \citet{Dubuc89} provides technical details for the subdivision scheme and the associated limit function  $\varphi(x)$. Similarly, the subdivision scheme can be applied on a two- and three-dimensional mesh, $\mathcal G^0$.  
Fig~\ref{fig:phi3d} presents such a scaling function, $\varphi_k(\bm x)$, on a $3$-dimensional mesh of the domain~$[-1,1]\times[-1,1]\times[-1,1]$, where $\varphi_k(\bm x)=1$ on $\bm x_k=(0,0,0)\in\mathcal G^0$. \Add{A one-dimensional plot of the function along the line intersecting $XY$ and $XZ$ planes is shown in Fig~\ref{fig:phi3d}$a$.} If $\varphi_k(\bm x)$ shown in Fig~\ref{fig:phi3d}$b$ is dilated by a factor of $1/2$ in each direction, we get a dilated $\varphi_k(\bm x)$ as shown in Fig~\ref{fig:phi3d}$c$. Similarly, a further dilation by a factor of $1/2$ leads to $\varphi_k(\bm x)$ that is shown in Fig~\ref{fig:phi3d}$d$. This shows how the present filtering method resolves multiscale features of a flow. %
For convenience we describe the following properties by restricting $\varphi_k(\bm x)$ on the $x$-axis~\cite{Dubuc89,Alam2014}.  
\begin{itemize}
\item As seen in Fig~\ref{fig:phi3d}$(a)$ for $p=2$, $\varphi_k(x)$ vanishes outside the interval $[x_{k-2p+1},x_{k+2p-1}]$, and has exactly $4p-2$ zeros in this interval. %

\item $\varphi_k(x)$ is an even polynomial of degree $2p$ and the basis $\{\varphi_k(x)\}$ reproduces polynomials up to degree $2p-1$. In other words, $\varphi_k(x)$ has $2p$ vanishing moments.
This property implies that the filtered solution given by~(\ref{eq:mra}) may be adequate to represent the subgrid scale flow on some nodes, $\bm x_k$'s, particularly when subgrid scale processes are localized. In addition, it helps control the propagation of spatial error in the time domain.

\item As discussed in details by~\citet{Dubuc89}, the above properties imply that at most $4p-1$ nearest $\varphi_k(\bm x)$'s are needed to approximate the derivative of $u(\bm x)$ on the node $\bm x_k$ using eq(~\ref{eq:mra}).
\end{itemize}
The following collocation method is developed with the help of above listed properties.
\subsection{The weighted residual collocation method}
To keep the terminology consistent with that of~\citet{Bruce}, we call~(\ref{eq:mra}) a trial solution. To discretize the action of partial differential operators on the trial solution~(\ref{eq:mra}), we follow the weighted residual collocation method presented by~\citet{Bruce} and demonstrate only the additional materials associated to the present approximation~(\ref{eq:mra}). 

Given the multiscale basis $\{\varphi_k(\bm x)\}$, one may choose the dual basis  $\{\tilde\varphi_j(\bm x)\}$ such that $\langle\varphi_k(\bm x),\tilde\varphi_j(\bm x)\rangle=\delta_{kj}$, where $\langle\cdot\rangle$ denotes the usual inner product. Thus, the resolved component $u^{\mathcal N}(\bm x)$ of $u(\bm x)$ is obtained on every node of a mesh using the multiscale decomposition~(\ref{eq:mra}) such that 
$$
\langle u(\bm x),\tilde\varphi_k(\bm x)\rangle  = \langle u^{\mathcal N}(\bm x),\tilde\varphi_k(\bm x)\rangle.
$$

A collocation method may choose $\tilde\varphi_k(\bm x) = \delta(\bm x-\bm x_k)$ with respect to a set of nodes $\{\bm x_k\}$ in the domain~$\Omega$~\cite{Bruce}, \Add{and assumes that the residual $\langle\frac{\partial u}{\partial x}-\frac{\partial u^{\mathcal N}}{\partial x},\tilde\varphi_k(\bm x)\rangle$ vanishes on every node $\bm x_k$. In the present work,  for $\mathcal N\rightarrow\infty$ the derivative of~(\ref{eq:mra}) ({\em e.g.} with respect to $x$) is the same as the exact derivative $\frac{\partial u}{\partial x}$ expanded in the form of~(\ref{eq:mra}). } Accordingly, first and second order derivatives of $u(\bm x)$ are approximated by the following expressions~(see {\em e.g.} \cite{Bruce,Alam2014}),
$$
\left\langle\frac{\partial u}{\partial x}(\bm x),
\tilde\varphi_k(\bm x)\right\rangle
\equiv
\left\langle\sum_{j=0}^{\infty}c'_j\varphi_j(\bm x),
\tilde\varphi_k(\bm x)\right\rangle
=
\sum_{j=k-2p+1}^{k+2p-1}c_j\varphi'_j(\bm x_k)
$$
and
$$
\left\langle\frac{\partial^2 u}{\partial x^2}(\bm x),
\tilde\varphi_k(\bm x)\right\rangle
\equiv
\left\langle\sum_{j=0}^{\infty}c''_j\varphi_j(\bm x)
\tilde\varphi_k(\bm x)\right\rangle
=
\sum_{j=k-2p+1}^{k+2p-1}c_j\varphi''_j(\bm x_k).
$$
\Add{For each of two equations above, the last equality appears from the fact that a derivative of $\varphi_k(\bm x)$ vanishes on all nodes outside the support of $\varphi_k(\bm x)$~\cite{Alam2014,Dubuc89}.}
Using the DD subdivision method~\cite{Dubuc89}, the procedure for computing $\varphi'_j(\cdot)$ and $\varphi''_j(\cdot)$ is given in detail by~\citet{Alam2014}. 
In the next section, we demonstrate some features of the developed method for approximating functions and their derivatives.
\subsection{A numerical verification of accuracy}
\Add{Three primary sources of numerical errors are now investigated. For all color filled contour plots used in the rest of the article red, blue, and the yellow represent positive, negative, and zero values, respectively unless it is stated otherwise.} We are interested to show how the quality of numerical simulations of a fluid flow can be improved using the proposed numerical method.  First, we consider filtering a discontinuous function using~(\ref{eq:mra}), and motivate toward the direction that a multiresolution method is superior to the most powerful Fourier spectral method when a solution involves isolated sharp changes on a relatively smooth large scale background. Second, we use an example to verify that the discretization error for the present method is $\mathcal O(\Delta^{2p})$. Third, we solve a partial differential equation and show that a moving sharp interface is resolved by the present method without artificial damping.  
\subsubsection{The efficiency of filtering by~(\ref{eq:mra}).} 
In order to compare some benefits of the present development with respect to classical Fourier spectral methods, consider the following function:
\begin{equation}
  \label{eq:sqr}
u(x) = \left\{
  \begin{array}{cc}
    0,& 0 \le x < 1/4\\
    1,& 1/4\le x \le 3/4\\
    0,& 3/4 < x \le 1,
  \end{array}
\right.
\end{equation}
a widely used example to demonstrate the performance of numerical methods. We have filtered the discontinuous function $u(x)$ using~(\ref{eq:mra}) with respect to the DD~subdivision of order $4$ ($p=2$) and repeated the same using the Fourier spectral filtering. We construct $\varphi(x)$ in $[0,1]$ starting with $5$ nodes, and form the basis $\{\varphi_k(x)\}$. We take $\mathcal N$ function evaluations of $u(x)$ such that $c_k = u(x_k)$, and evaluate $u^{\mathcal N}(x)$ using~(\ref{eq:mra}). 

\begin{figure}
  \centering
  \begin{tabular}{cc}
    Present method & Fourier method\\
    \includegraphics[height=4cm]{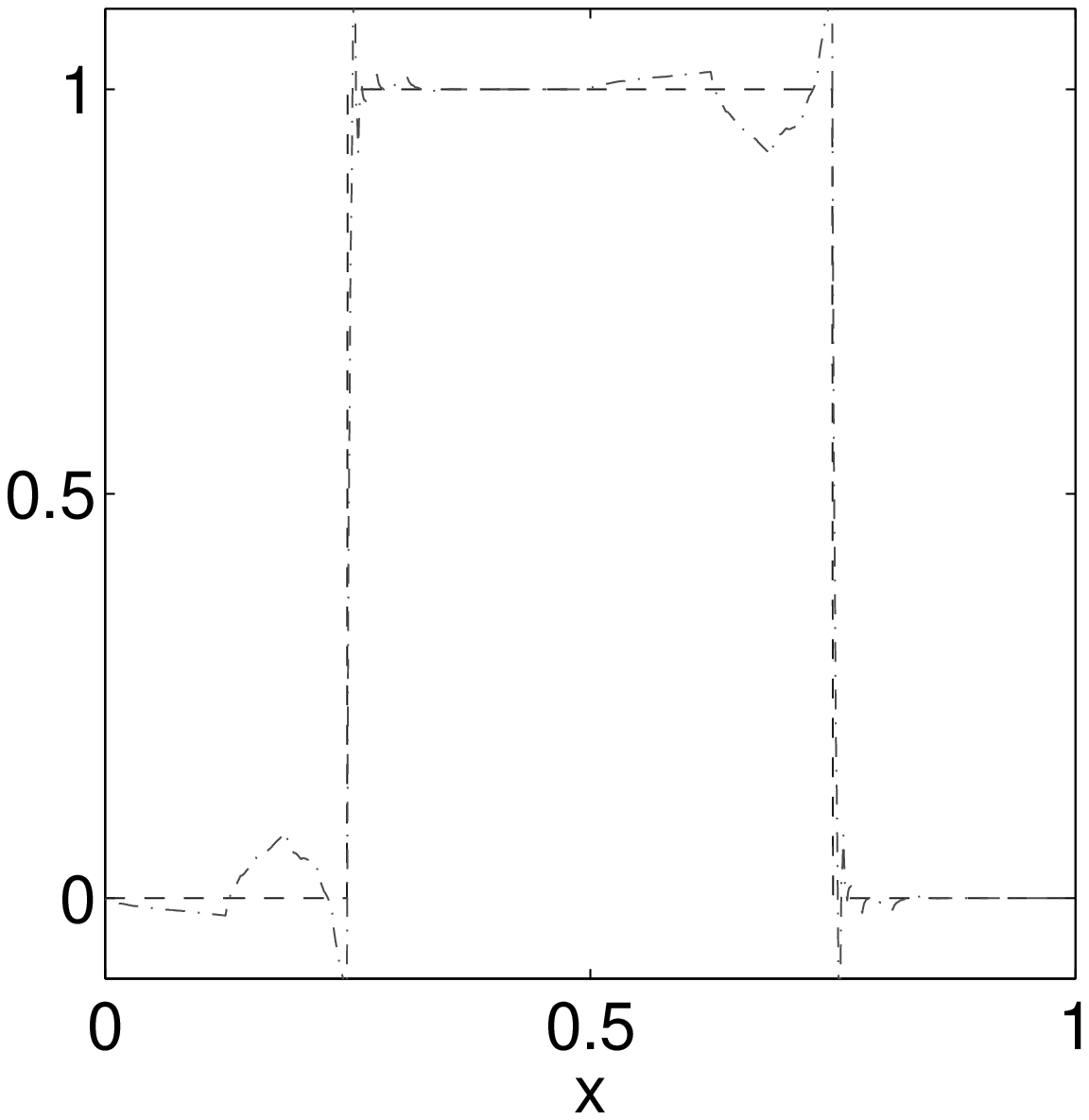}&
    \includegraphics[height=4cm]{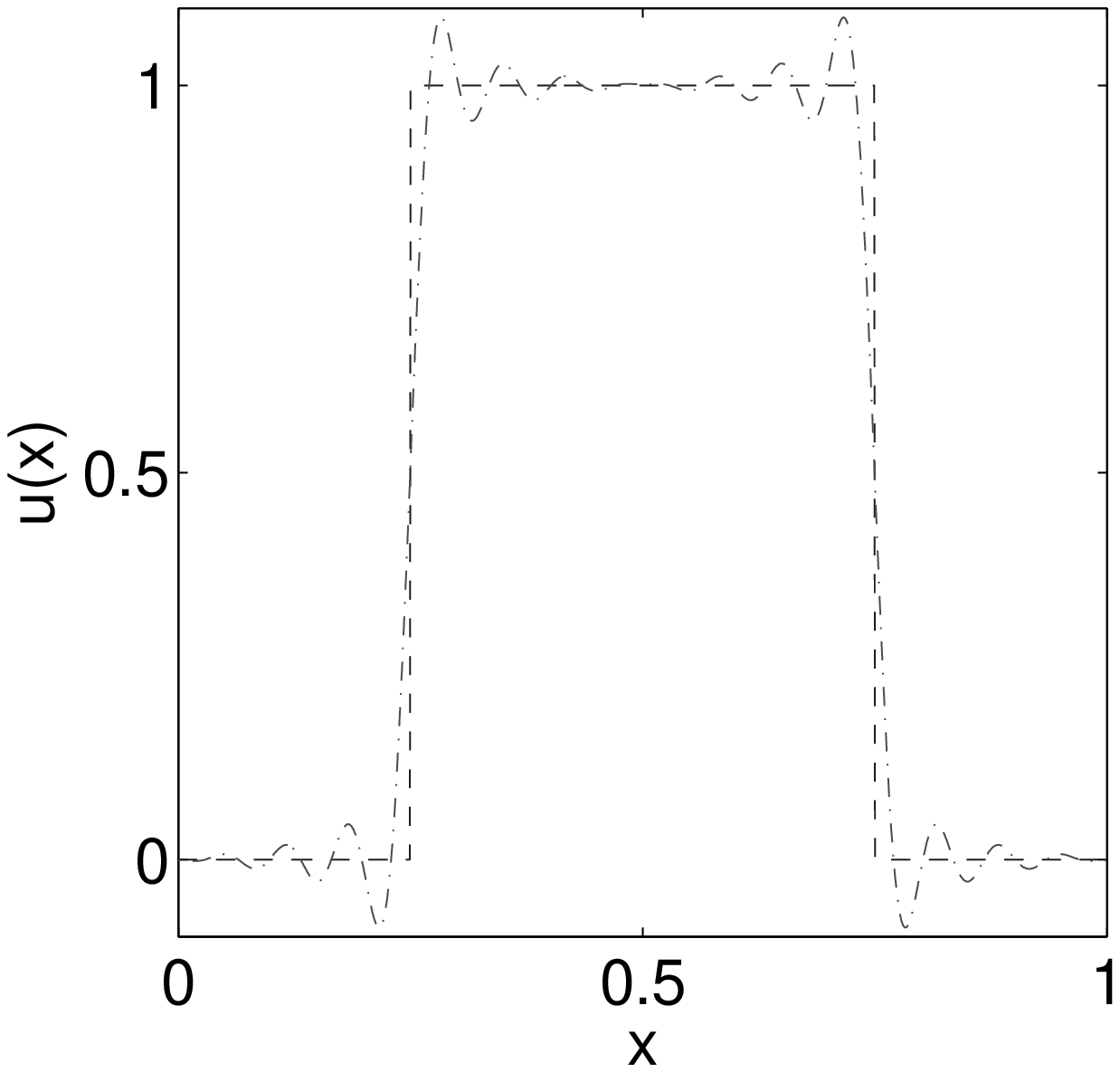}\\
    $(a)\,\mathcal N=17$ & $(d)\,\mathcal N=17$ \\
    \includegraphics[height=4cm]{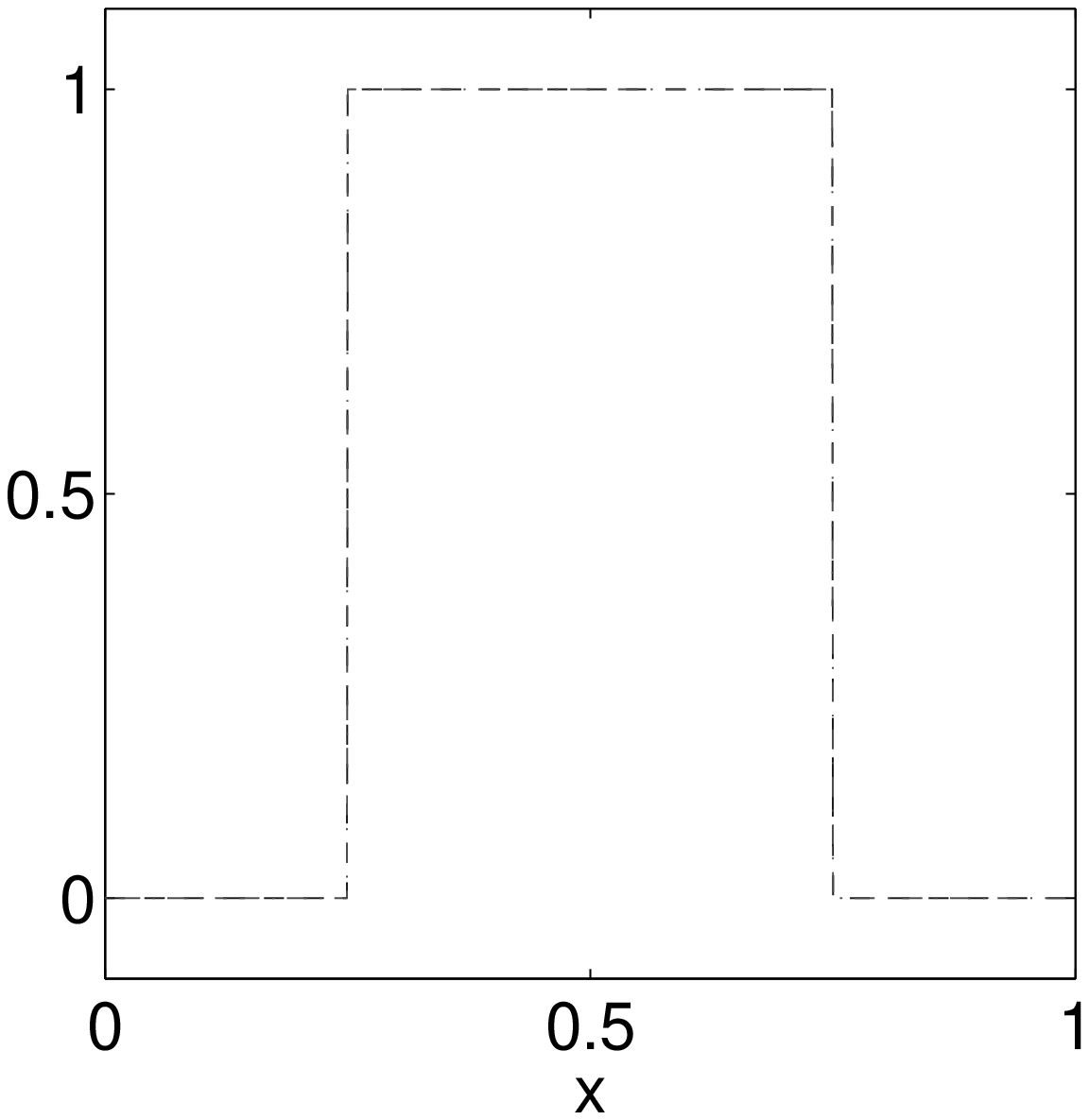}&
    \includegraphics[height=4cm]{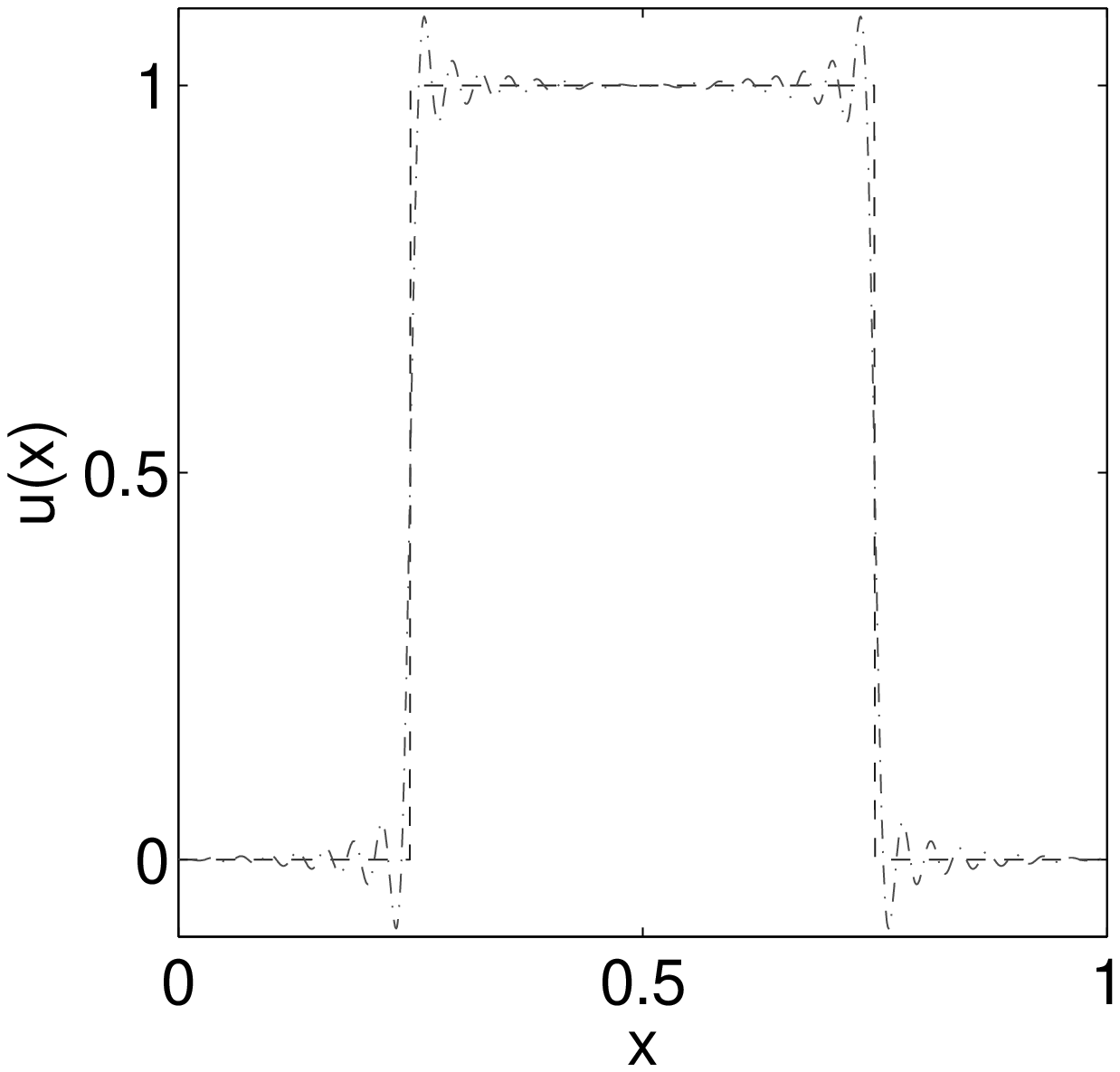}\\
    $(b)\,\mathcal N=33$ & $(e)\,\mathcal N=33$ \\
    \includegraphics[height=4cm]{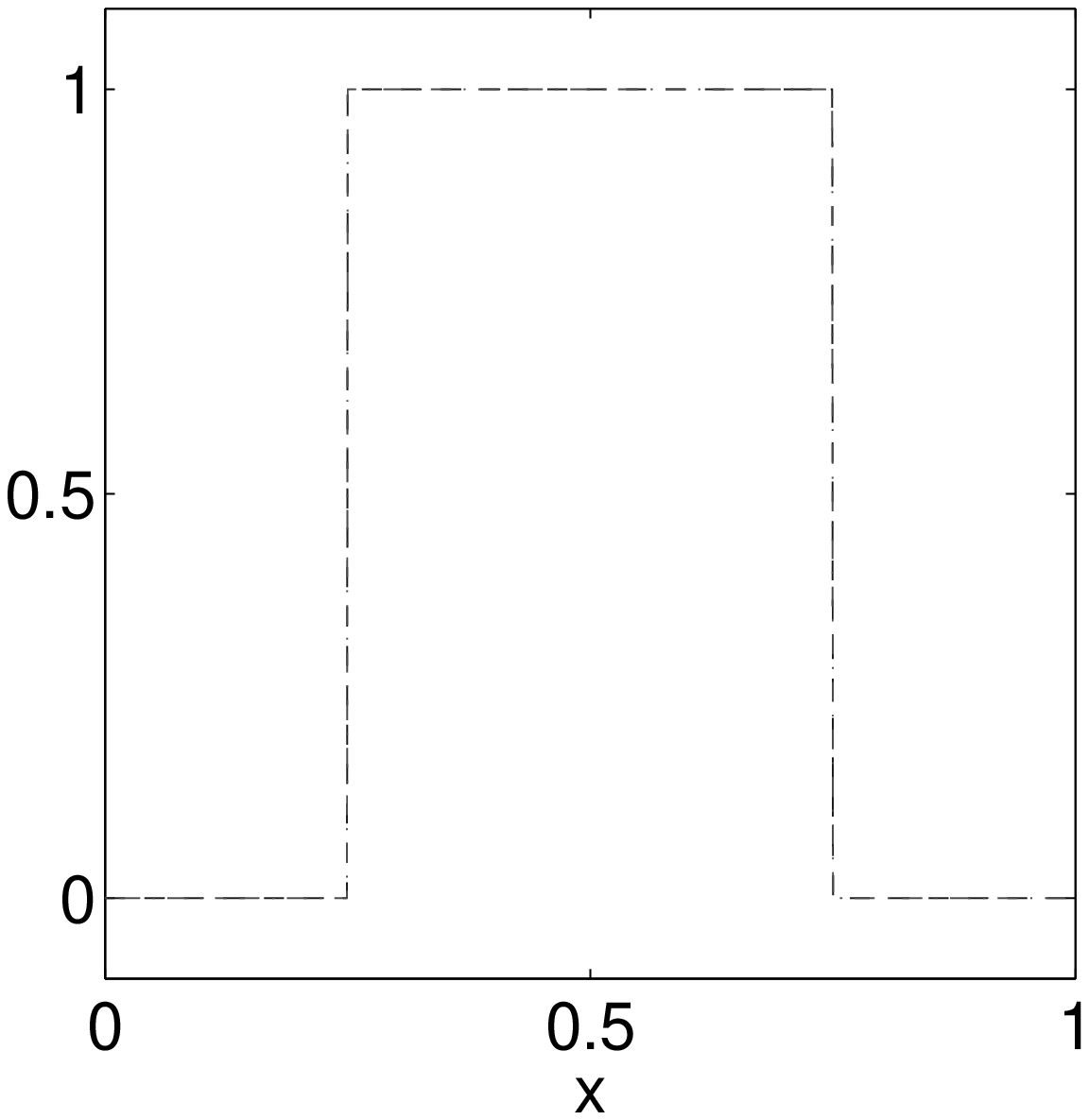}&
    \includegraphics[height=4cm]{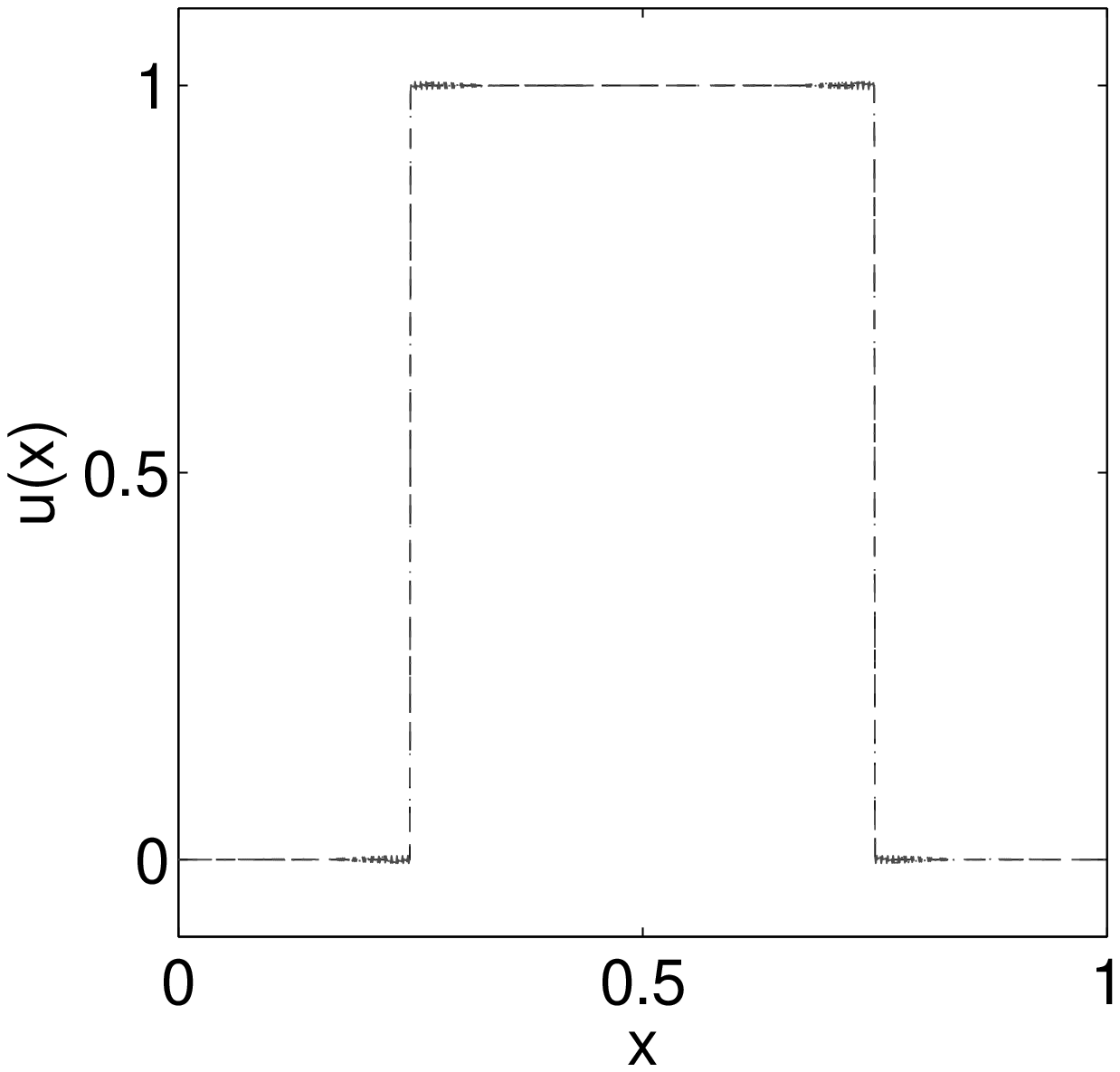}\\
    $(c)\,\mathcal N=513$ & $(f)\,\mathcal N=513$ \\
  \end{tabular}
  \caption{The approximation of a discontinuous function $u(x)$ with wavelet filtering $(a,b,c)$, as well as with Fourier spectral filtering $(d,e,f)$.}
  \label{fig:wf}
\end{figure}

Fig~\ref{fig:wf}($a,b,c$) compares $u^{\mathcal N}(x)$ with the exact function $u(x)$, {\em i.e.}~(\ref{eq:sqr}), for $\mathcal N=17$,~$33$, and~$513$, respectively. As noted above, the multiresolution approximation captures isolated small scale details. Thus, extraneous peaks of oscillations -- as seen in Fig~\ref{fig:wf}$a$ -- are sufficiently reduced by~(\ref{eq:mra}) with $\mathcal N=33$, see Fig~\ref{fig:wf}$b$. The Gibbs oscillations are a step response of low-pass filter \Add{and lead to what is also called the ringing artifacts}. For this example, the resolved frequencies with $N\ge 33$ are necessary in the sense that the Gibbs oscillations are not visible with the naked eye when the DD~subdivision method is used. 

On the other hand, one notices the oscillations in Fig~\ref{fig:wf}$(d,e,f)$, where the Fourier spectral method is used. In this case, we apply the FFT (Fast Fourier Transform) on $\mathcal N$ function evaluations to get the Fourier coefficients $\{\hat u_k\}$ and compute the Fourier series for $u^{\mathcal N}(x)$ using the coefficients $\{\hat u_k\}$. \Add{Clearly, the present choice of the basis functions and the multiresolution approximation is more effective when the solution exhibits an isolated discontinuity or a sharp change. Although we deal with continuous solutions in this work, we have adopted this widely used example to demonstrate the promise of the proposed multiresolution method.}  In~\citet{Mallat89,Mallat2009}, readers will find the theoretical details on how adding a new approximation node in the neighbourhood of a coarser scale node adds additional local details to the approximation.
\subsubsection{The accuracy of discretization:}
\begin{figure}
   \centering
   \begin{tabular}{cc}
     \includegraphics[height=6cm]{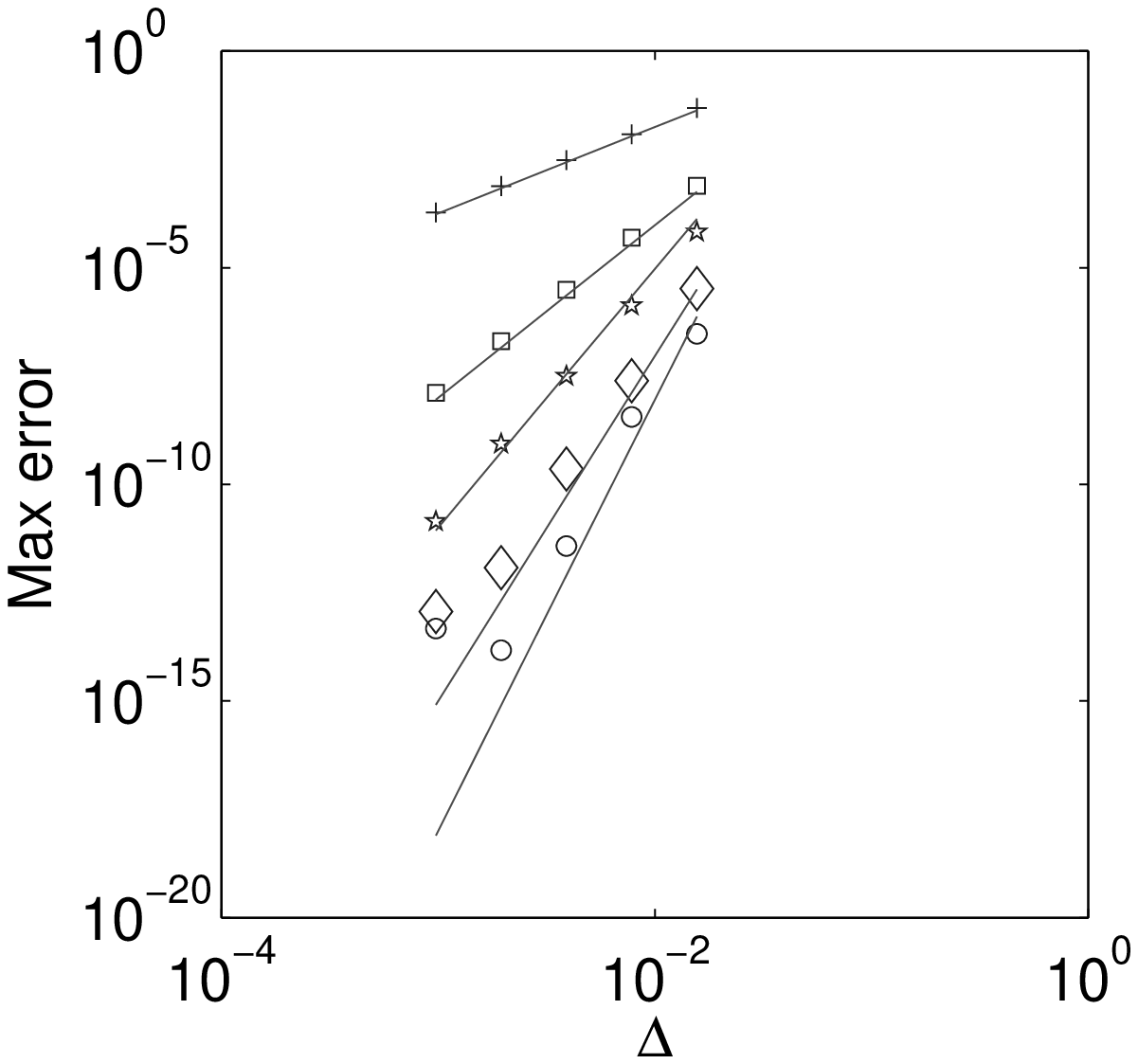}&
     \psfrag{N}{$\mathcal N$}
      \includegraphics[height=6cm]{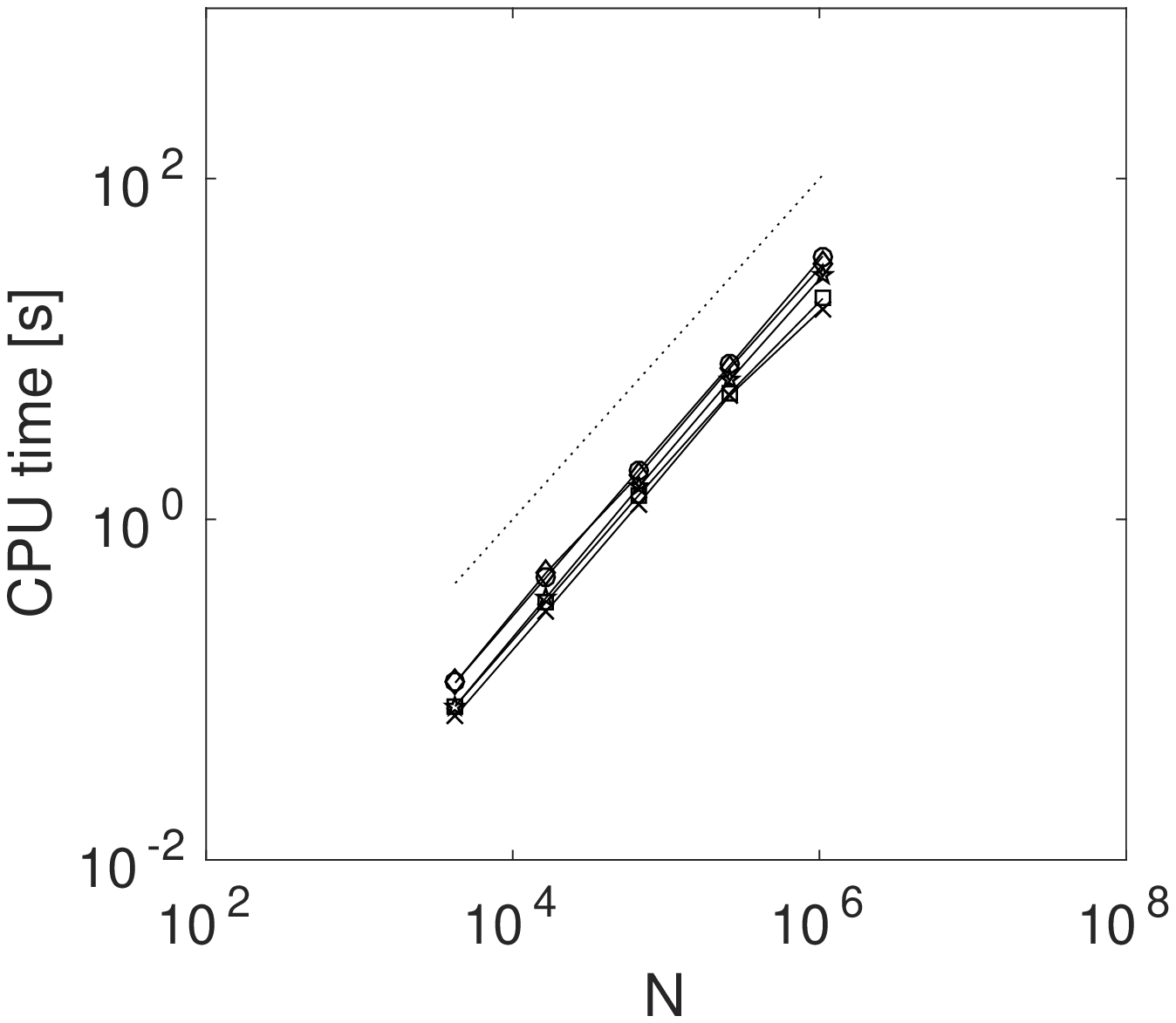}\\
     $(a)$ & $(b)$
   \end{tabular}
   \caption{$(a)$~Each row of table~\ref{tab:err} is compared logarithmically with $\Delta^{2p}$. $\times (p=1)$, $\Box (p=2)$, $\star (p=3)$, $\diamond (p=4)$, and $\circ(p=5)$. $(b)$ The CPU time [s] associated to the calculation in each row of table~\ref{tab:err} is compared logarithmically with $\mathcal N$. The dotted line indicates a slope of $1$.}
   \label{fig:df}
 \end{figure}

The interpretation of error for approximating derivatives by the present method requires additional care because there are two parameters involved. Let us consider a uniformly refined 2D mesh with $\mathcal N$ nodes for which $\Delta\sim\mathcal N^{-1/2}$. When the mesh is refined uniformly, {\em i.e} $\Delta$ decreases with fixed $p$ or $p$ is increased on a fixed mesh, {\em i.e.} with a fixed $\Delta$, the error decreases like $\mathcal O(\Delta^{2p})$, as expected in a collocation method. 
The accuracy of approximating derivatives has been demonstrated by considering two functions
$$
u(x,y) = \frac{-y(1-e^{-x^2-y^2})}{2\pi(x^2+y^2)}
\quad\hbox{and}\quad
v(x,y) = \frac{ x(1-e^{-x^2-y^2})}{2\pi(x^2+y^2)}.
$$
Both functions have a singularity at $(0,0)$ and are differentiable otherwise. Writing $\bm u = \langle u(x,y), v(x,y), 0\rangle$, we can work with only one component of $\bm\nabla\times\bm u$. We now approximate $\nabla^2(\bm\nabla\times\bm u)$ numerically, where the numerical error is computed with respect to the following exact differentiation:
$$
\nabla^2(\bm\nabla\times\bm u) = 
-\frac{\partial^3u}{\partial x^3}
+\frac{\partial^3v}{\partial y^3}
-\frac{\partial^3u}{\partial x^2\partial y}
+\frac{\partial^3v}{\partial x\partial y^2}.
$$
Note, we have computed first derivatives to find $\bm\nabla\times\bm u$ before computing second derivatives to find $\nabla^2(\bm\nabla\times\bm u)$.  
Here, we want to understand how accuracy is improved as $\mathcal N\rightarrow\infty$ or $p\rightarrow\infty$. Table~\ref{tab:err} demonstrates the error for $\mathcal N=65^2,\,129^2,\,257^2,\,513^3,$ and $1\,025^2$ and $2p=2,\,4,\,6,\,8$, and $10$. \Add{Clearly, the error decreases as $\mathcal N$ or $p$ increases. To verify the theoretical rate of convergence, Fig~\ref{fig:df}$(a)$ represents graphically each row of table~\ref{tab:err}. For each row we have compared the data with a monomial to verify how closely the logarithmic decay of the error follows $\mathcal O(\Delta^{2p})$ if $\Delta$ decreases for each $p$. Fig~\ref{fig:df}$(a)$ shows that the effect of the floating point arithmetic is noticeable for $p=4$ and $p=5$ as $\Delta\rightarrow 0$ when the error is close to the machine precision.}

We have analyzed the CPU time. Since the basis has a compact support, the increase of CPU time with respect to an increase of $p$ is primarily influenced by the implementation of the quad-tree data structure. In view of scientific computing, our objective is to have the overall performance of the method asymptotically optimal if the mesh is refined. Fig~\ref{fig:df}$(b)$ demonstrates that the CPU time increases approximately linearly with $\mathcal N$. As expected, the CPU time increases slightly for an increase of $p$. 
\begin{table}
  \centering
  %\begin{spacing}{1.5}
    \begin{tabular}{|llllll|}
      \hline
      $2p\downarrow$ & $\mathcal N\rightarrow 65^2$ & $129^2$ &
      $257^2$ & $513^2$ & $1025^2$ \\
      \hline
      2 &
      $4.86\times 10^{-2}$ &
      $1.21\times 10^{-2}$ &
      $3.05\times 10^{-3}$ &
      $7.62\times 10^{-4}$ &
      $1.90\times 10^{-4}$ \\
      4 &
      $7.82\times 10^{-4}$ &
      $4.94\times 10^{-5}$ &
      $3.10\times 10^{-6}$ &
      $2.02\times 10^{-7}$ &
      $2.39\times 10^{-8}$ \\
      6 &
      $6.74\times 10^{-5}$ &
      $1.33\times 10^{-6}$ &
      $3.17\times 10^{-8}$ &
      $8.61\times 10^{-10}$ &
      $4.40\times 10^{-11}$ \\
      8 &
      $3.27\times 10^{-6}$ &
      $2.47\times 10^{-8}$ &
      $2.27\times 10^{-10}$ &
      $1.19\times 10^{-11}$ &
      $5.16\times 10^{-11}$ \\
      10 &
      $2.96\times 10^{-6}$ &
      $3.62\times 10^{-9}$ &
      $3.77\times 10^{-12}$ &
      $1.48\times 10^{-11}$ &
      $5.69\times 10^{-11}$ \\
      \hline
    \end{tabular}
  %\end{spacing}
  \caption{Maximum error for numerical approximation of $\nabla^2(\bm\nabla\times\bm u)$.}
  \label{tab:err}
\end{table}

\subsubsection{Numerical simulation of water vapor condensation:}
Simulating a moving interface between moist and dry air is a challenging endeavor~(see~\cite{Pielke2002}). In the present model,  the moist and dry air is treated as a binary fluid~\cite{Bannon2002,Bryan2002}.  To solve~(\ref{eq:vpr}) for the mixing ratio, $r_v=\rho_v/\rho_a$, several approaches exist for 
the parameterization of terms on the right hand side, a discussion of which is beyond the scope of the present article. Here, we consider the loss of water vapor due to condensation and diffusion, such that 
$$
\dot{r}_{\hbox{cond}} + \dot{r}_{\hbox{dep}} + \dot{r}_{\hbox{diff}}=
\frac{\delta}{\delta r_v}\int(0.5r_v^2(r_v-1)^2 + D_v|\nabla r_v|^2 dV.
$$
The first term of the integrand parameterizes condensation through the air-vapor interface, the second term parameterizes the diffusion of vapor relative to the dry air, and the deposition term is neglected for simplicity, where \citet{Bannon2002} adopted $D_v=2.11\times 10^{-5}$~m$^2$/s. Thus, convection is initiated due to condensation where diffusion occurs only on the interface. We want to see if such a physical situation -- in the absence of mixing -- is resolved by our numerical method. Note that this is an idealization for the condensation where the condensed liquid is removed from the system. 

\begin{figure}
   \centering
   \begin{tabular}{cc}
     \includegraphics[height=5cm]{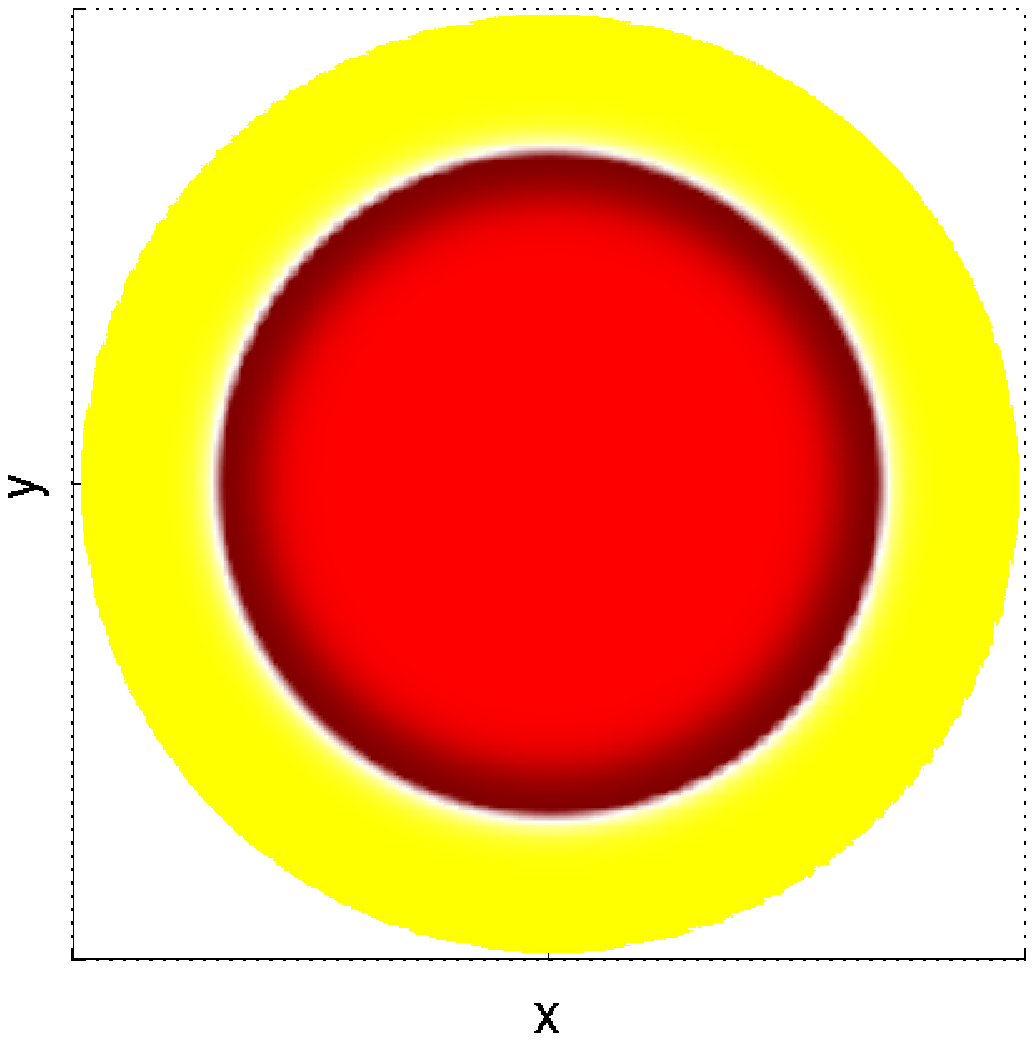}&
     \includegraphics[height=5cm]{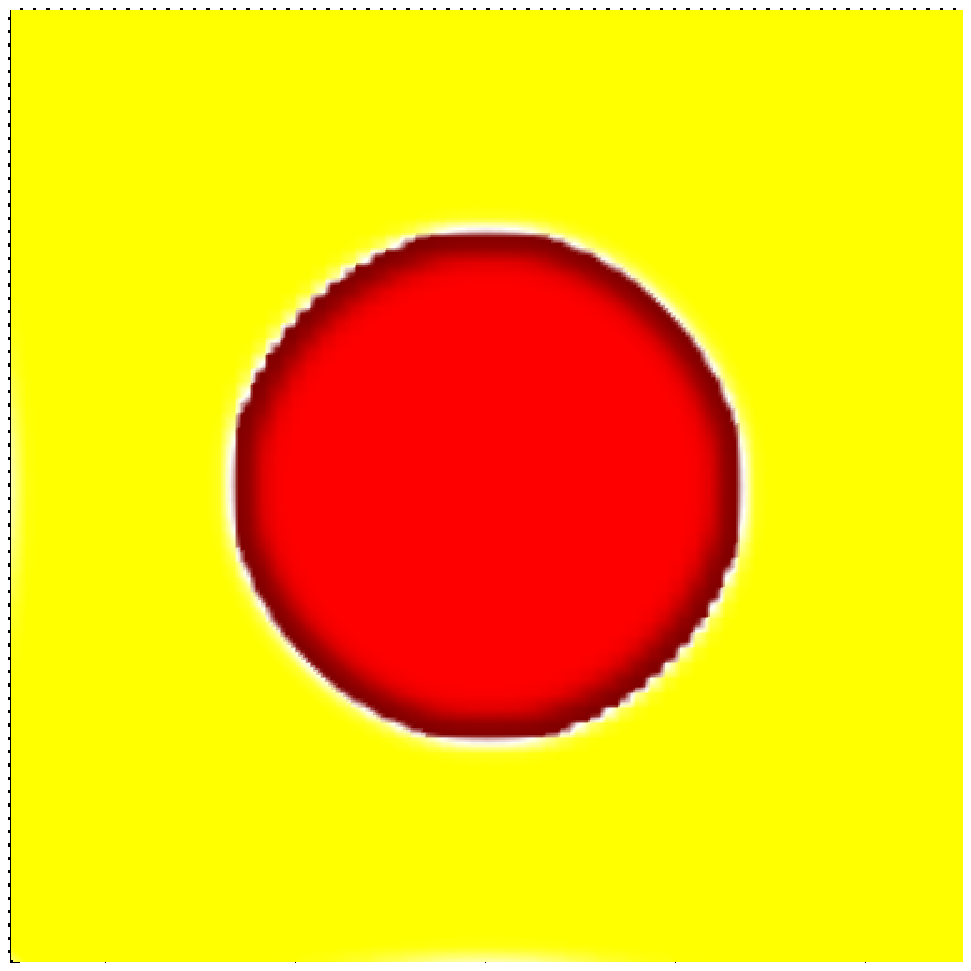}\\
     $(a)$ & $(b)$\\
     \includegraphics[height=5cm]{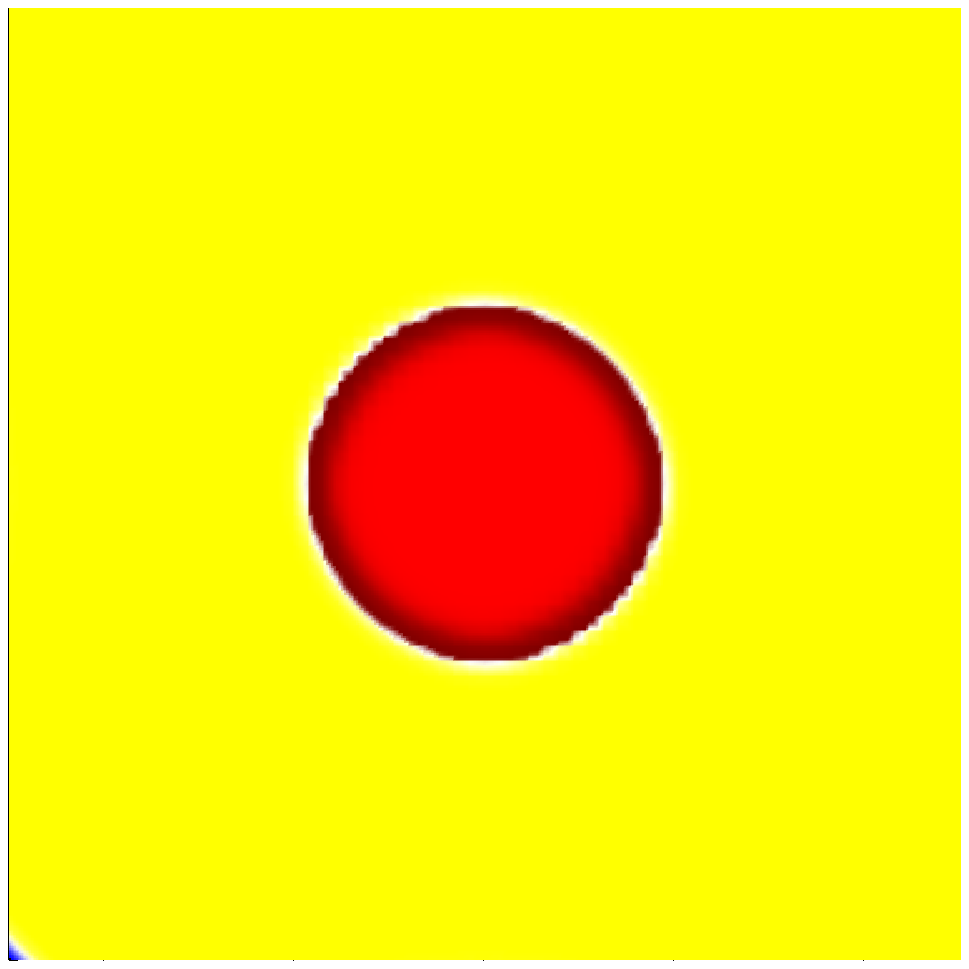}&
     \includegraphics[height=5cm]{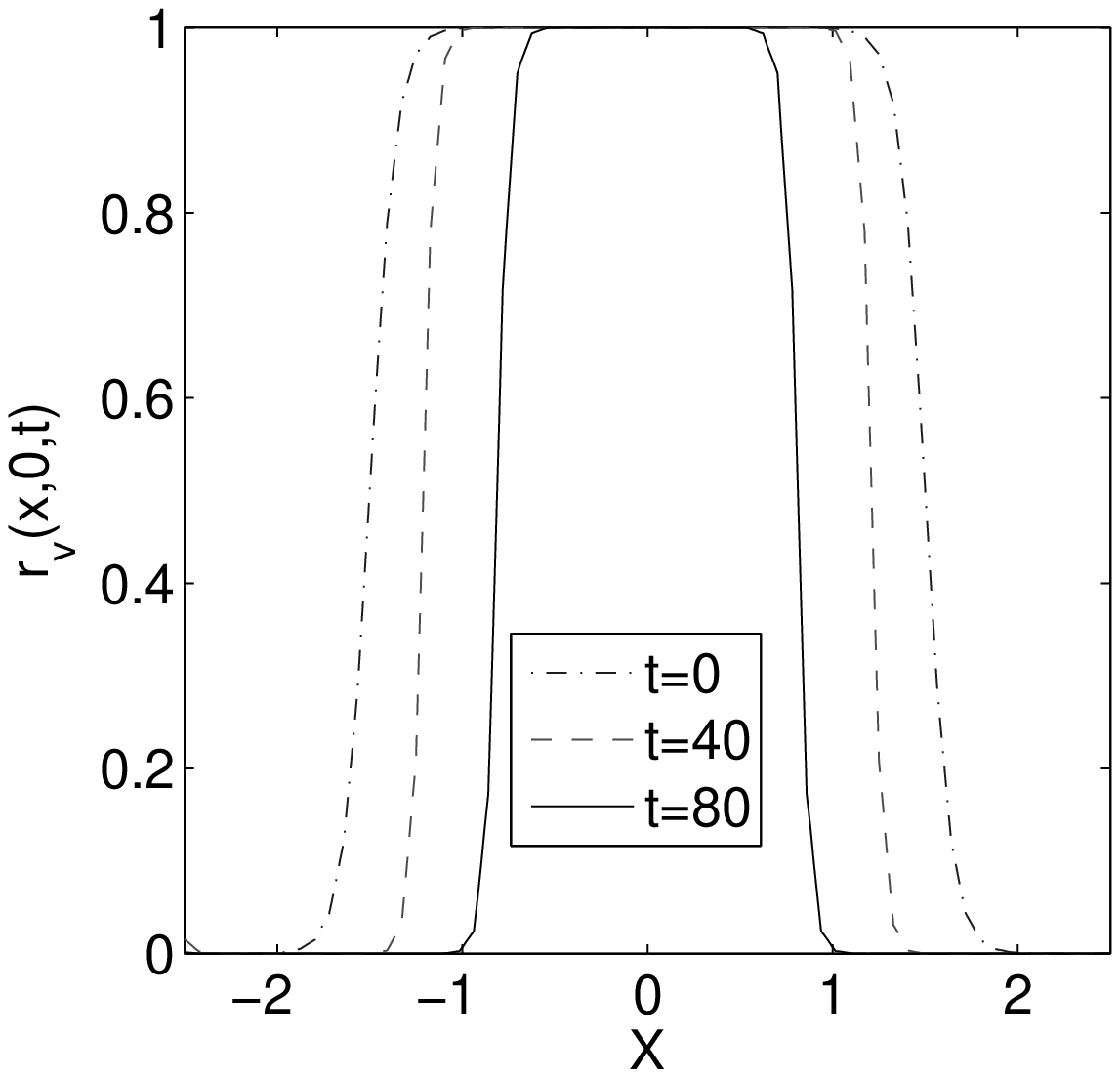}\\
     $(c)$ & $(d)$
   \end{tabular}
   \caption{Moving interface due to the condensation of water vapor. $(a)\,t=0$, $(b)\,t=40$, $(c)\,t=80$, and $(d)$ a comparison of $r_v(x,0,t)$ for $t=0,\,40,\,80$.}
   \label{fig:vpr}
 \end{figure}

The initial circular patch of water vapor $r_v(x,y,0)$ -- as shown in Fig~\ref{fig:vpr}$(a)$ -- would shrink in size due to the condensation. This means that an interfacial wave moves toward the center. Eq~(\ref{eq:vpr}) has been solved for a calm condition. Here, the purpose is to demonstrate a non-oscillatory transport of the air-vapor interface. The numerical solution is presented for dimensionless times, $t=40$ and $t=80$ in Fig~\ref{fig:vpr}$(b,c)$. For clarity, profiles of $r_v(x,0,t)$ along the line $y=0$ is computed, and compared for $t=0,\,40,\,80$ in Fig~\ref{fig:vpr}$(d)$, showing no visible oscillations, as expected. %There is no visible oscillations in the solution, which supports the approximation property discussed with Fig~\ref{fig:wf}.  
\Add{The continuous sharp interfaces are resolved by the proposed method with a finite number of basis using eq~(\ref{eq:mra}). Note the advantage of the present method for approximating a sharp interface is understood from the example shown in Fig~\ref{fig:wf}. For this example, the dynamics of the moving interface is simulated where the proposed discretization method controls dissipation and dispersion errors so that the wave moves without having numerical oscillations and there are no noticeable amplitude errors.} 

This test clearly demonstrates the desired performance of the method. \Add{Authors also note the challenges of capturing a moving sharp interface with classical higher order numerical methods, which is also discussed by~\citet{Tannehill97} using several examples.} We will demonstrate and discuss further on the solution of~(\ref{eq:vpr}) after demonstrating some examples with cyclone interactions.

\section{Binary interaction of two tropical cyclones}\label{sec:bi}
In this section, we study the deformation of concentric vortex structures when two tropical cyclones interact each other. \Add{Here, by a `cyclone' we mean a simplified vortex pattern of a cyclone (see~\cite{Kundu,Moon2010,Prieto2003}).} One objective of this study is to understand how accurately such interactions can be simulated using the present numerical model. We analyze the evolution of two cyclones under the $f$-plane approximation, and compare the results with simplified vortex dynamics and previously reported simulations. %For all such simulations, we have used a fixed Rossby number $\mathcal Ro=10$, and a fixed $p=3$ to get a $6$~th order accurate spatial trunction error.
\subsection{Initial condition}
The velocity field of a tropical cyclone may be modeled with a flow of solid body rotation embedded in nearly irrotational flow. A mathematical form for the complex potential of such a model of cyclone is~\cite{Kundu} 
$$
\mathbf w = \frac{\Gamma i}{2\pi r}\left(1-e^{-r^2/a^2}\right),
$$
were $\Gamma$ is the circulation along a closed path. Clearly,
if $r < a$, we have the flow like a solid body rotation% 
$$u_{\theta} = \frac{\Gamma r}{2\pi a^2},\quad u_r=0$$ 
with a peak wind at $r=a$,
and for $r > a$, we have the irrotational vortex
$$u_{\theta} = \frac{\Gamma }{2\pi r},\quad u_r=0.$$ 
This choice of initial condition is qualitatively equivalent to several other approaches with a slightly modified $\mathbf w$, which are discussed in refs~\cite{Moon2010,Kuo2004,Prieto2003,Nomura2007,Dritschel92}.
\subsection{A validation with Fujiwhara interaction of two convective vortices}
Basic physical mechanisms of binary interactions between cyclones in the form of a complete merger is strongly correlated with the stratified turbulence {\em via} the synoptic scale cascade of kinetic energy of cyclones where the enstrophy propagates downscale in the form of vorticity filaments~\cite{Kundu,Kuo2004}. Thus, the overall quality of the numerical simulation depends largely on how accurately this enstrophy cascade has been modeled. \Add{To test the above hypothesis, we have setup a numerical simulation based on the parameters from a previously analyzed set of satellite data~\cite{Davis2007}, have compared the results with that of the experiments of~\citet{Fujiwhara21}, and finally, we have compared our results with that of a DNS model~\cite{Nomura2007}.}

\subsubsection{Observational study of atmospheric vortices.}
A \Add{mesoscale convective vortex (MCV) is a low-pressure system within a thunderstorm that pulls winds into a circling pattern, or vortex.}
\citet{Davis2007} documented observational data of a spectrum of MCVs with characteristic length scales between $50$~km and $200$~km. In principle, the assumptions used by~\citet{Davis2007} to model their data are similar to what we have adopted. Thus, the simulation in this section help understand the characteristic interaction between MCVs.

\subsubsection{Comparison with the experimental investigation of Fujiwhara.}
Based on a series of laboratory experiments in a water tank by~\citet{Fujiwhara21,Fujiwhara23,Fujiwhara31}, a theory on the interaction between thunderstorms was developed. This theory is fully valid with atmospheric observations, for example, the interaction between hurricanes Gil and Henriette. As it was observed in the water tank experiment~\cite{Fujiwhara21}, if two cyclones are in a close proximity of each other, they start rotating one another, and depending on various physical situations, both cyclones merge into one. \Add{Thus, present simulations with a simplified physics of MCVs help validate the proposed numerical model.} 

\subsubsection{Comparison with direct numerical simulation (DNS).}
There are two representative DNSs on the Fujiwhara interactions. One is done by~\citet{Dritschel92} who investigated the inviscid dynamics using a Lagrangian numerical method. The other DNS is reported by~\citet{Nomura2007} who investigated the interaction of vortices in a viscous fluid. Our numerical model reproduces results from both of these DNS database. For a more quantitative understanding, we now present a specific comparison result.

\subsubsection{Present simulation of MCVs.}
A comparison is done with the direct numerical simulation of the vortex merger presented by~\citet{Nomura2007}.

\begin{figure}
  \centering
  \begin{tabular}{ccc}
    \includegraphics[height=3.75cm]{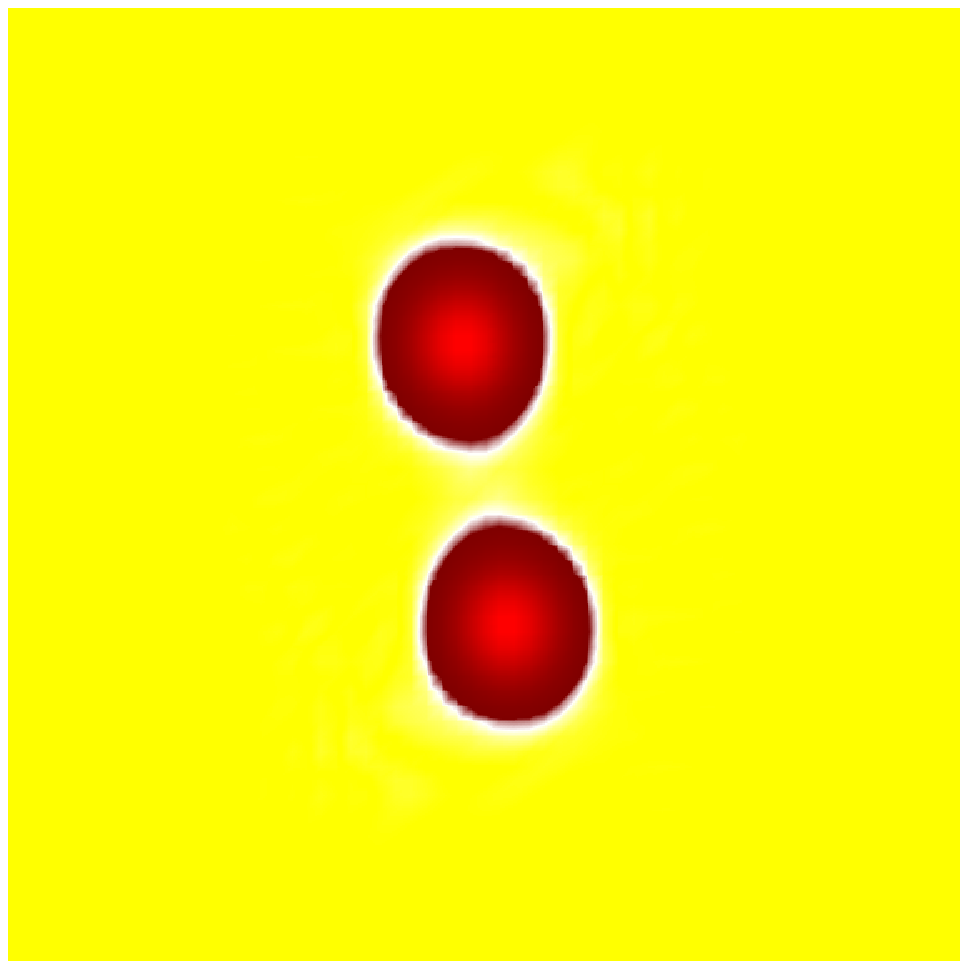}&
    \includegraphics[height=3.75cm]{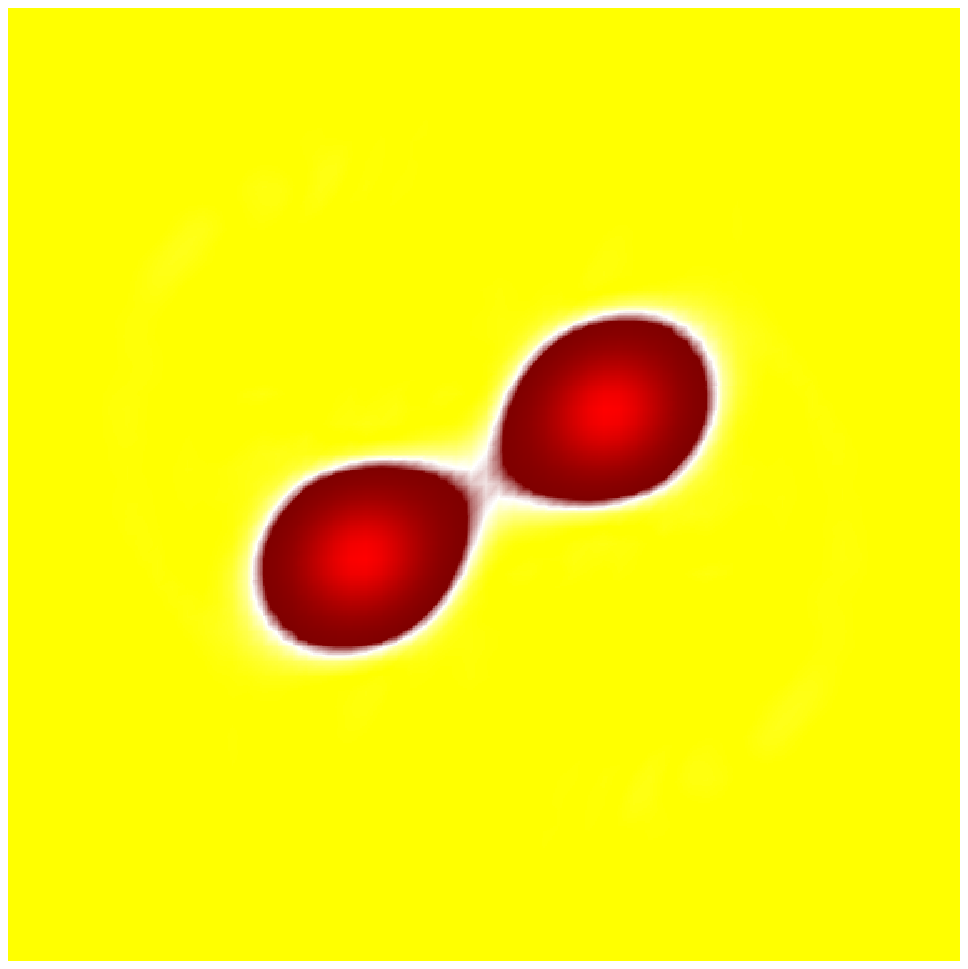}&
    \includegraphics[height=3.75cm]{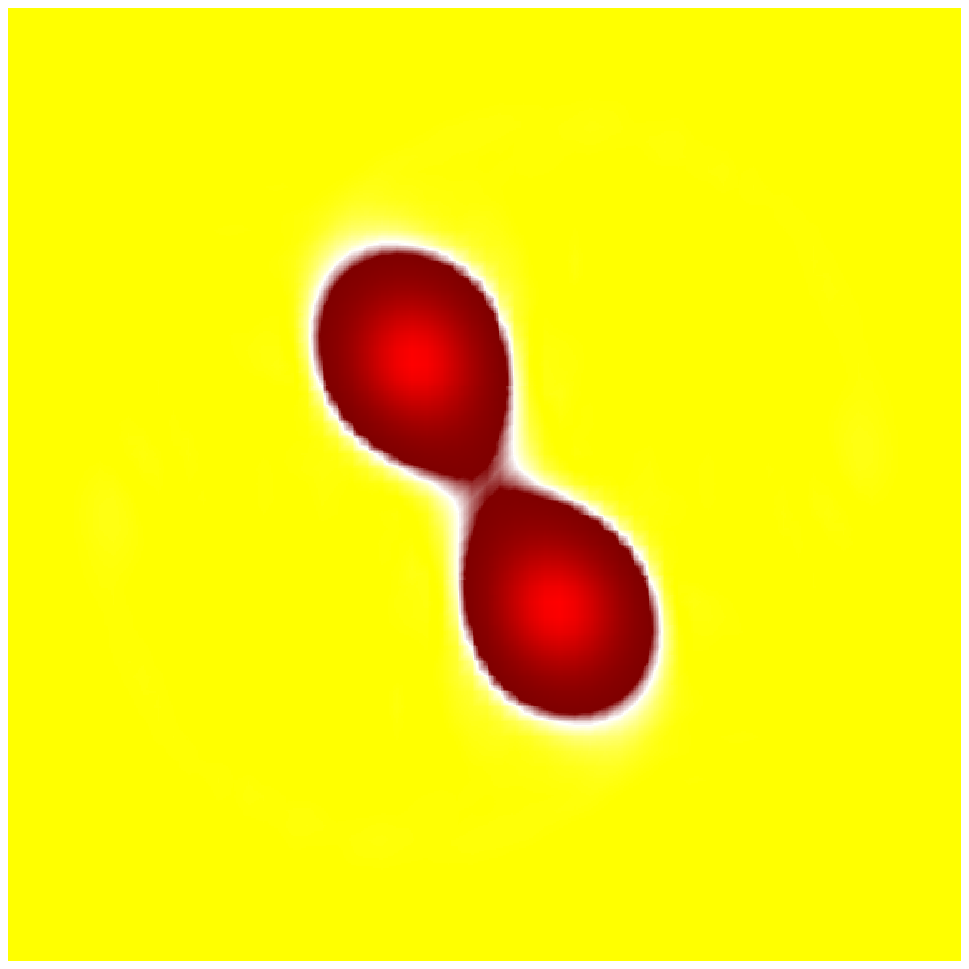}\\
    $(a)~ t^*=0.26 (0.26)$ & $(b)~ t^*=0.55 (0.55)$ & $(c)~t^*=0.78 (0.78)$\\
    \includegraphics[height=3.75cm]{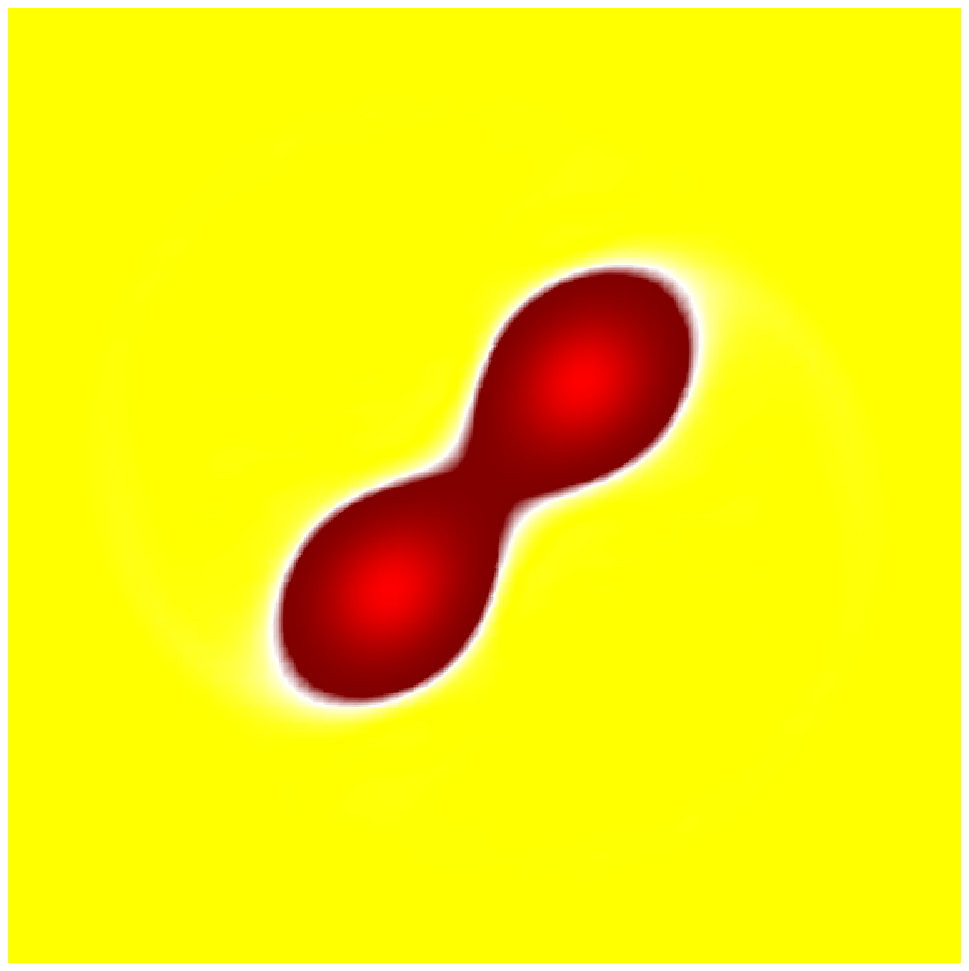}&
    \includegraphics[height=3.75cm]{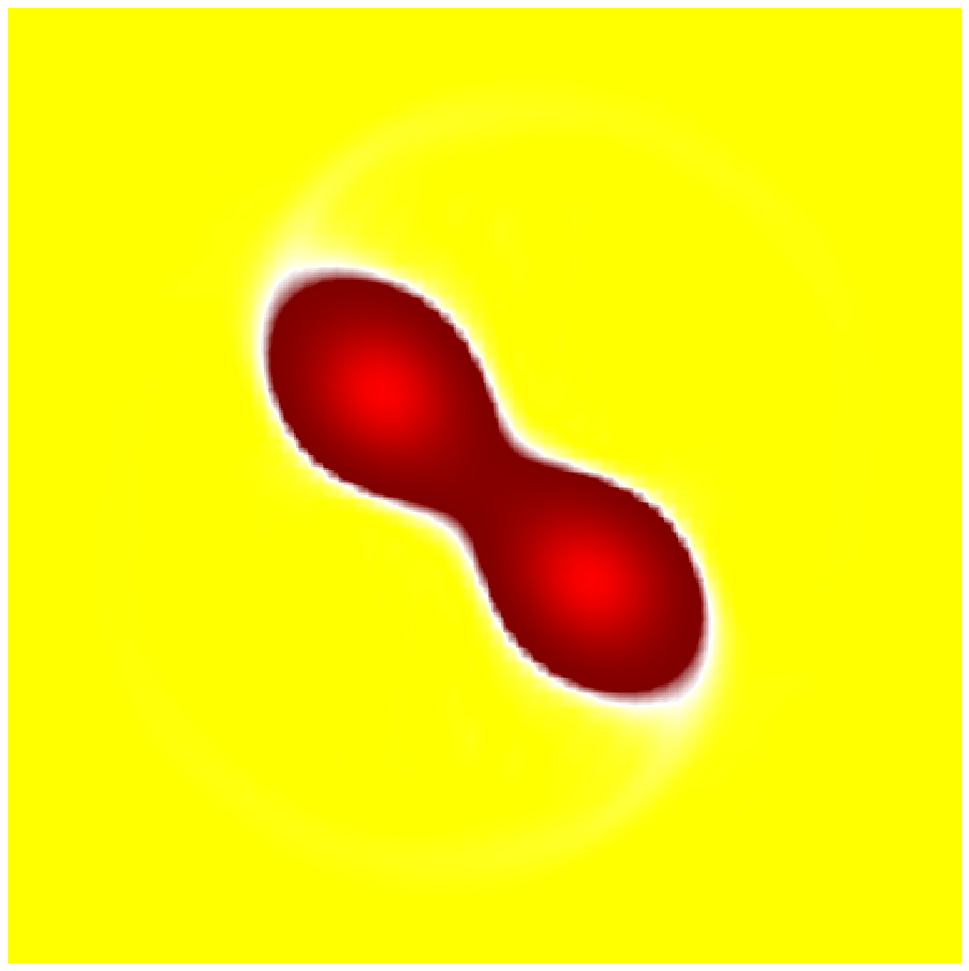}&
    \includegraphics[height=3.75cm]{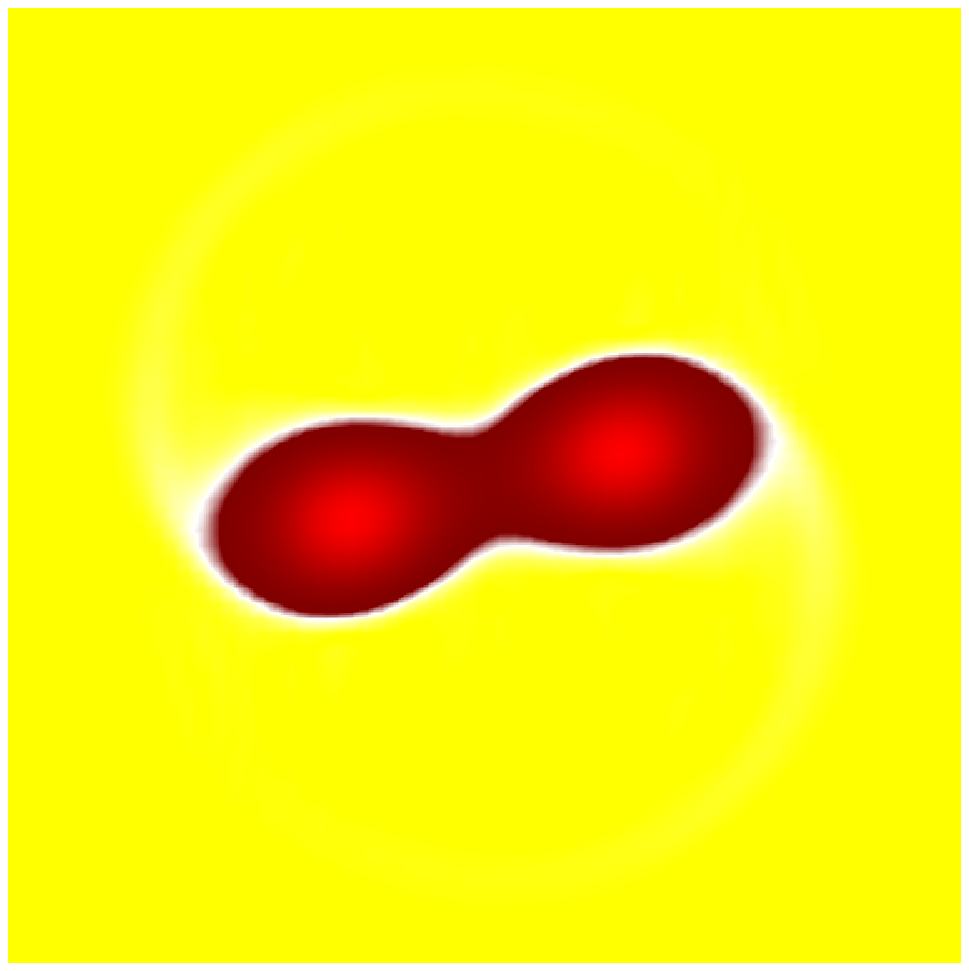}\\
    $(d)~ t^*=1.05 (1.07)$ & $(e)~ t^*=1.28 (1.31)$ & $(f)~t^*=1.42 (1.45)$\\
    \includegraphics[height=3.75cm]{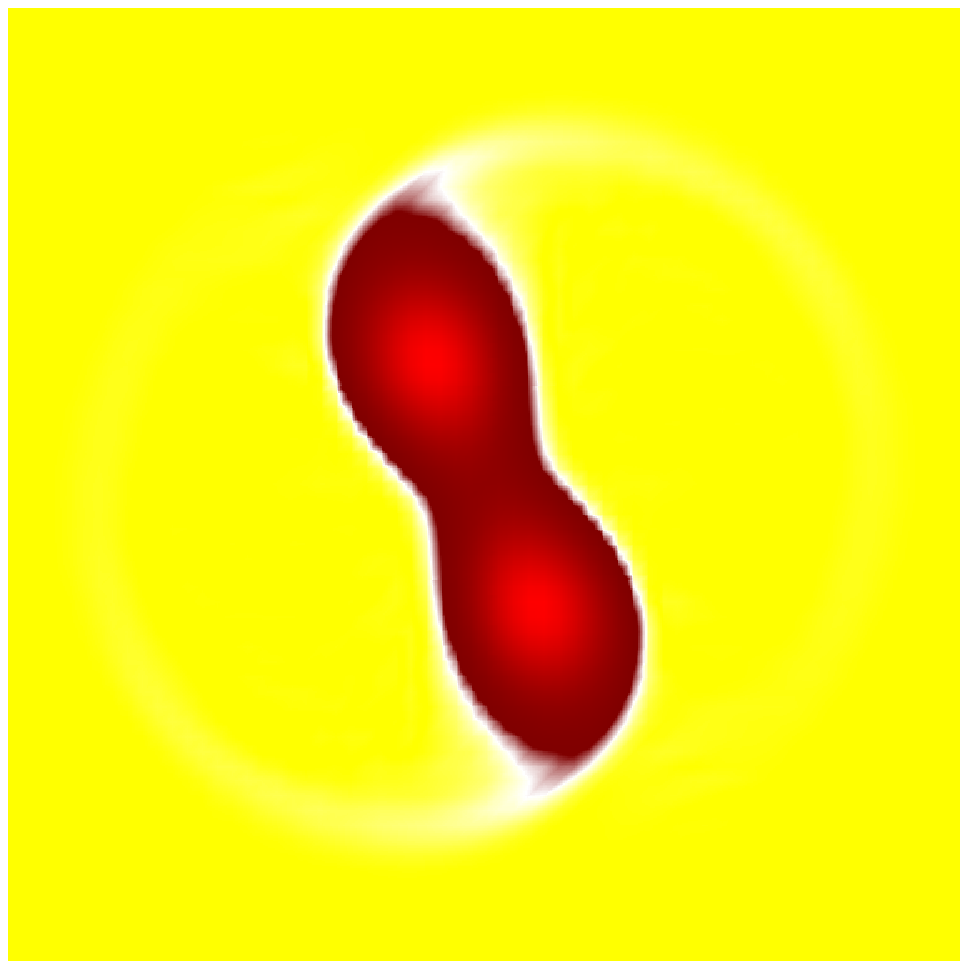}&
    \includegraphics[height=3.75cm]{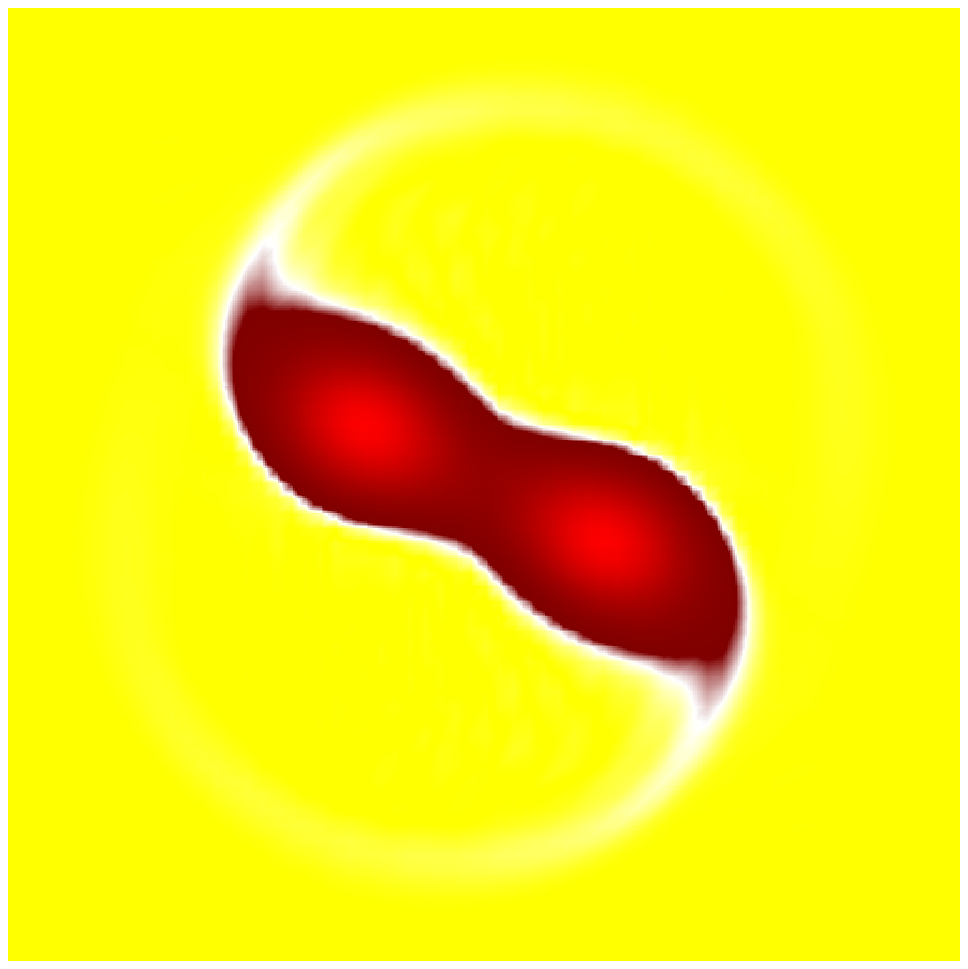}&
    \includegraphics[height=3.75cm]{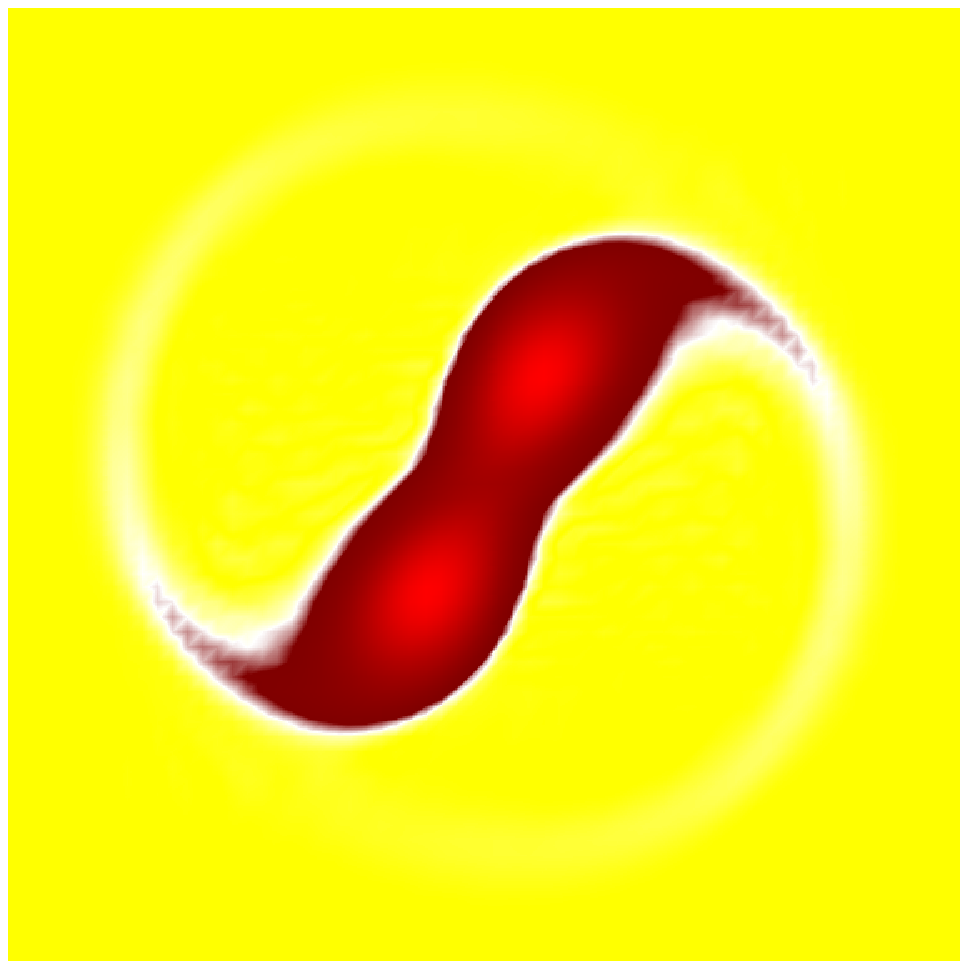}\\
    $(g)~ t^*=1.65(1.69)$ & $(h)~ t^*=1.75(1.78)$ & $(i)~t^*=1.93 (1.97)$\\
    \includegraphics[height=3.75cm]{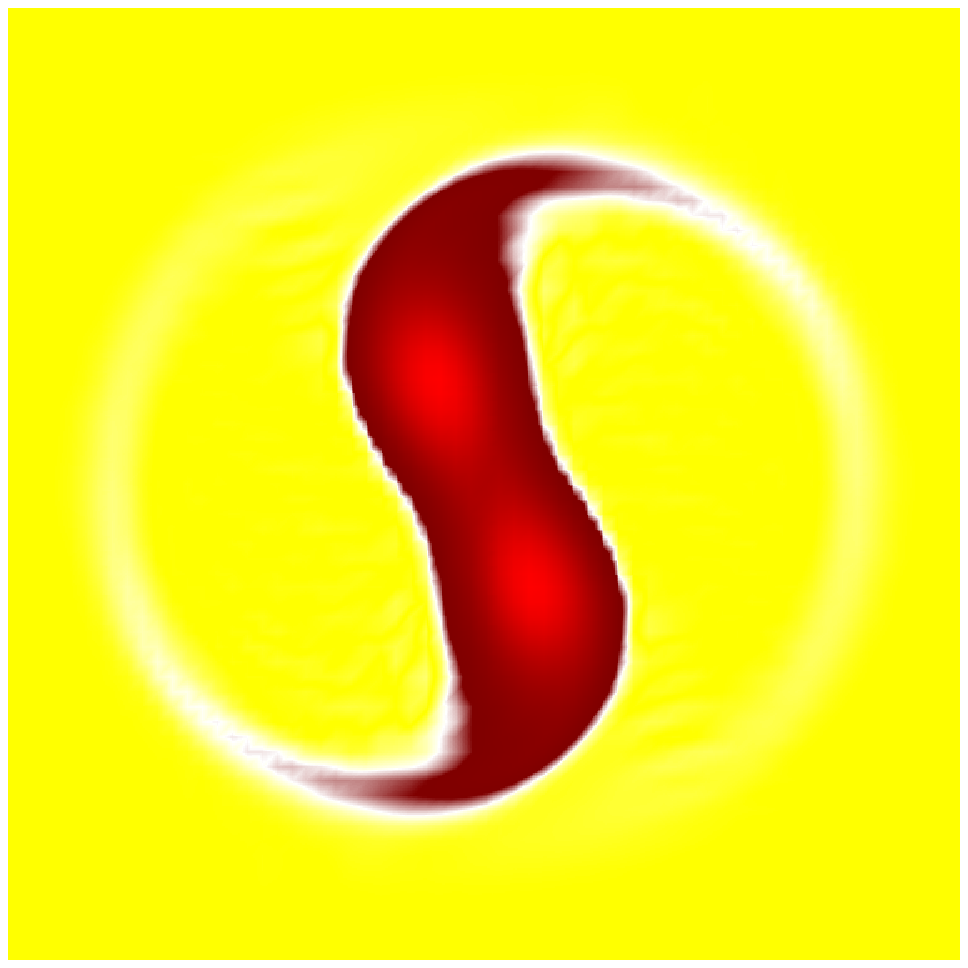}&
    \includegraphics[height=3.75cm]{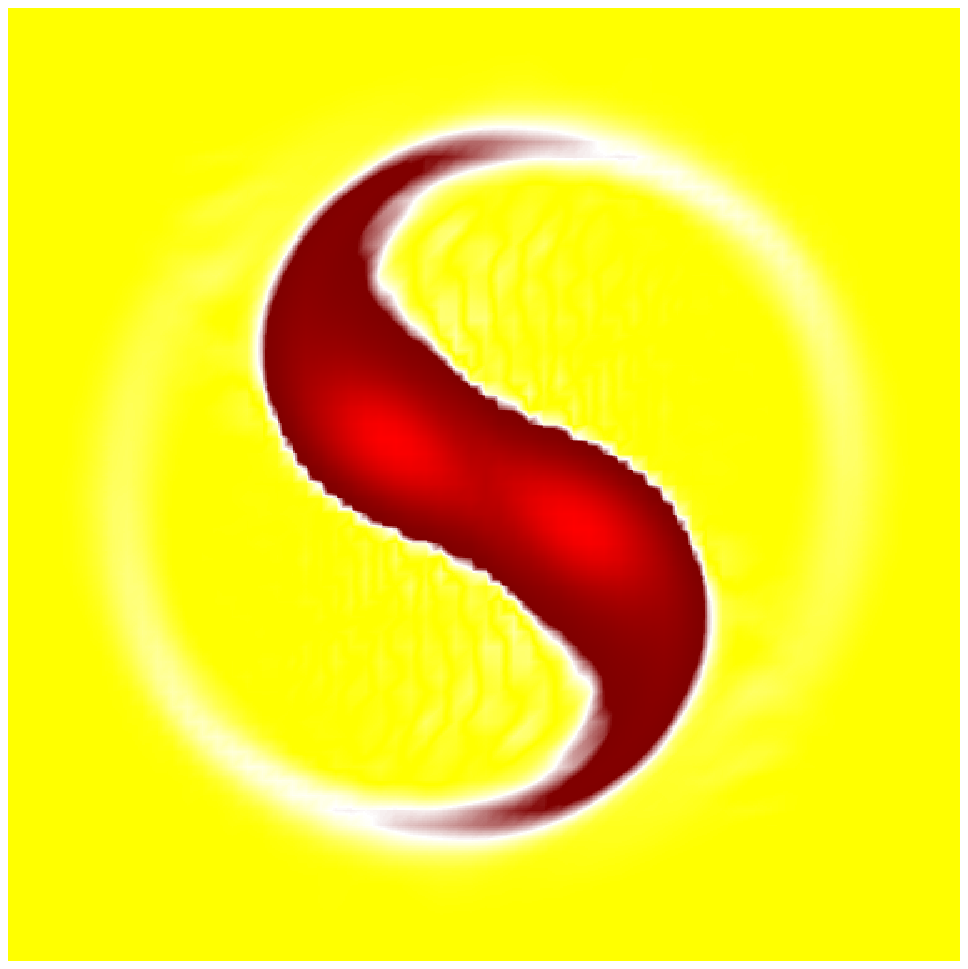}&
    \includegraphics[height=3.75cm]{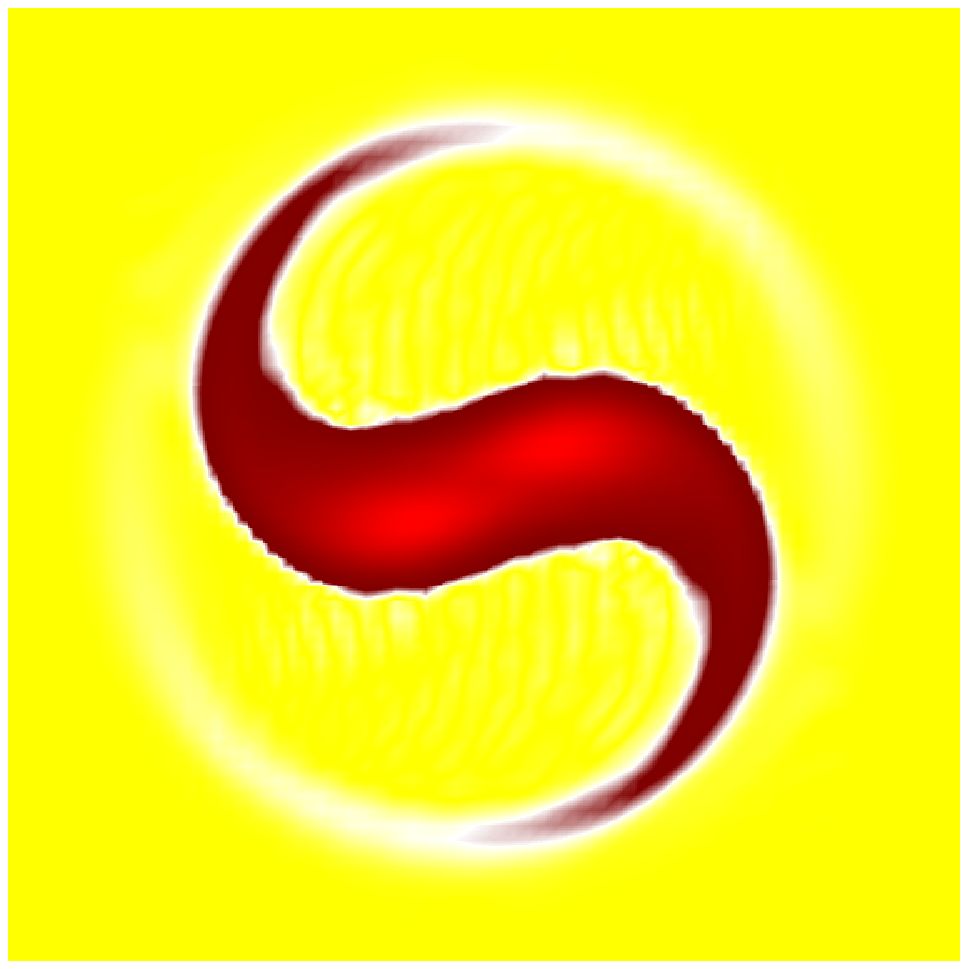}\\
    $(j)~ t^*=2.03 (2.05)$ & $(k)~ t^*=2.10(2.12)$ & $(l)~t^*=2.17(2.16)$\\
  \end{tabular}
  \caption{Fujiwhara interaction between two MCVs. The dimensionless time $t^*$ is compared with that of~\citet{Nomura2007}, where the value within the parenthesis is the reference value.}
  \label{fig:fbi}
\end{figure}

For the present simulation, the initial vorticity of each MCVs is $\omega_0 = 2.5\times 10^{-4}~\hbox{s}^{-1}$, and the initial radii of both MCVs are kept equal ($a=R_1=R_2$) where $a$ is chosen between $50$~km and $200~\hbox{km}$ ({\em e.g.}~\cite{Davis2007}). The range of the radii is chosen according to the data from Table~1 of~\cite{Davis2007}, and a set of $4$ simulations ($a=50$~km, $100$~km, $150$~km, $200$~km) with a fixed separation distance $b=670$~km is considered to analyze the Fujiwhara interaction. For the results presented in Fig~\ref{fig:fbi}, the initial distance between the centers of two MCVs is $670$~km and $a=R_1=R_2=100$~km. Thus, $a/b=0.149$ which is very close to the value $0.157$ used by~\citet{Nomura2007}. It is useful to define the distance $\Delta_c = b-R_1-R_2$ between the edges of two MCVs, where the wind takes a maximum value. The separation ratio -- as defined in~\cite{Kuo2004} -- is $\Delta_c/R=4.7$. Note that $\Delta_c/R$ is about twice as large, and $\omega_0$ is about twice as small as what was used by~\citet{Kuo2004}. Thus, the parameters are selected in a way that the present simulation is an idealization of known satellite observations~\cite{Prieto2003,Kuo2004,Davis2007}.

The turbulent eddy viscosity, $\nu_{\tau}=250~\hbox{m}^2/\hbox{s}$, is estimated with the Smagorinsky model~(\ref{eq:tau}) where the magnitude of the rate of deformation is approximately $|S|\sim 2.5\times 10^{-5}$~s$^{-1}$. In other words, the rate of energy dissipation is about $2\times 10^{-2}$~m$^2$/s$^3$. The initial velocity is set to $U=12.5$~m/s ($45$~km/h)~\cite{Davis2007}. The Reynolds number $\mathcal Re=UR/\nu_{\tau}$ takes approximately a value $5\,000$, and thus, the present simulation is qualitatively comparable to the observed MCVs of~\citet{Davis2007} (in the sense of length and time scales), as well as to the DNS of simplified vortex dynamics of~\citet{Nomura2007} (in the sense of vortex merger). Fig~\ref{fig:fbi} shows the time evolution of the vorticity field. A visual comparison indicates an excellent agreement between the present result with that of~\cite{Nomura2007} (their Figure 1). 

\begin{figure}
  \centering
  \includegraphics[height=5cm]{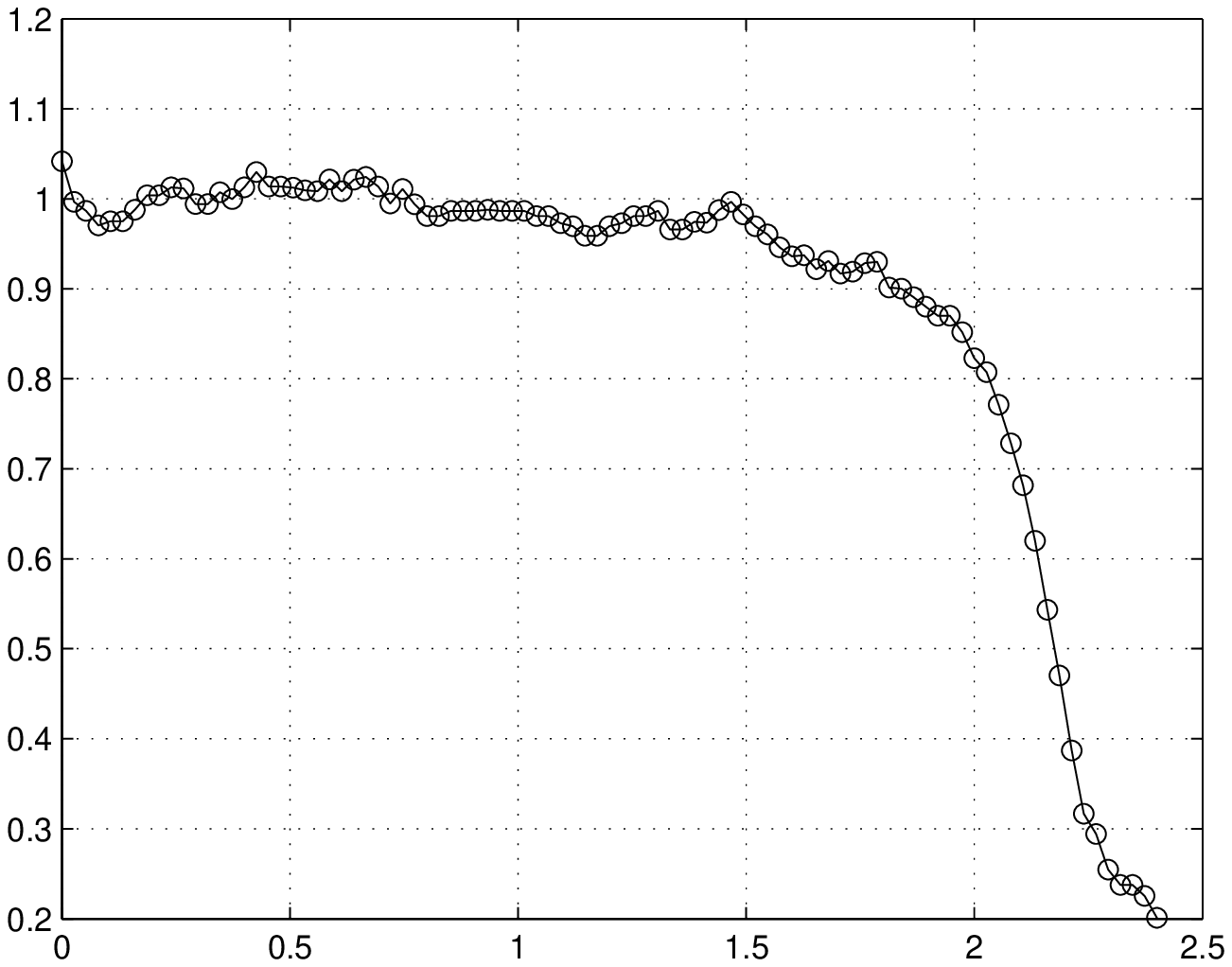}
  \caption{Normalized distance (along the vertical axis) between the center of two MCVs as a function of normalized time (along the horizontal axis).}
  \label{fig:dfbi}
\end{figure}

To aid a more quantitative comparison with the DNS, let us compute the characteristic convective time scale, $t_c$. Based on the rotational period of a corresponding point vortex system, $t_c = 2\pi b^2/\Gamma$. In Fig~\ref{fig:fbi}, the dimensionless time, $t^*=t/t_c$, is displayed for each plot, where the value reported by~\citet{Nomura2007} is also given in parenthesis.  An excellent agreement on the rotational period of the present simulation with that of the DNS clearly justifies the accuracy of the present time integration. 
 In order to provide a further comparison, we have displayed the normalized distance $b(t^*)/b(0)$ between the center of two cyclones in Fig~\ref{fig:dfbi}. Clearly, the computations are in \Add{very good} agreement with that of~\cite{Nomura2007} when one compares the present plot in Fig~\ref{fig:dfbi} with the plot~({\em e.g} $\circ$) from Fig~5(a) of~\cite{Nomura2007}. 

Note however, that the highest numerical resolution for the results in Fig~\ref{fig:fbi} is given by $\Delta x=\Delta y = 12.35$~km, which means that there are only $16$ grid points across the core of each vortex ({\em i.e.} core~$\approx 2a$). Coarsening the resolution to $8$ grid points across the vortex core ($\Delta\approx 25$~km) does not affect the calculated value of $t^*$ significantly. \citet{Nomura2007} used a high resolution grid along with $54$ grid points across the vortex core, which resolves the atmospheric microscale ($\Delta\approx 3.5$~km) approximately. % and solved the incompressible Navier-Stokes equation. In meteorology, the phenomena at a length scale~$\sim 2$~km are microscale processes.  
Meteorological simulations typically employ a resolution given by $\Delta\approx 20$-$30$~km where the subgrid scale physics is parameterized~(see {\em e.g.}~\cite{Skamarock2008,Pielke2002}). One does not expect to model the microscale ($<1$-$2$~km) physics sufficiently accurately using a meteorological model at $\Delta\sim 20$~km~\cite{Skamarock2008}. Thus, the present LES approach where resolved scales are complemented with a parameterization scheme based on the Smagorinsky model provide some useful feedback in this direction, at least for the simulation of interactions between MCVs.

\subsection{Decoupled inner eyewall during the interaction between two cyclones of unequal strength}\label{sec:vref}
Concentric eyewall patterns are commonly observed in the intense hurricanes. The pattern refers to two or more eyewalls nearly concentric to the center, in which deep convection within the primary, inner eyewall of intense tropical cyclones is surrounded by an outer partial or complete ring of deep convection. The inner eyewall may be idealized by the radius of maximum sustained wind. In the case of Fujiwhara interactions, previous numerical simulations revealed that the inner eyewall decouples from the outer eyewall, resulting in concentric ring structures~\cite{Kuo2004,Moon2010}. Here, we study the binary interaction between hurricanes of unequal strength, and demonstrate how the secondary eyewall is robbed of the primary eyewall.

According to the Saffir-Simpson scale, hurricane winds $\le 17$~m/s, $18\hbox{-}32$~m/s, $33\hbox{-}42$~m/s, and $43\hbox{-}49$~m/s represent a tropical depression, tropical storm, category $1$ and category $2$ hurricanes, respectively. 
The interaction between a strong category 2 hurricane at maximum wind speed $U_1=46.8$~m/s with a cyclone of maximum wind speed $U_2$ is simulated, where $U_2$ is a variable parameter. Both cyclones are assumed to have the same core size; {\em i.e.} $R_1=R_2=100$~km, and the separation distance between them is kept fixed at $b=450$~km. 

\begin{figure}
  \centering
  \begin{tabular}{cccc}
     \includegraphics[trim=2cm 0cm 2cm 0cm,clip=true,height=3.5cm]{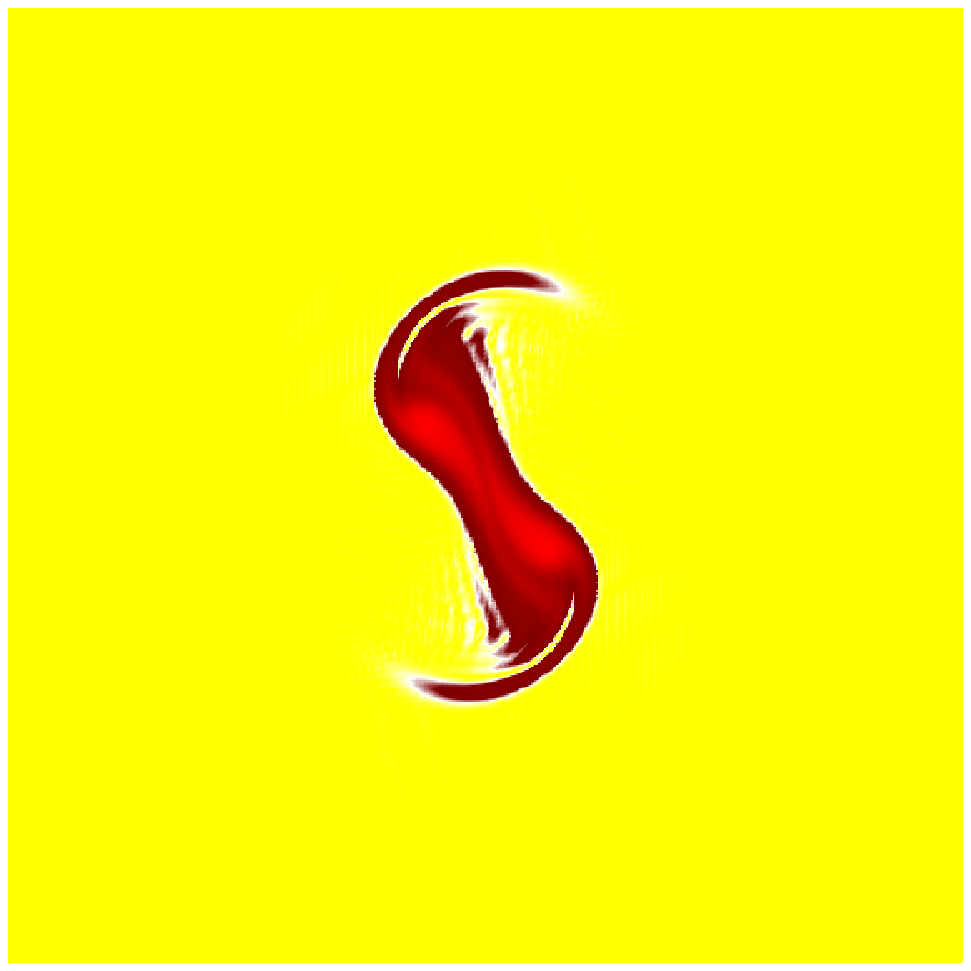}&
     \includegraphics[trim=2cm 0cm 2cm 0cm,clip=true,height=3.5cm]{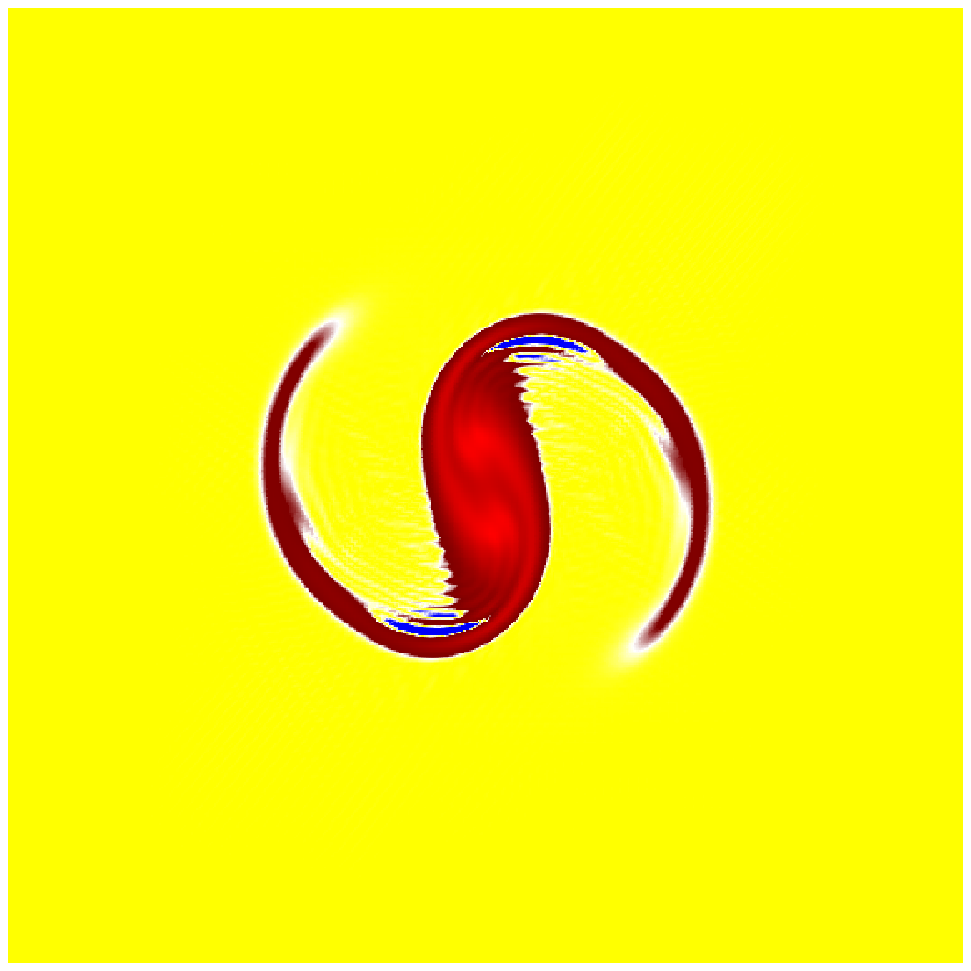}&
     \includegraphics[trim=2cm 0cm 2cm 0cm,clip=true,height=3.5cm]{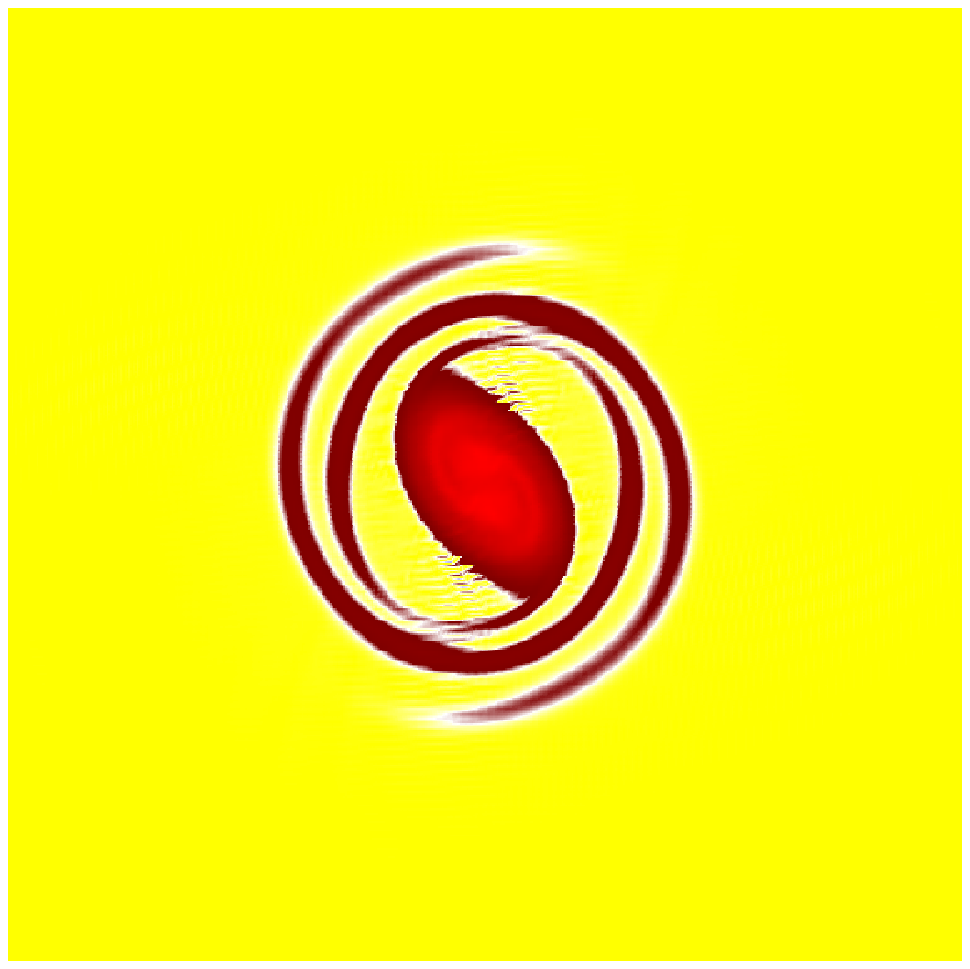}\\
    $U_2=46.8$~m/s, 9.375 h & 18.75 h & 37.5 h\\
    \includegraphics[trim=1cm 0cm 1cm 0cm,clip=true,height=3.5cm]{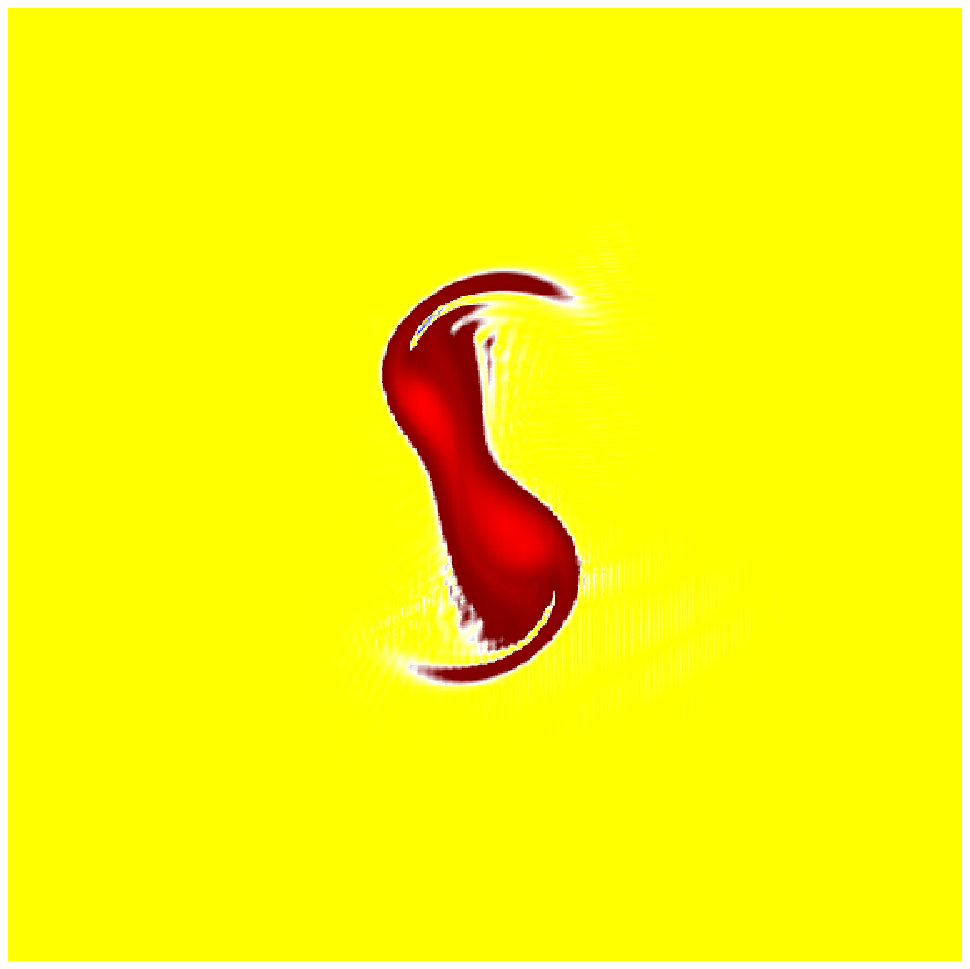}&
    \includegraphics[trim=1cm 0cm 1cm 0cm,clip=true,height=3.5cm]{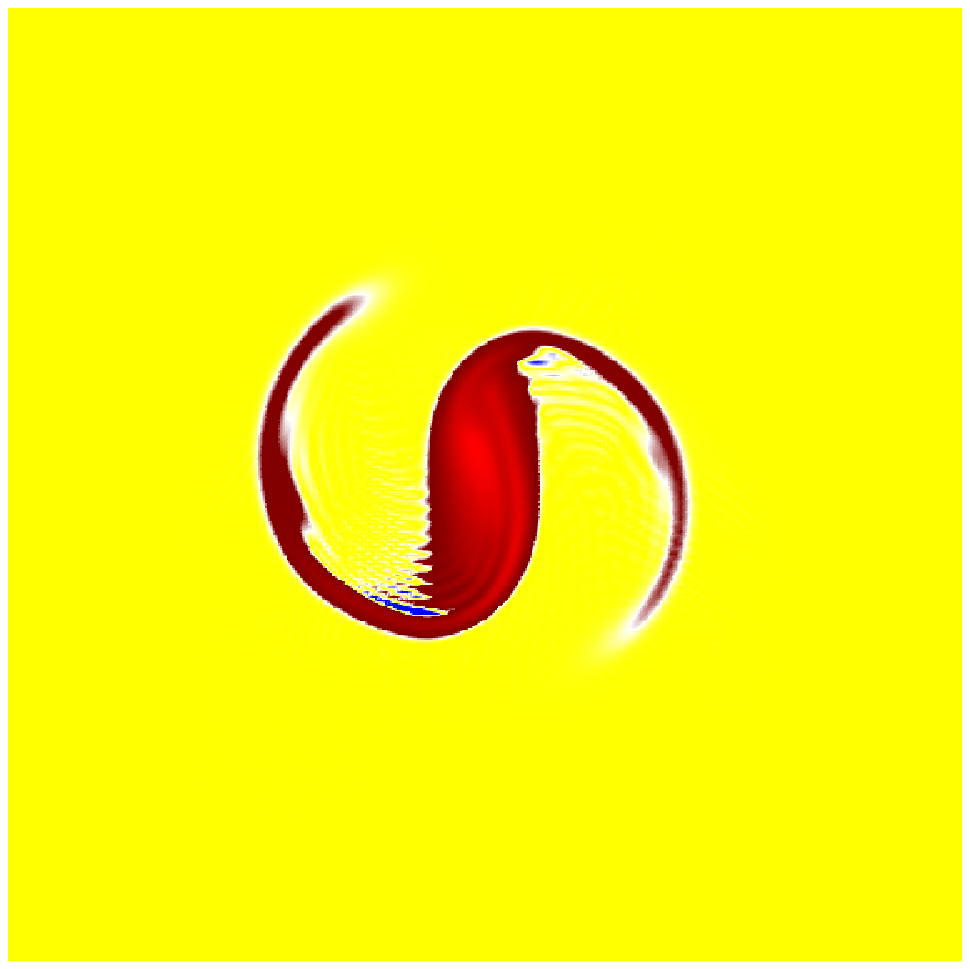}&
    \includegraphics[trim=1cm 0cm 1cm 0cm,clip=true,height=3.5cm]{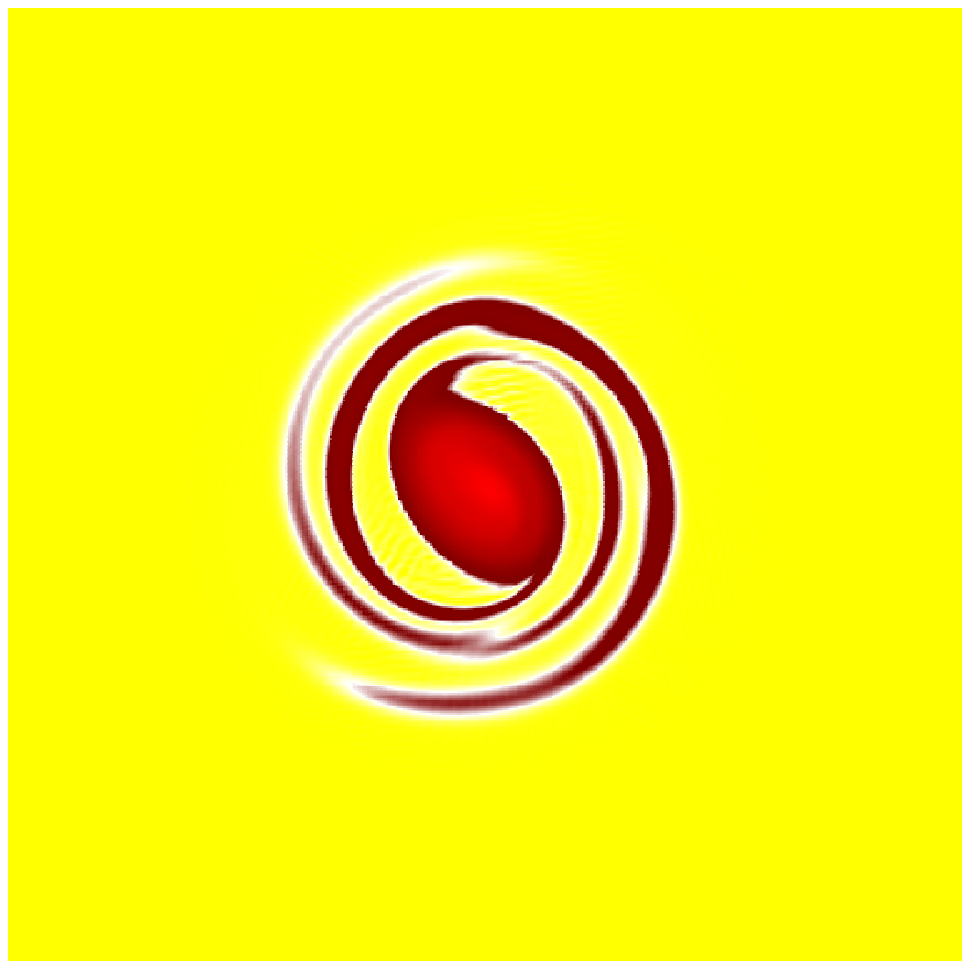}\\
    $U_2=37.5$~m/s, 9.375 h & 18.75 h & 37.5 h\\
    \includegraphics[trim=1cm 0cm 1cm 0cm,clip=true,height=3.5cm]{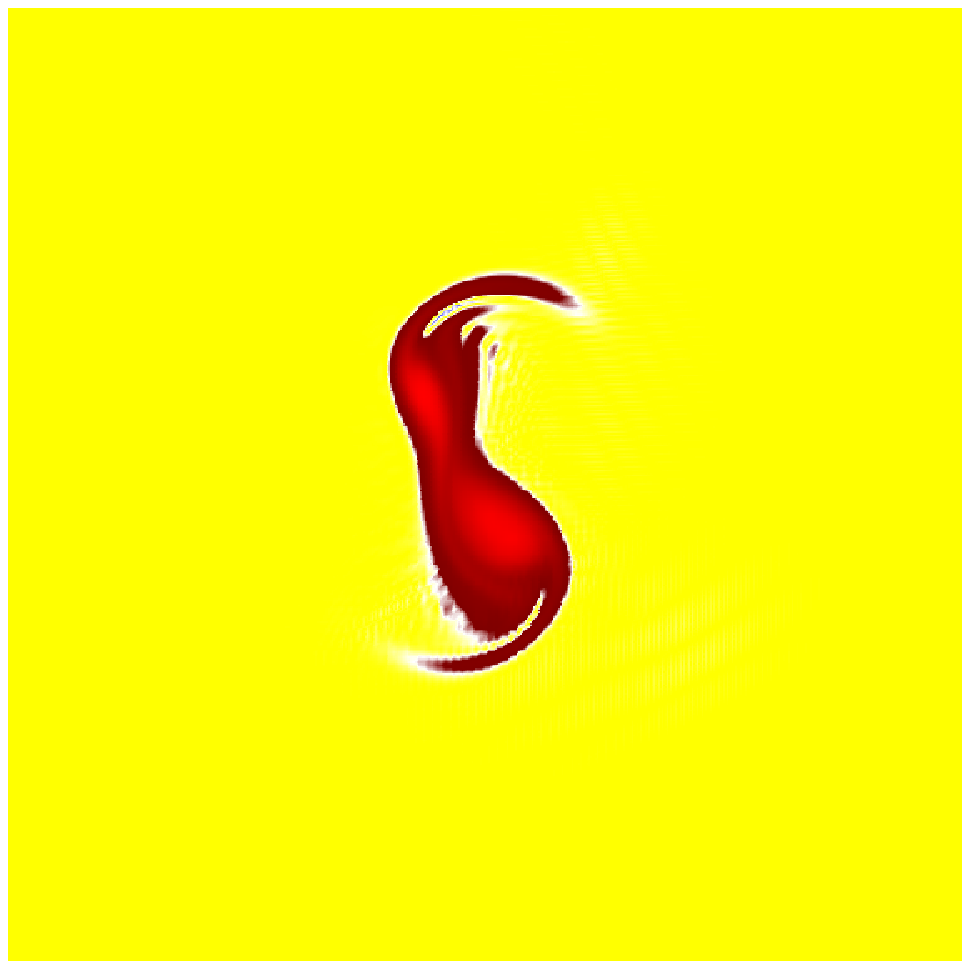}&
    \includegraphics[trim=1cm 0cm 1cm 0cm,clip=true,height=3.5cm]{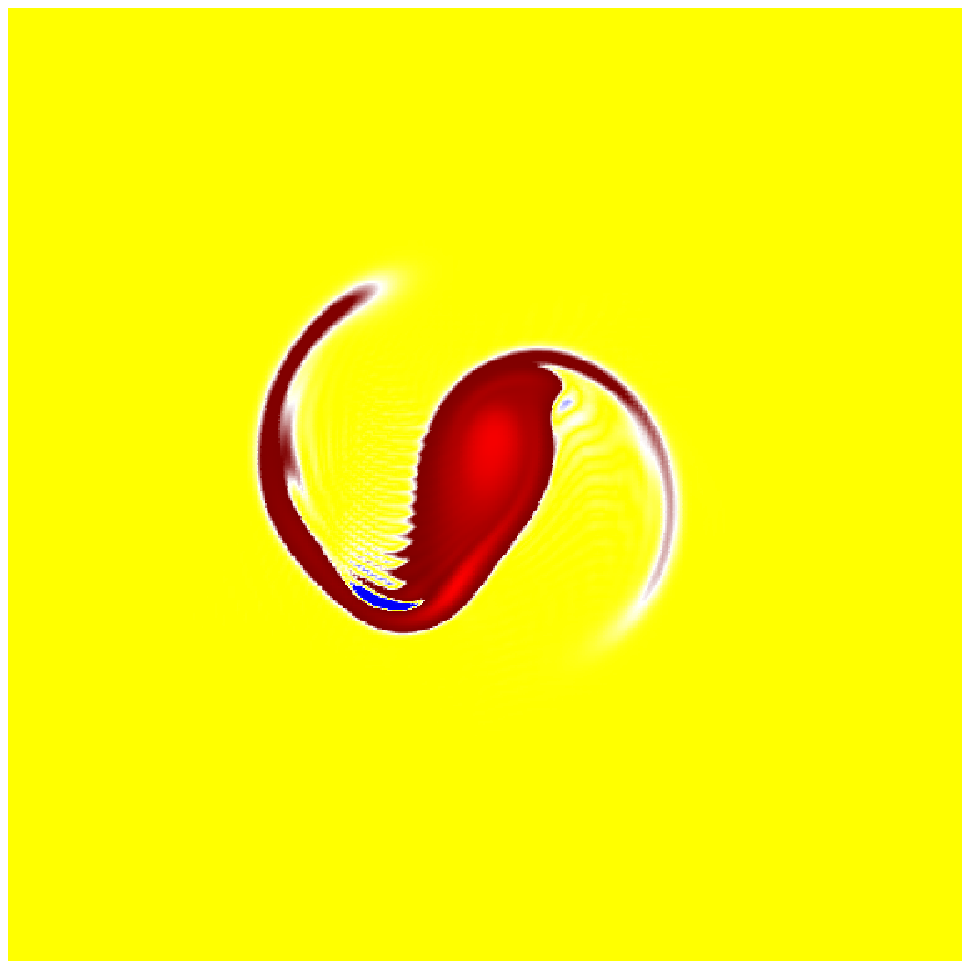}&
    \includegraphics[trim=1cm 0cm 1cm 0cm,clip=true,height=3.5cm]{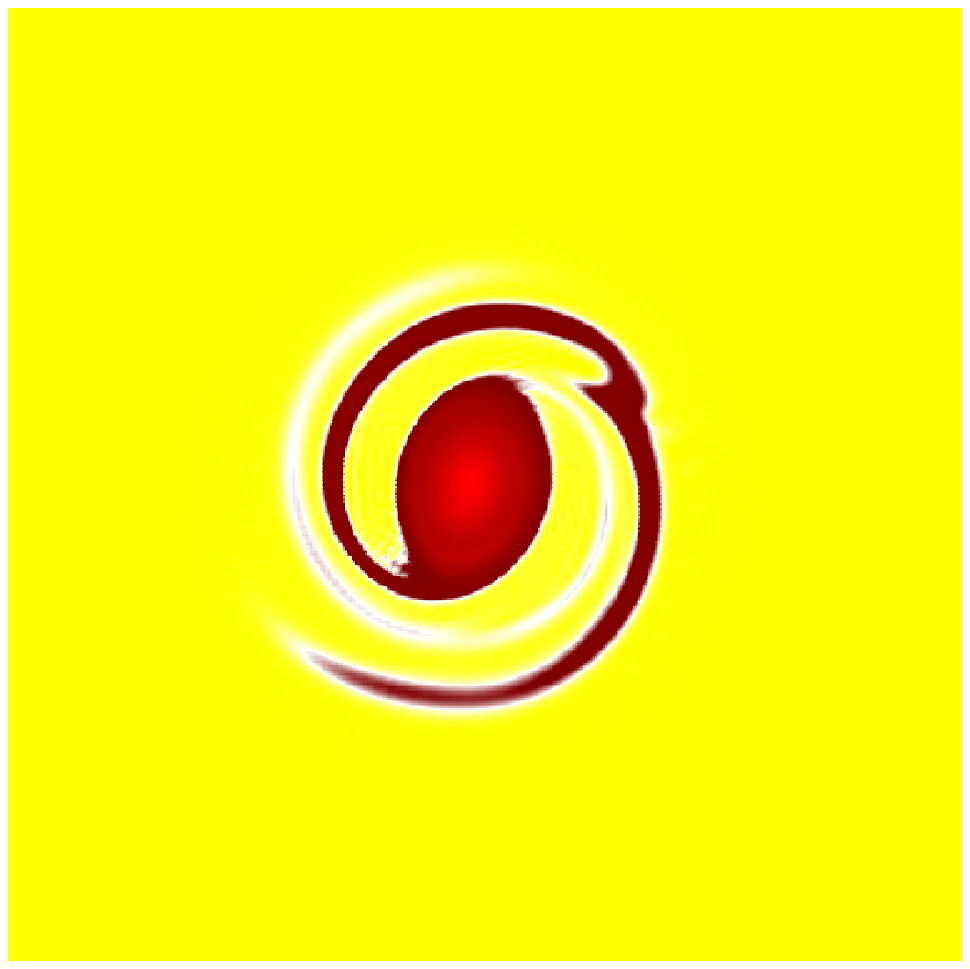}\\
    $U_2=32.8$~m/s, 9.375 h & 18.75 h & 37.5 h\\
    \includegraphics[trim=1cm 0cm 1cm 0cm,clip=true,height=3.5cm]{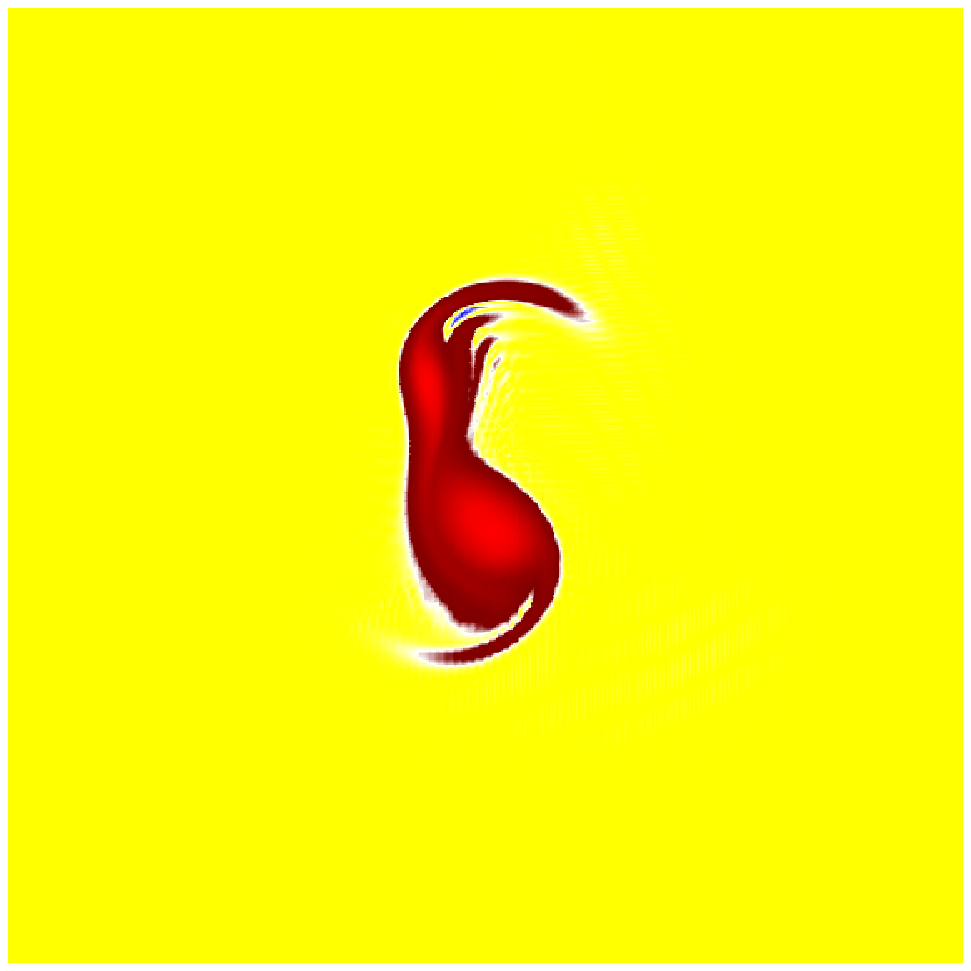}&
    \includegraphics[trim=1cm 0cm 1cm 0cm,clip=true,height=3.5cm]{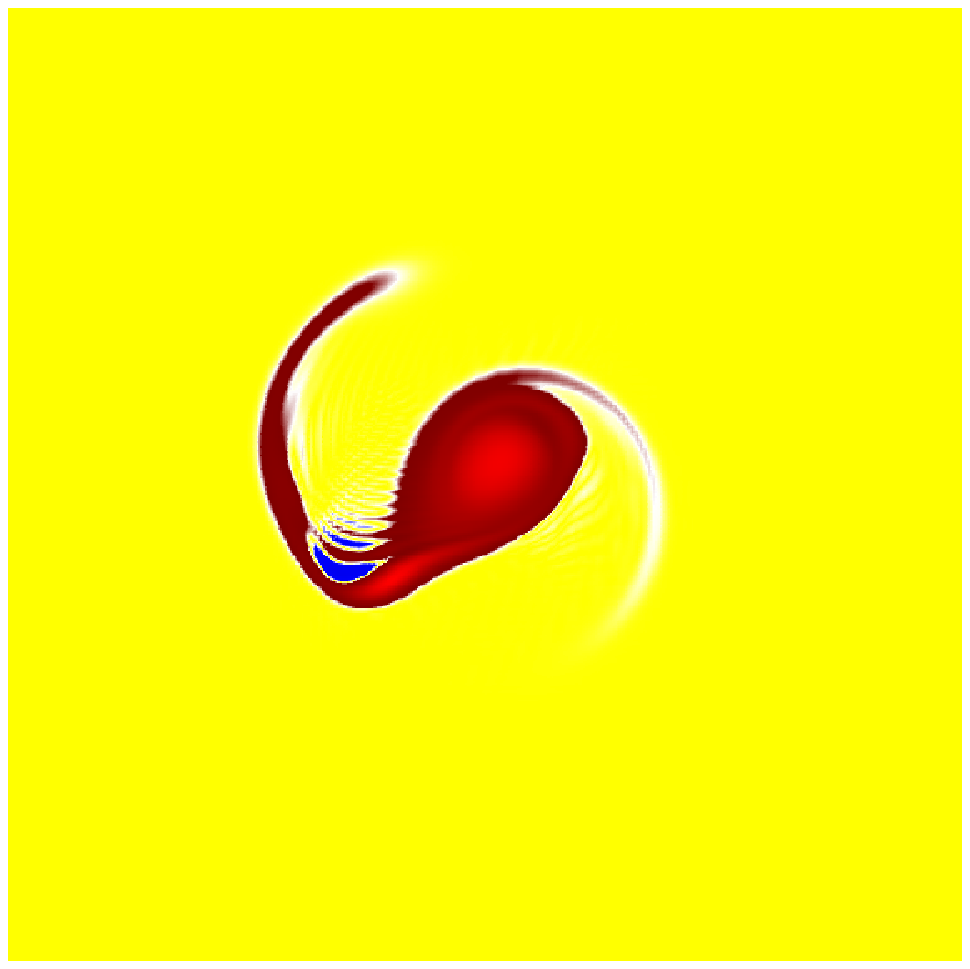}&
    \includegraphics[trim=1cm 0cm 1cm 0cm,clip=true,height=3.5cm]{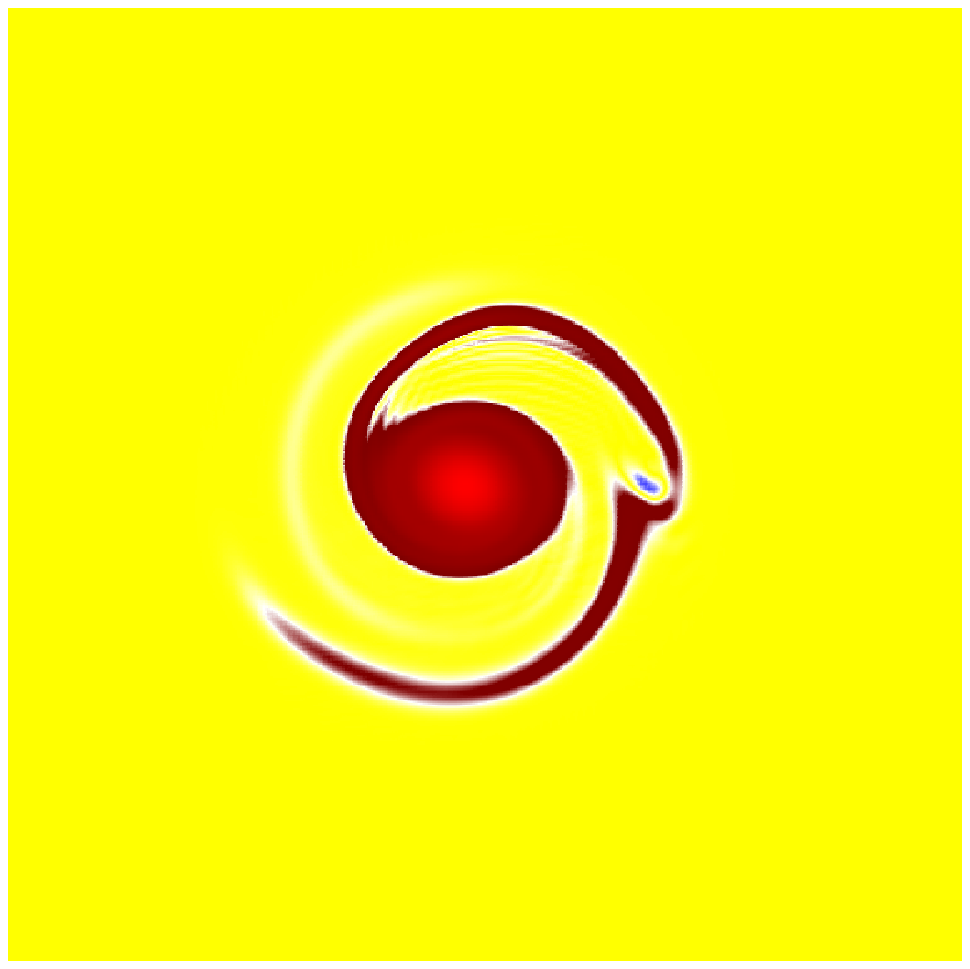}\\
    $U_2=28.1$~m/s, 9.375 h & 18.75 h & 37.5 h\\
    \includegraphics[trim=1cm 0cm 1cm 0cm,clip=true,height=3.5cm]{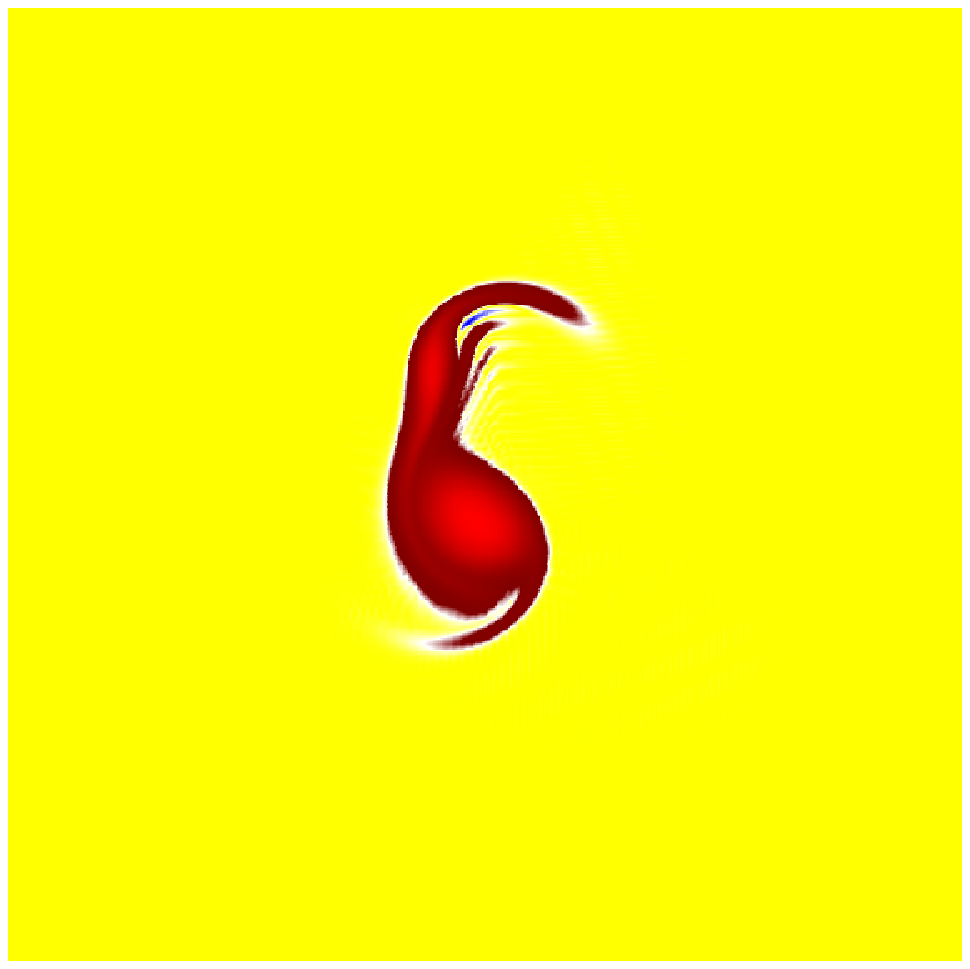}&
    \includegraphics[trim=1cm 0cm 1cm 0cm,clip=true,height=3.5cm]{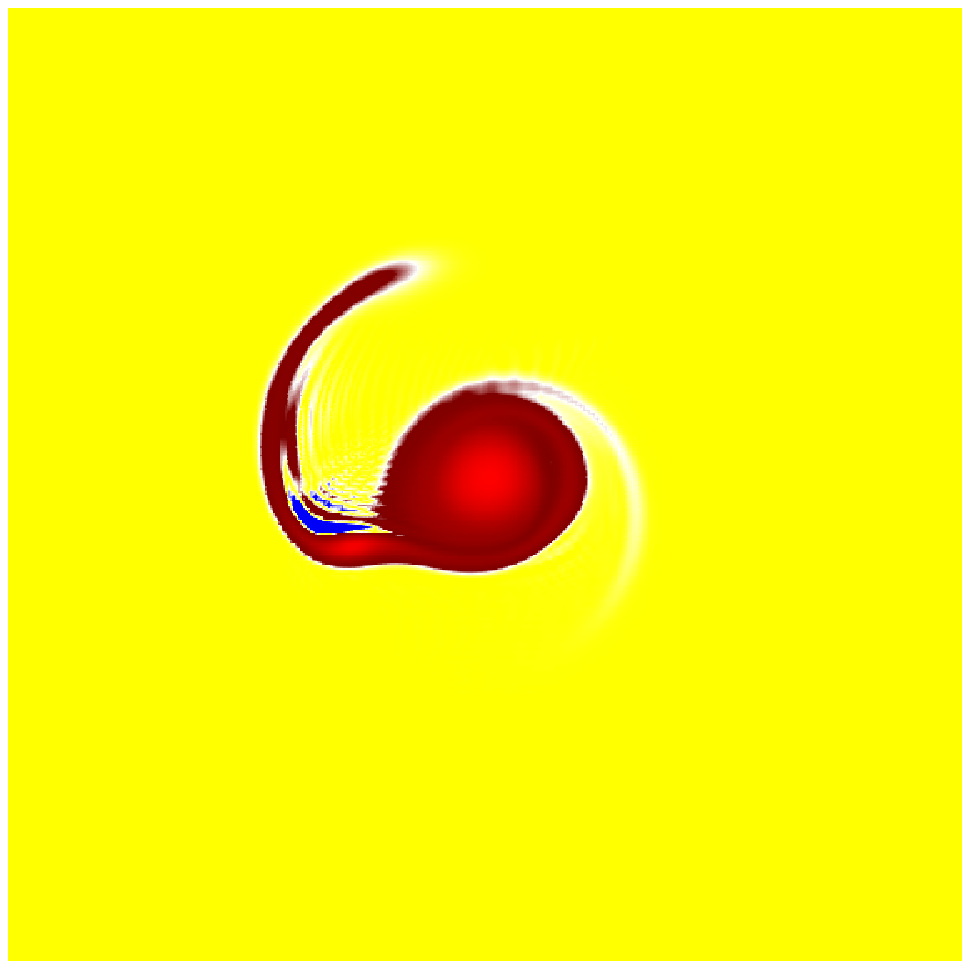}&
    \includegraphics[trim=1cm 0cm 1cm 0cm,clip=true,height=3.5cm]{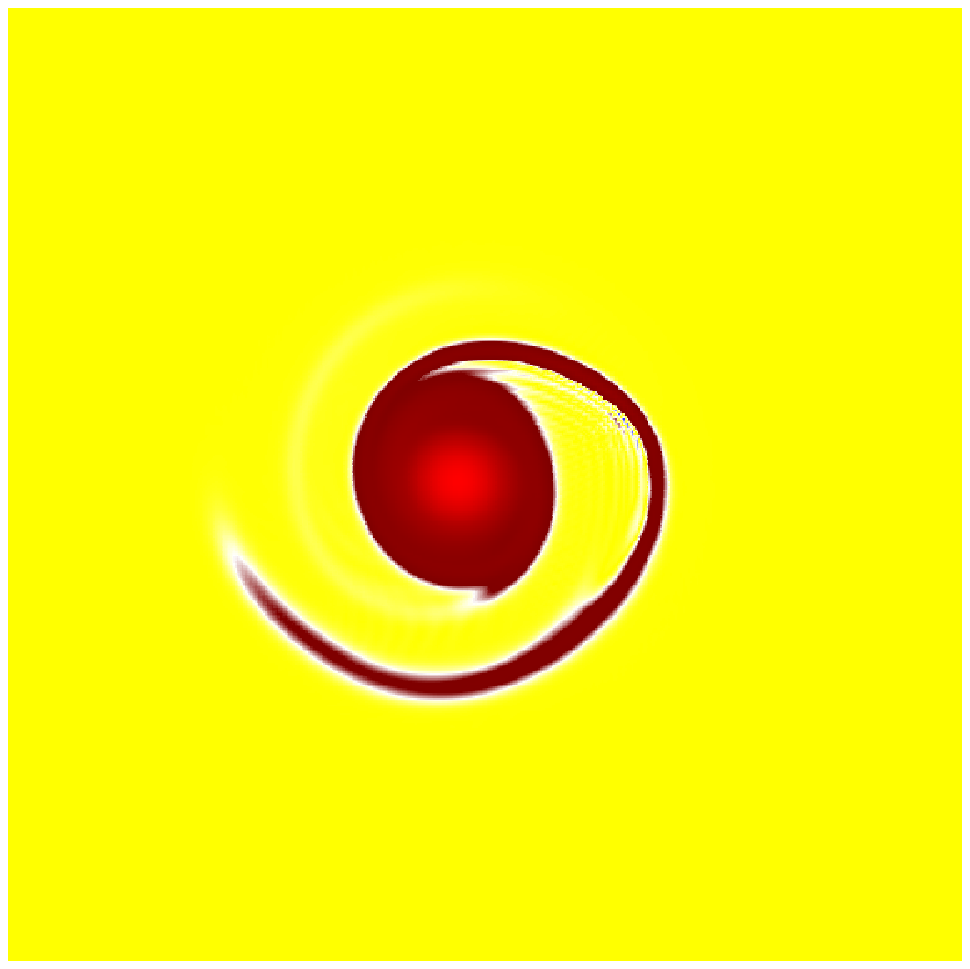}\\
    $U_2=23.4$~m/s, 9.375 h & 18.75 h & 37.5 h
  \end{tabular}
  \caption{Binary interactions between two cyclones where the initial strength of one cyclone is varied as indicated by the value of $U_2$. Columns are for times $t=9.375$~h, $18.75$~h, and $37.5$~h as indicated. (h stands for hour.)}
  \label{fig:vref}
\end{figure}

Fig~\ref{fig:vref} shows the time sequence of vorticity contour plots. When both cyclones are equal in strength ($U_1=U_2$, top row of Fig~\ref{fig:vref}), the inner eyewall of each cyclone deforms elliptically; however, the outer eyewall of each cyclone deforms into filaments. As time increases, deep convection within the primary eyewall tends to induce centripetal acceleration reducing the separation distance, as well as entrainment, developing a `rotor' in the inner eyewall, which joins the filamentary outer eyewall. At $t=37.5$~h, rotors in the inner eyewall collapse with each other. A close inspection indicates that inner eyewalls form a pair of `tadpoles' at $t=18.75$~h and $t=37.5$~h, which is surrounded by outer eyewall in the form of a filament. 

Fig~\ref{fig:vref} also compares the above interaction at $U_2=37.5$~m/s (category 1), $U_2=32.8$~m/s (tropical storm), $U_2=28.1$~m/s, and $U_2=23.4$~m/s (tropical depression).  
Clearly, if $U_2$ is reduced, the deformation rate of the weaker cyclone becomes increasingly greater in comparison with that of the stronger one. In the case the weaker cyclone is a tropical storm, it deforms relatively rapidly and encircles the stronger cyclone, eventually becoming the outer eyewall of the stronger cyclone. As can be seen, rotors do not form in the interaction between two unequal cyclones. 

A complete understanding of the secondary eyewall formation -- as depicted in Fig~\ref{fig:vref} -- is critical in the numerical prediction of hurricanes. The weaker cyclone in this set of simulations represents qualitatively the circulation induced by moist convection outside the hurricane core~\cite{Moon2010}. \citet{Kuo2004} also observed that a stronger cyclone would shear apart the weaker cyclone into thin filaments that encircle the stronger cyclone. Although the details of the numerical method, governing equations, and parameters are not exactly identical between the present simulation and that of~\cite{Kuo2004}, the results obtained in both simulations are qualitatively equivalent. \Add{More specifically, we have used $\Gamma_1/\Gamma_2 = 1,\,1.6,\,2.0,\,2.7,\,4$  and  $R_1/R_2 = 1,\,1.25,\,1.4,\,1.7,\,2$, where \citet{Kuo2004} used several other values in the ranges $1\le\Gamma_1/\Gamma_2\le 10$ and $1\le R_1/R_2\le 4$. }

\begin{figure}
  \centering
    \includegraphics[trim=0cm 0cm 1cm 0cm,clip=true,height=5cm]{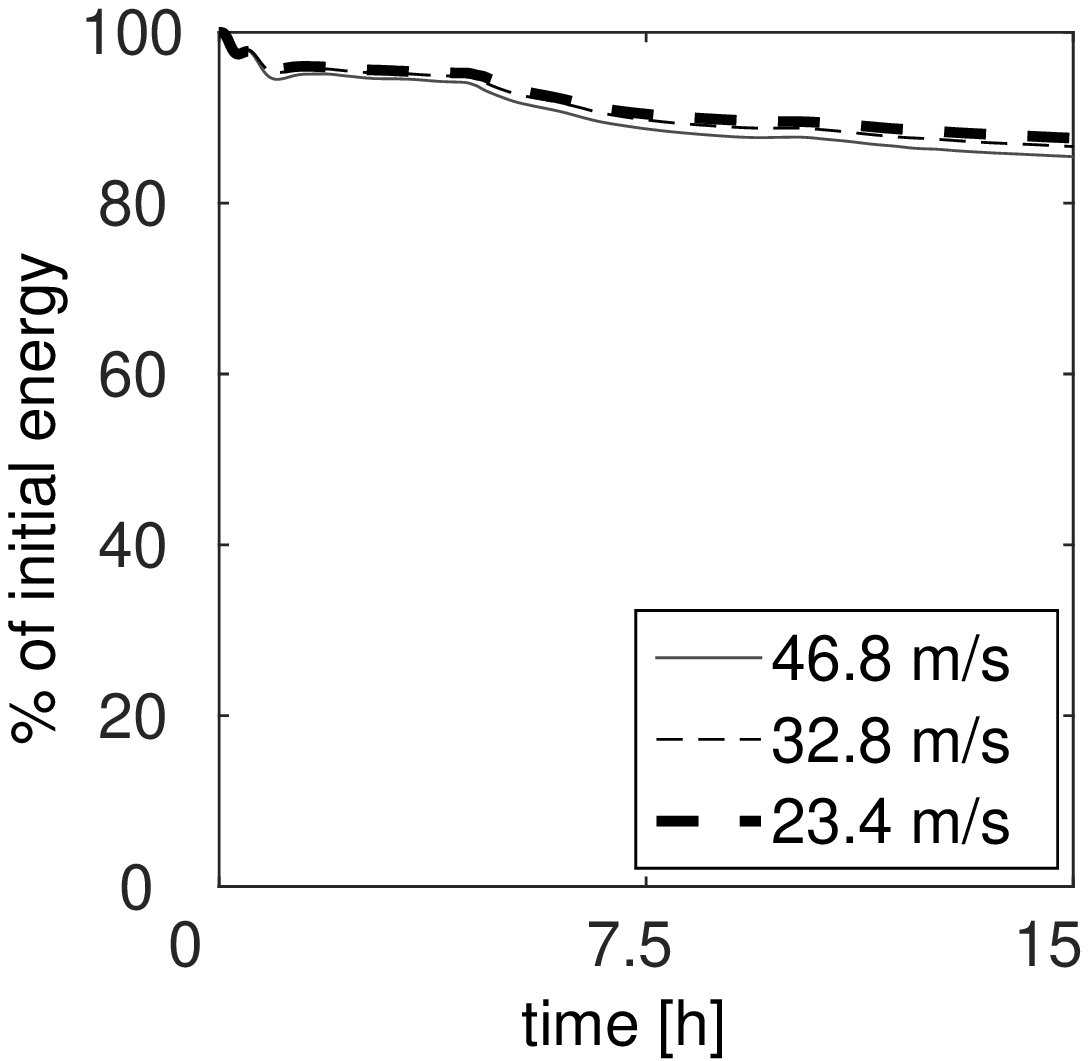}
  \caption{For the $6$ cases of cyclone interactions shown in Fig~\ref{fig:vref}, the time evolution of kinetic energy of the pair of cyclones for three cases are shown in this figure as indicated by the value of $U_2$.}
  \label{fig:enrgy}
\end{figure}

With the Smagorinsky model~(\ref{eq:tau}) the estimated value for the coefficient of eddy viscosity is $\nu_{\tau}=125$~m/s$^2$ for this set of experiments, which is based on the mean value for the rate of strain, $|S|\sim 10^{-5}\hbox{s}^{-1}$. With respect to the fixed maximum wind speed $U_1\sim 46.8$~m/s, the value of the \Add{effective} Reynolds number is $Re=37\,440$. It can be shown that 
$$
\frac{dE}{dt} = -2\mathcal Z/\mathcal Re,
$$
where $\mathcal Z$ is the enstrophy. Clearly, the kinetic energy is nearly conserved. 
The time evolution of kinetic energy for the experiments depicted in Fig~\ref{fig:vref} is shown in Fig~\ref{fig:enrgy}. The kinetic energy is normalized by the initial kinetic energy for each simulation. The near conservation of kinetic energy possess the characteristics of two-dimensional turbulence. In the case of a nearly inviscid flow (where $\mathcal Re\rightarrow\infty$) closely packed vorticity contours increase the palinstrophy, and thus, $\mathcal P/\mathcal Re$ may not be small although $\mathcal Re$ is large. It can be shown that 
$$
\frac{d\mathcal Z}{dt} = -2\mathcal P/\mathcal Re.
$$
Hence, there is a significant enstrophy cascade, which leads to a near conservation of kinetic energy. This is a phenomena of selective decay of enstrophy and kinetic energy in two-dimensional turbulence.  
\subsection{The interaction between a cyclone  and a convection induced vortex}
A binary interaction between a cyclone scale vortex and a convection induced weaker vortex was simulated by some authors where a convection induced vortex is modeled as a large vortex of relatively weak vorticity that also rotates cyclonically~\cite{Kuo2004,Moon2010}. \citet{Moon2010} shows that there exist a physical situation where the interaction between a cyclone and a convection induced vortex may lead to barotropically unstable vorticity filaments. 
\citet{Dritschel92} observed a `complete straining out' phenomena, when two cyclones are not equal in size, where one cyclone deforms completely and wraps around the other cyclone in the form of a concentric ring~\cite{Kuo2004}.
The formation of the concentric ring structure was also observed by previous numerical studies~\cite{Kuo2004,Dritschel92}. In the simulations of~\citet{Kuo2004}, the stronger vortex representing a tropical cyclone is smaller in size compared to the weaker vortex representing a vorticity field induced by moist convection. \citet{Prieto2003} simulated a binary interaction with $\omega_2/\omega_1=0.6$ and $b/R_1=3$, where the stronger cyclone is larger in size compared to the weaker cyclone. Clearly, the parameter regime varies in a way that is not straight forward enough to go with a one-to-one comparison, unless all of the above simulations are repeated. %However, we can test the present model by simulating interaction between We want to understand if the present model similates the physical phenomena that is observed by other authors.

\begin{figure}
  \centering
  \begin{tabular}{ccc}
    \includegraphics[height=3.75cm]{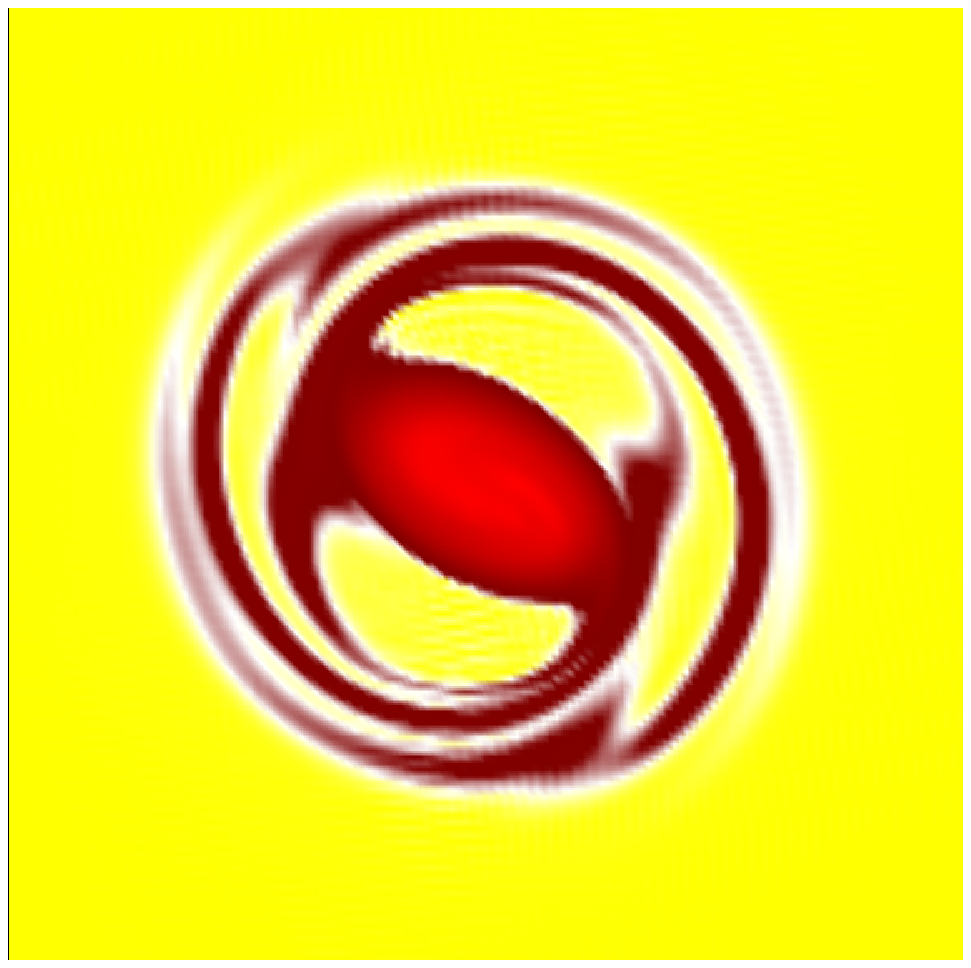}&    
    \includegraphics[height=3.75cm]{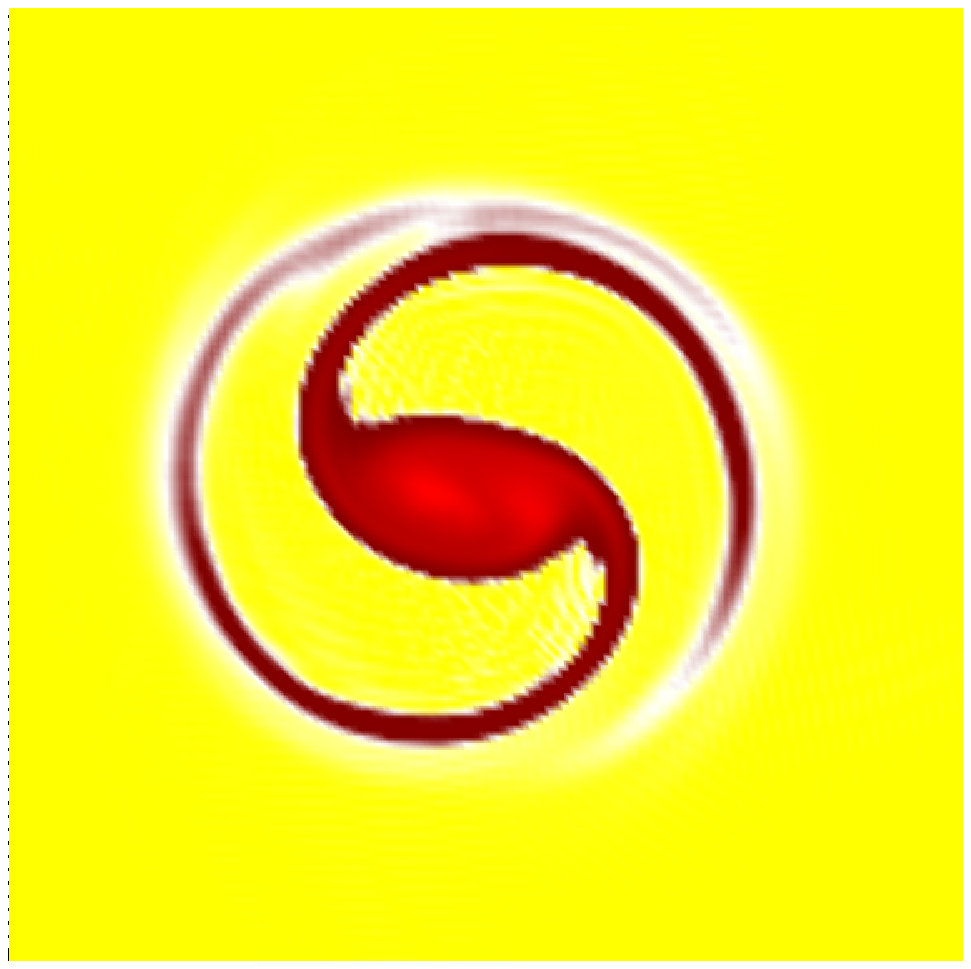}&    
    \includegraphics[height=3.75cm]{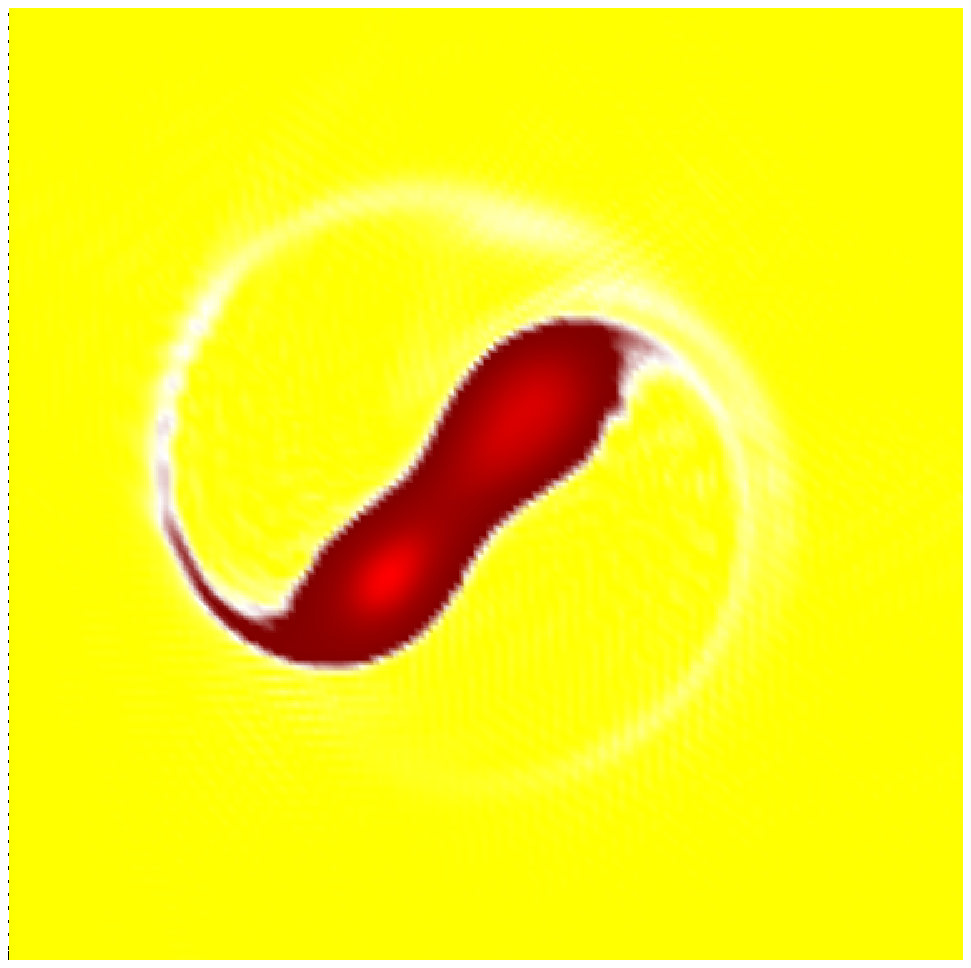}\\    
    $(a)~R_2/R_1=0.9$ & $(b)~R_2/R_1=0.7$ & $(c)~R_2/R_1=0.5$ 
  \end{tabular}
  \caption{$(a)$~As shown by~\citet{Kuo2004}, the interaction between a large cyclone scale vortex with a relatively weak vortex that is slightly smaller in size; $(b,c)$~the effect on such a binary interaction if the size of the weaker vortex is reduced.}
  \label{fig:ueqw}
\end{figure}

Here, we study the interaction between a cyclone scale vortex with a another vortex of the same vorticity and different size; {\em i.e.} $\omega_1=\omega_2$ and $R_1\ne R_2$.   The radius and maximum sustained wind of the cyclone scale vortex is taken $R_1=100$~km and $U_1=46.8$~m/s, respectively. The vorticity ($\omega_2$) of the other vortex is computed from its strength
$$
\Gamma_2 = \oint\bm u\cdot d\bm r = \iint\underbrace{\bm\nabla\times\bm u}_{\bm\omega_2}\cdot d\bm S,
$$
where its radius ($R_2$) of maximum wind is reduced within the range $0.5\le R_2/R_1\le 1.0$ such that $\omega_2 = \omega_1$. For $R_2/R_1=0.9$ and $t=37.5$~h, Fig~\ref{fig:ueqw}$a$ shows that there is an elliptic inner vortex along with two distinctive minima in the moat. \citet{Kuo2004} called similar patterns a tripole formation (see their Fig 3, second row). Figs~\ref{fig:ueqw}$(b,c)$ show the vortex structure at $t=37.5$~h for $R_2/R_1=0.7$ and $R_2/R_1=0.5$, respectively. As discussed by~\citet{Kuo2004}, the tripole pattern is absent in the interaction between a cyclone and a convection induced vortex.

\subsection{Barotropic instability of tropical cyclones}

Observational concentric patterns of tropical cyclones suggest that barotropic instability may be a possible mechanism for destructive suction spots embedded within cyclones ({\em e.g.} simulated vertical velocity pattern of Hurricane Bill (2009) as shown in Fig~1 of~\cite{Moon2010}). In their study, \citet{Moon2010} demonstrated that convection induced vortices in the outer region of a cyclone core lead to barotropic instability. The interaction of the inner core of a cyclone with convection induced anticyclone in the outer core was modeled by a positive negative vorticity dipole.  \citet{Chan84} suggested that vorticity advection and the initial wind profile may play an important role in Fujiwhara interactions of cyclones through the introduction of barotropic instability.  Barotropic models of tropical cyclones with a distribution of positive vorticity in a region surrounded by convection induced weakly negative vorticity show the formation of polygonal eyewalls as a result of barotropic instability~\cite{Kossin2001}.  However, when two such cyclones -- surrounded by convection induced weak vorticity -- come in close proximity, they exhibit Fujiwhara interactions in the early stage, but barotropic instability eventually dominates. The instability across the outer ring of enhanced vorticity may occur when the outer ring is sufficiently narrow and the circulation of the cyclone core is sufficiently weak. The sign change of the vorticity gradient in the outer region is sensitive to tropical cyclone intensity forecast, which is necessary for barotropic instability. 

\begin{figure}
  \centering
  \begin{tabular}{ccc}
    \includegraphics[height=4cm]{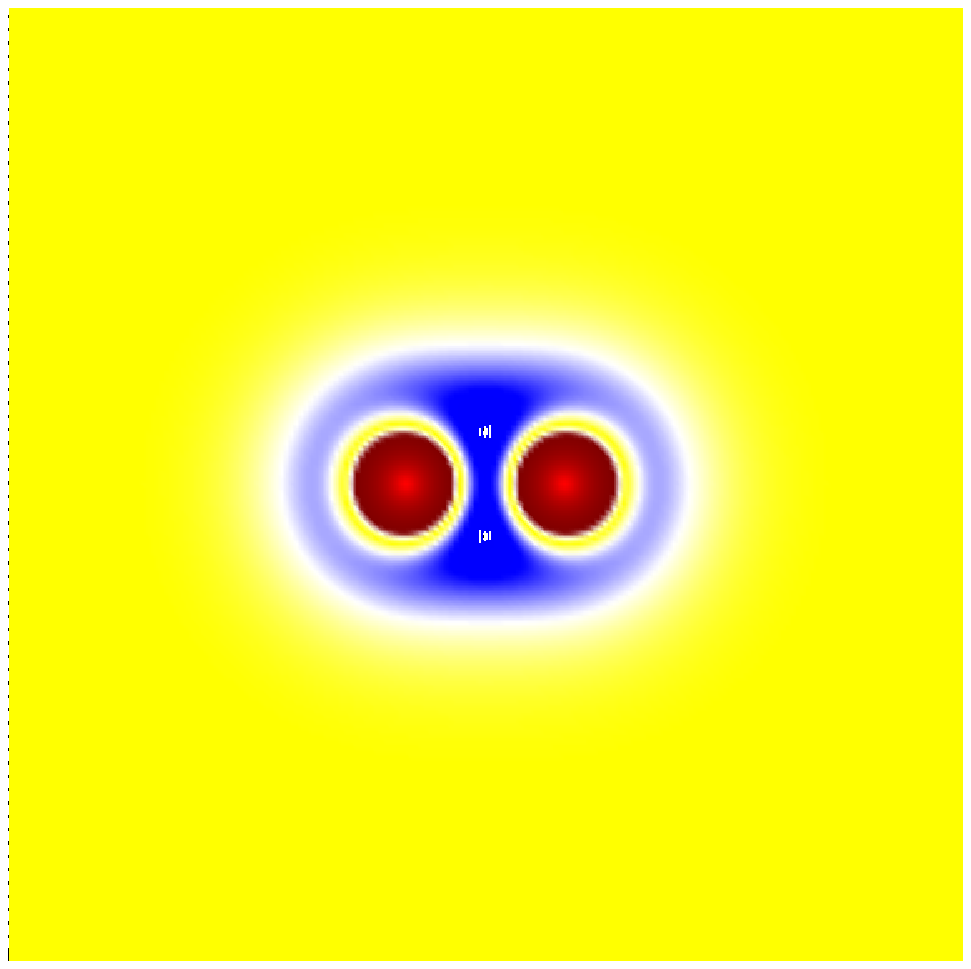}&
    \includegraphics[height=4cm]{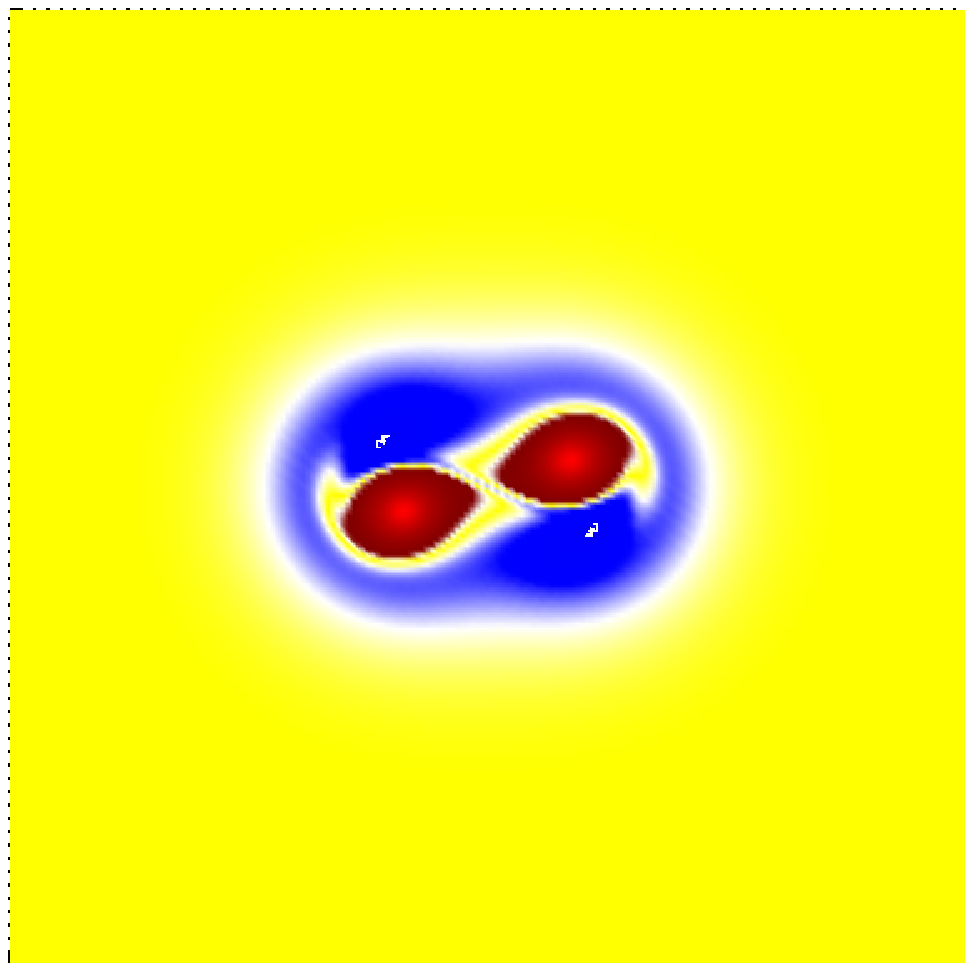}&
    \includegraphics[height=4cm]{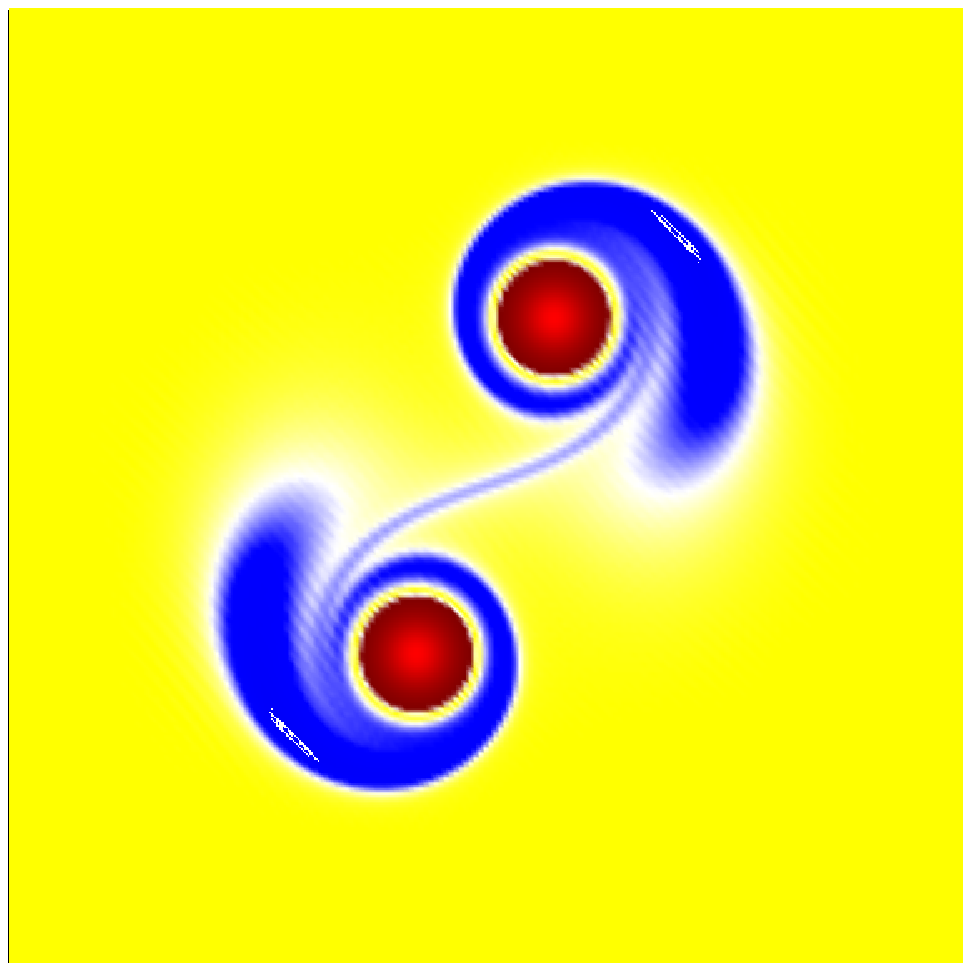}
    \\
    $(a)$ & $(b)$ & $(c)$\\
    \includegraphics[height=4cm]{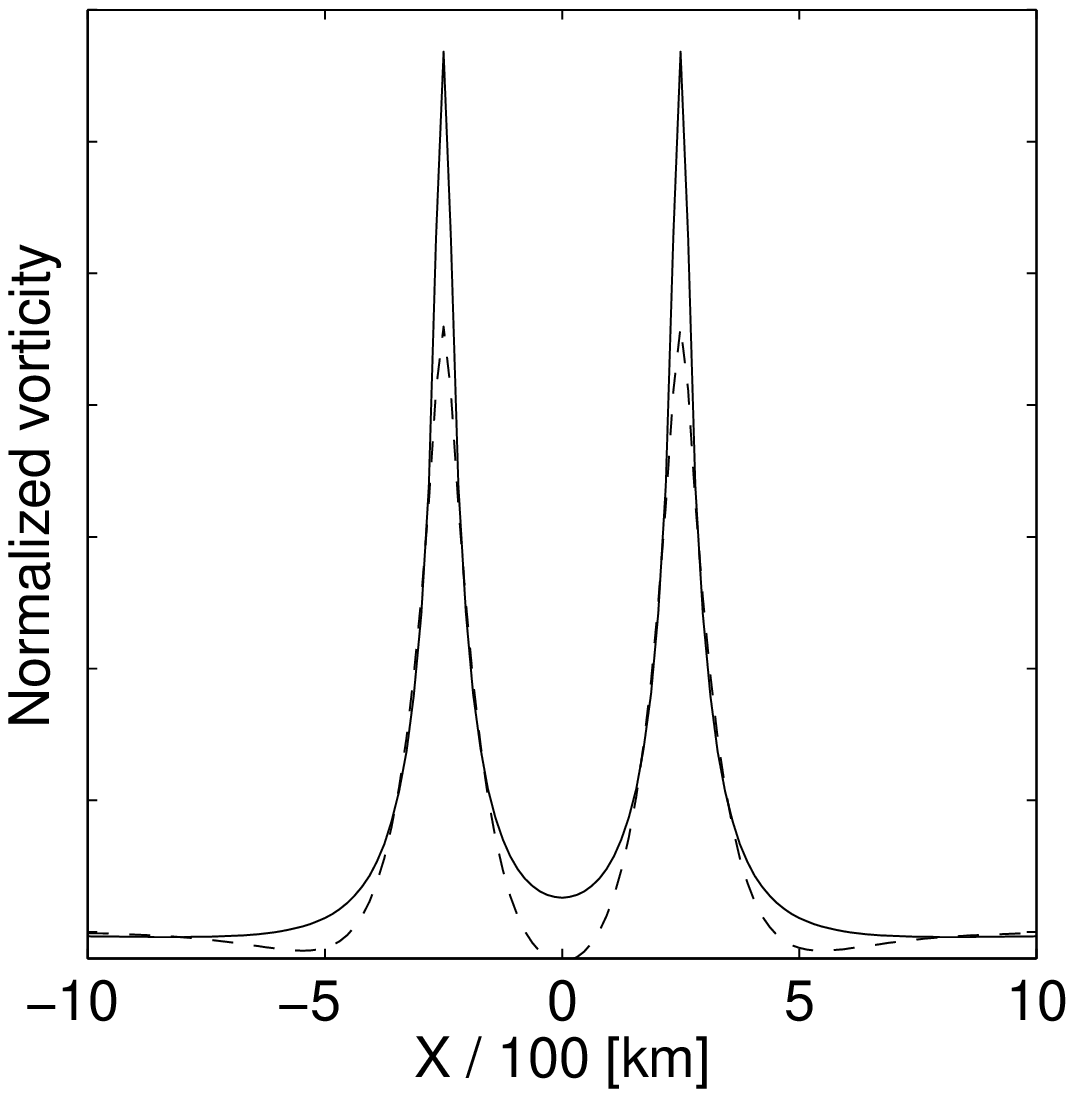}&
    \includegraphics[height=4cm]{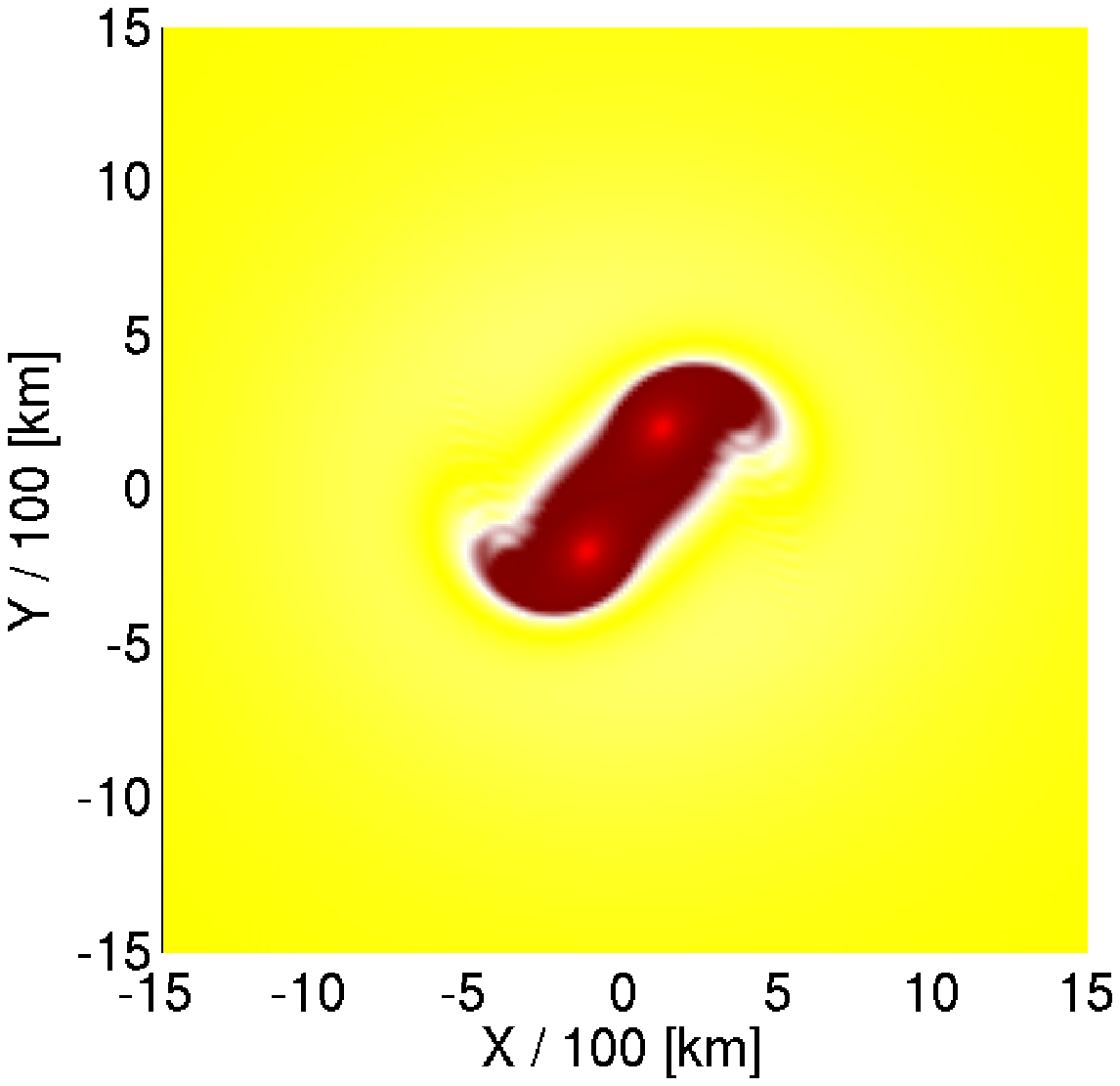}&
    \includegraphics[height=4cm]{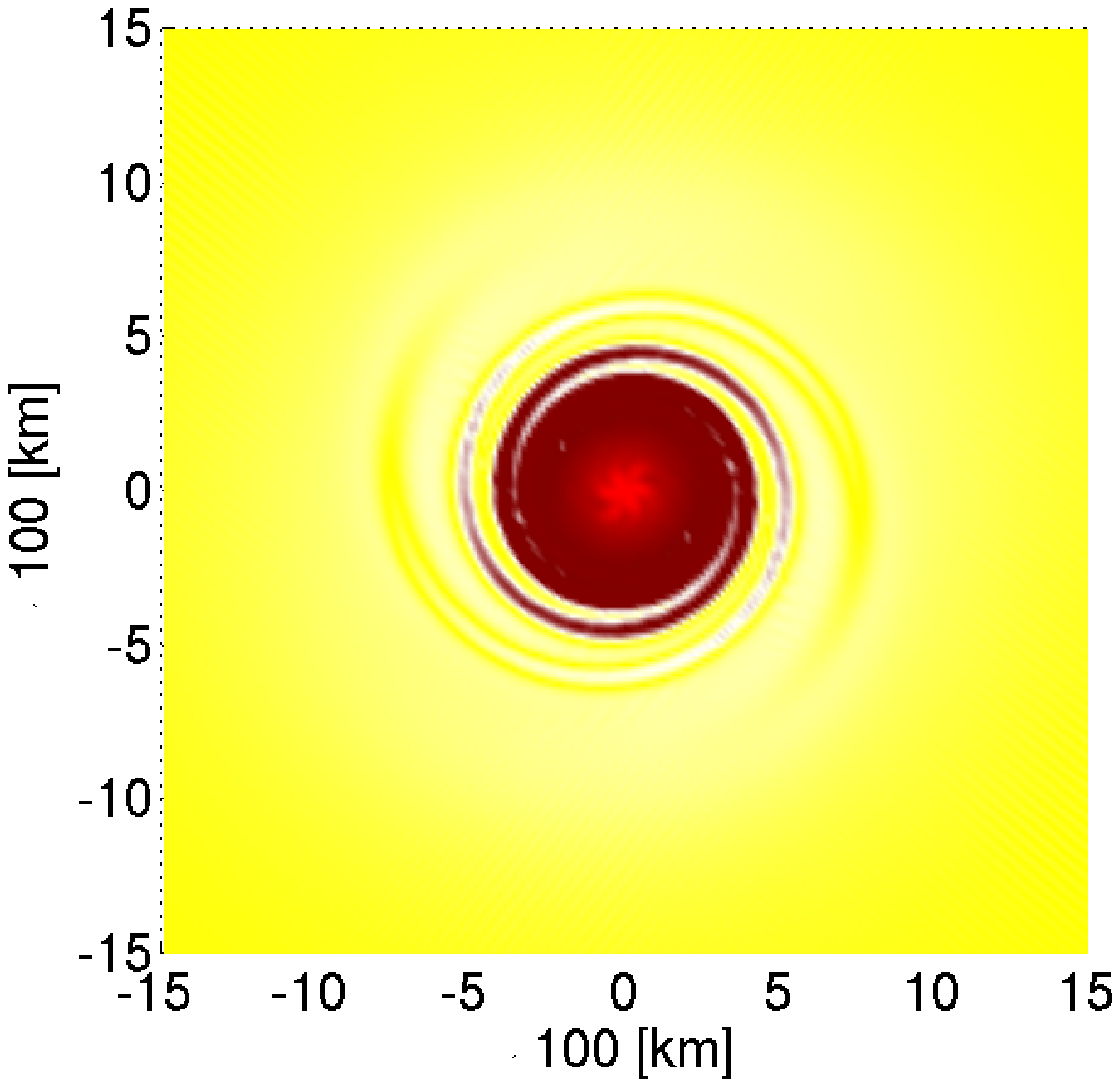}
    \\
    $(d)$ & $(e)$ & $(f)$\\
  \end{tabular}
  \caption{Barotropic instability due vorticity advection. Here, red, blue, and yellow are used to represent positive, negative, and zero vorticity. $(a)$~Initial vortex pattern; $(b)$~Fujiwhara type attraction in the early stage; $(c)$~repulsion of the vortex pair due to barotropic instability in a later stage; $(d)$~a comparison of two initial configurations: one is the same as $(a)$ and the other is the same as $(a)$ when the negative vorticity is removed, `$-\,-$' the vorticity profile from $(a)$ along the line $y=0$ and `\textemdash\textemdash'~the same from $(a)$ when the negative vorticity is removed, $(e,f)$ Fujiwhara type interaction in the absence of negative vorticity pattern in $(a)$. }
  \label{fig:demaria}
\end{figure}

The first order effects of the vorticity advection~({\em e.g.}~\cite{Chan84}) can be modeled by introducing a relatively faster decay of tangential velocity outside the inner core of the cyclone. Consider a cyclone of maximum wind $30$~m/s, where the radius of maximum wind is $100$~km, and the velocity decays to $3$~m/s at a distance of $500$~km from the center of the cyclone. Under this initial velocity field, a pair of initial cyclones with a center to center separation distance $500$~km is shown in Fig~\ref{fig:demaria}($a$). Now, a second pair of such cyclones where the velocity decays to $10$~m/s at a distance of $500$~km from the center has been considered. The vorticity field for both pairs has been shown in Fig~\ref{fig:demaria}$(d)$ as a function of $x$ along the line $y=0$. Clearly, the strength of the anticyclonic narrow band around the cyclone is only marginal. 
 
Fig~\ref{fig:demaria}$(b)$ shows that the cyclones have an initial tendency to interact through an elliptic instability and exchange of vorticity. However, the opposite sign weak vorticity in the outer region is redistributed to form two lateral poles with each of them forming a dipolar structure with one of the original cyclones. Later, rolls of negative vorticity patch are formed around the original cyclones, and they eventually move away from each other.

The above simulation has been repeated where the decay of the initial velocity is adjusted so that there is not negative vorticity patch around the original cyclones. The Fujiwhara type interaction is evident from Fig~\ref{fig:demaria}$(e,f)$.  

In the following section, we continue with the vortex interaction discussed in section~\ref{sec:vref} where we simulate the transport of moisture by a pair of interacting cyclones.  
\subsection{Transport of water vapor as a passive scalar}
\begin{figure}
  \centering
  \begin{tabular}{cc}
    \includegraphics[trim=1cm 0cm 2cm 0cm,clip=true,height=5cm]{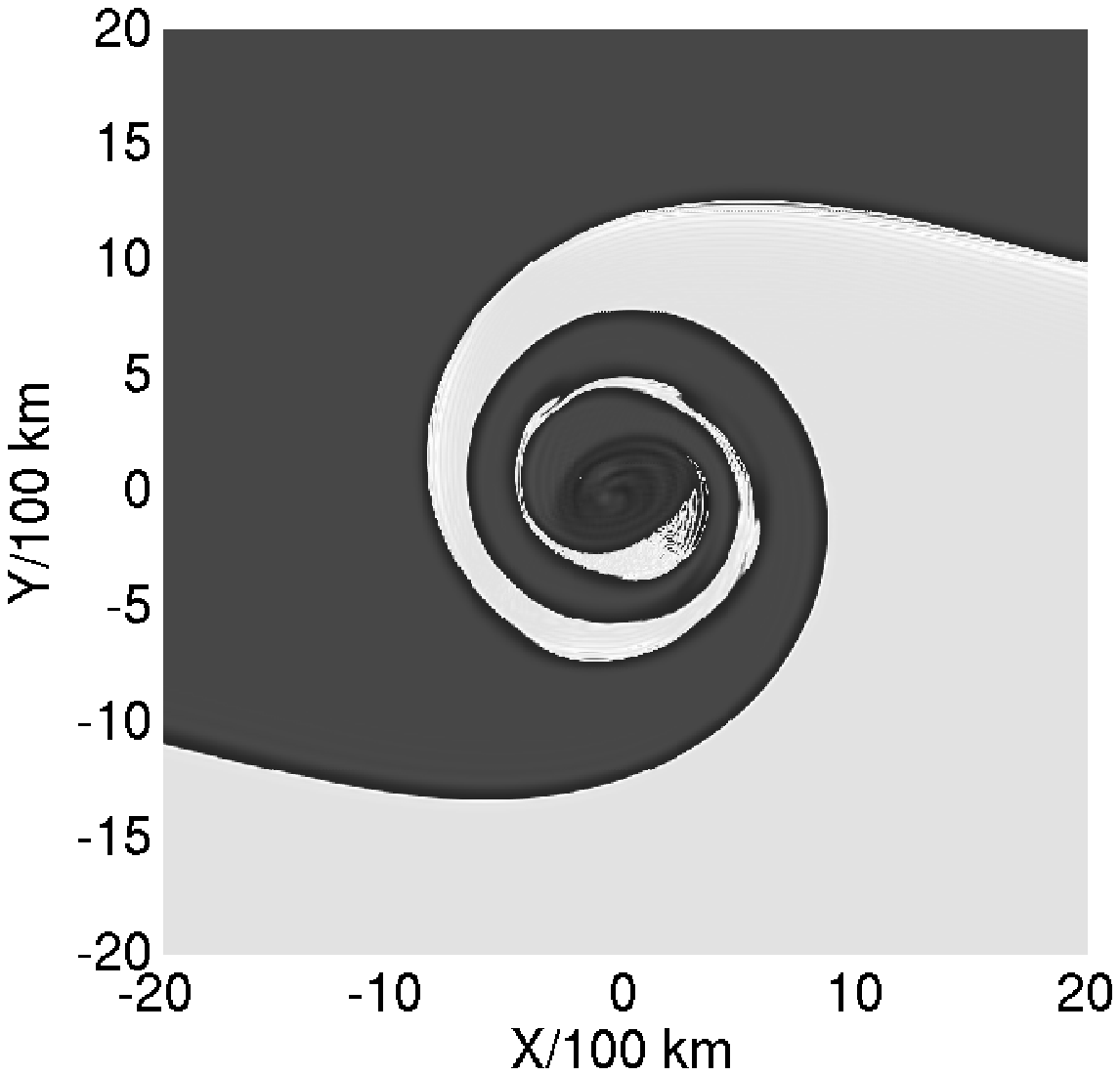}&
    \includegraphics[trim=1cm 0cm 2cm 0cm,clip=true,height=5cm]{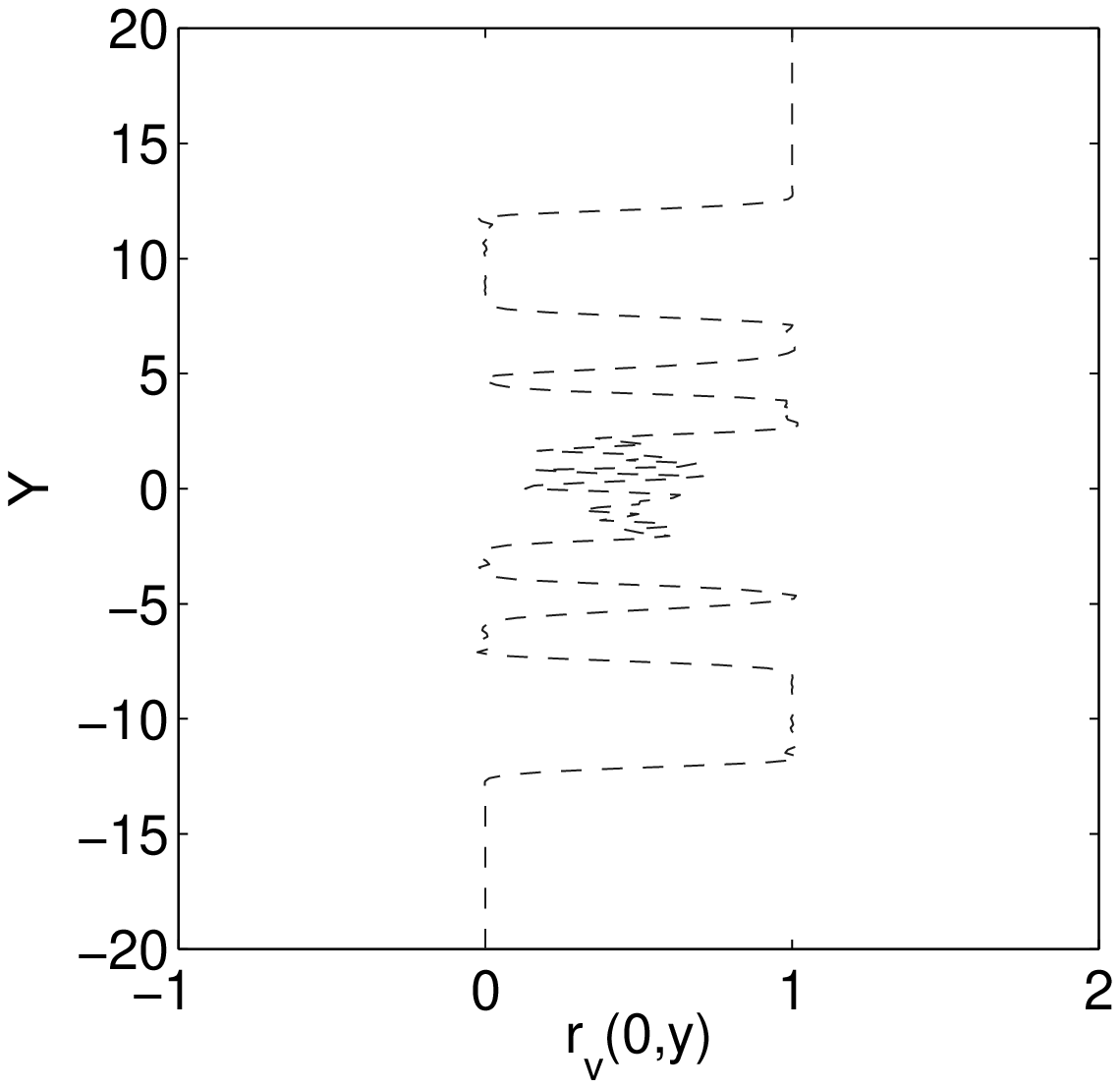}
  \end{tabular}
  \caption{A numerical solution of the equation for the vapor mixing ratio, $(a)$ the contour plot of $r_v (x,y,t)$ showing a chaotic mixing of the vapor and $(b)$ the line plot of $r_v(0,y,t)$, both are at $t=37.5$ h.}
  \label{fig:rv}
\end{figure}

For each simulation associated with the result in Fig~\ref{fig:vref}, the vapor equation~(\ref{eq:vpr}) has been solved using the same initial condition, $r_v(x,y,0)=1$ for $y > 0$ otherwise $r_v(x,y,0)=0$. In order to see the numerical solution of the deformational advection of moisture, only the diffusion term is used on the right hand side of~(\ref{eq:vpr}). The solution in Fig~\ref{fig:rv} corresponds to the $3$rd row of Fig~\ref{fig:vref} with $U_2=32.8$~m/s. A cross section of the data in Fig~\ref{fig:rv}($a$) along the line $x=0$ is shown in Fig~\ref{fig:rv}$(b)$. A fixed time step of $\Delta t=5.625$~minutes is used. Clearly, the solution does not exhibit any artificial damping after $400$ time steps. Note that the parameterization for the subgrid scale condensation/deposition/diffusion on the right hand side of~(\ref{eq:vpr}) is replaced with a simple diffusion term, thereby making~(\ref{eq:vpr}) a advection-diffusion equation. Thus, the solution in Fig~\ref{fig:rv}$b$ represents the multiresolution approximation of the transport phenomena without using a dedicated avection algorithm ({\em e.g.}~\cite{Pielke2002,Skamarock2008}) appropriate for meteorological simulations. It is important to note that our numerical method does not exhibit artificial numerical diffusion. The maximum and the minimum values of $r_v$ between the initial and the final time of this simulation remain approximately the same where the effect of chaotic mixing -- as expected -- is noticeable in Fig~\ref{fig:rv}. 

\section{Concluding remarks}\label{sec:cr}
In this work, we have developed a numerical model for simulating the interactions between a pair of tropical-cyclone-like vortices. 
Note also that there are a number of documented numerical models for simulating  the interaction between tropical cyclones~\cite{Prieto2003,Kuo2004,Nomura2007,Moon2010}, which involve solving the nondivergent barotropic vorticity equation. \Add{However, we demonstrate a novel numerical approach for solving the momentum equations, the equations for Exner function and water vapor mixing ratio.} More specifically, this article outlines a weighted residual collocation methodology~(AWCM++) for simulating interactions between tropical cyclones using a set of compressible equations based on the velocity and the Exner function. In this system, a balance among surface friction, geostrophic pressure gradients, Coriolis forces, and vertical mixing of momentum has been assumed, which reduces the nonhydrostatic physics to the single layered barotropic physics. \Add{The large scale dynamics is resolved at the typical meteorological resolution $\mathcal O(10\hbox{ km})$. However, an efficient multiresolution approximation method~({\em e.g.}~\cite{Mallat89}) is proposed to seperate the resolved scales from the unresolved ones in contrast to classical grid volume averaging used in meteorological models~\cite{Pielke2002}. The subgrid scale unresolved turbulence has been parameterized with the Smagorinsky model.} %We present a numerical methodology for solving the compressible atmospheric model, and take advantage of some existing barotropic models for a validation of the present model.

For testing the proposed numerical model, we have idealized some satellite data on kinematic and thermodynamic structure of MCVs ({\em e.g.} see~\cite{Kuo2004,Davis2007}). \citet{Davis2007} provides details on such idealization where the data was analysed with a model of Poisson equation satisfying by stream function and potential vorticity. In the present model, we have used such observational data to characterize the interaction between two MCVs. Comparing our results with a dynamically equivalent DNS model of a vortex merger~\cite{Nomura2007}, we have quantified the accuracy of the proposed model. The results show that the proposed multiresolution approach adequately captures the Fujiwhara interaction between two MCVs, although the cut-off scale $\Delta$ is about $7$-$8$ time larger than that of the DNS model of~\citet{Nomura2007}. However, further research would help to understand whether $\Delta$ be adjusted to the elliptical deformation of cyclones' core or to the vorticity filamentation outside the core~\cite{Nomura2007}. Note also the effective dissipation for $\mathcal Re=5\,000$ as seen in Fig~\ref{fig:fbi}. 

Although closely packed vorticity contours may increase palinstrophy~\cite{Kuo2004}, the rate of enstrophy dissipation is lower for $\mathcal Re=37\,440$ with respect to $\mathcal Re=5\,000$. Unless the cut-off scale is lowered toward the scale of enstrophy cascade ({\em e.g.} Fig~\ref{fig:vref}), the Smagorinsky model may not be sufficient and the search for a more accurate subgrid scale parameterization remains an open challenge. Inadequate representation of the subgrid scale phenomena would exhibit unrealistic oscillations in the vorticity field~\cite{Kuo2004}. Findings of this research suggest a possible avenue for the development of a multiscale LES for modeling the dynamics of cyclones. In addition to such modelling questions, the present authors have not discussed the computational benefits of the Krylov method for meteorological simulations -- a topic requires independent investigations. 

\section*{Acknowledgements}
JMA acknowledges the discovery grant and RPW acknowledges the USRA scholarship from the National Science and Research Councill~(NSERC), Canada. Suggestions from two anonymous reviewers have improved the manuscript significantly. This work was benefited by the computing facility of the Shared Hierarchical Academic Research Computing Network ({\tt SHARCNET:www.sharcnet.ca}) and Compute/Calcul Canada.

\bibliographystyle{apalike}
\bibliography{bibrefs}

\begin{thebibliography}{}

\bibitem[Aechtner et~al., 2014]{Kevlahan2014}
Aechtner, M., Kevlahan, N. K.-R., and Dubos, T. (2014).
\newblock A conservative adaptive wavelet method for the shallow-water
  equations on the sphere.
\newblock {\em Quarterly Journal of the Royal Meteorological Society}, pages
  n/a--n/a.

\bibitem[Alam, 2006]{Alamphd2006}
Alam, J. (2006).
\newblock {\em A space-time adaptive wavelet method for turbulence}.
\newblock PhD thesis, McMaster University.

\bibitem[Alam, 2011]{Alam2011}
Alam, J. (2011).
\newblock Towards a multi-scale approach for computational atmospheric
  modelling.
\newblock {\em Monthly Weather Review}, 139(12).

\bibitem[Alam and Islam, 2015]{Alam2015}
Alam, J. and Islam, M.~R. (2015).
\newblock A multiscale eddy simulation methodology for the atmospheric ekman
  boundary layer.
\newblock {\em Geophysical \& Astrophysical Fluid Dynamics}, 109(1):1--20.

\bibitem[Alam et~al., 2006]{Alam2006}
Alam, J., Kevlahan, N. K.-R., and Vasilyev, O. (2006).
\newblock Simultaneous space--time adaptive solution of nonlinear parabolic
  differential equations.
\newblock {\em Journal of Computational Physics}, 214:829--857.

\bibitem[Alam et~al., 2014]{Alam2014}
Alam, J.~M., Walsh, R.~P., Alamgir~Hossain, M., and Rose, A.~M. (2014).
\newblock A computational methodology for two-dimensional fluid flows.
\newblock {\em International Journal for Numerical Methods in Fluids},
  75(12):835--859.

\bibitem[Bannon, 2002]{Bannon2002}
Bannon, P.~R. (2002).
\newblock Theoretical foundation for models of moist convection.
\newblock {\em Journal of the Atmospheric Sciences}, 59.

\bibitem[Brand, 1970]{Brand70}
Brand, S. (1970).
\newblock Interaction of binary tropical cyclones of the western north pacific
  ocean.
\newblock {\em J. Appl. Meteor.}, 9:433~441.

\bibitem[Brandt and Nomura, 2007]{Nomura2007}
Brandt, L.~K. and Nomura, K.~K. (2007).
\newblock The physics of vortex merger and the effects of ambient stable
  stratification.
\newblock {\em J. Fluid Mech.}, 592:413--446.

\bibitem[Bryan and Fritsch, 2002]{Bryan2002}
Bryan, G.~H. and Fritsch, J.~M. (2002).
\newblock A benchmark simulation for moist nonhydrostatic numerical model.
\newblock {\em Mon. Wea. Rev.}, 130.

\bibitem[Chan and Jackson, 1984]{Chan84}
Chan, T.~F. and Jackson, K.~R. (1984).
\newblock Nonlinearly preconditioned krylov subspace methods for discrete
  newton algorithms.
\newblock {\em SIAM J. Sci. Stat. Comput.}, 5:533--542.

\bibitem[Davis and Trier, 2007]{Davis2007}
Davis, C.~A. and Trier, S.~B. (2007).
\newblock Mesoscale convective vortices observed during bamex. part i:
  Kinematic and thermodynamic structure.
\newblock {\em Monthly Weather Review}, 135:2029--2049.

\bibitem[Deardorff, 1970]{dear70}
Deardorff, J.~W. (1970).
\newblock A three-dimensional numerical investigation of idealized planetary
  boundary layer.
\newblock {\em Geophys. Fluid Dyn.}, 1:377--410.

\bibitem[Deardorff, 1972]{dear72}
Deardorff, J.~W. (1972).
\newblock Numerical investigation of nutral and unstable planetary boundary
  layer.
\newblock {\em J. Atmospheric Science}, 29:91--115.

\bibitem[Deslauriers and Dubuc, 1989]{Dubuc89}
Deslauriers, G. and Dubuc, S. (1989).
\newblock Symmetric iterative interpolation process.
\newblock {\em Constructive Approximation}, 5:49--68.

\bibitem[Dong and Neumann, 1983]{Dong83}
Dong, K. and Neumann, C.~J. (1983).
\newblock On the relative motion of binary tropical cyclones.
\newblock {\em Mon. Wea. Rev.}, 111:945~953.

\bibitem[Dritschel and Waugh, 1992]{Dritschel92}
Dritschel, D. and Waugh, D. (1992).
\newblock Quantification of the inelastic interaction of unequal vortices in
  two--dimensional vortex dynamics.
\newblock {\em Phys. Fluids A}.

\bibitem[Edwards et~al., 1994]{Edwards94}
Edwards, W., Tuckerman, L., Friesner, R., and Sorensen, D. (1994).
\newblock Krylov methods for the incompressible navier-stokes equations.
\newblock {\em Journal of Computational Physics}, 110:82--102.

\bibitem[Finlayson, 1972]{Bruce}
Finlayson, B.~A. (1972).
\newblock {\em The Method of Weighted Residuals and Variational Principles}.
\newblock Academic Press.

\bibitem[Fujiwhara, 1921]{Fujiwhara21}
Fujiwhara, S. (1921).
\newblock The natural tendency towards symmetry of motion and its application
  as a principle in meteorology.
\newblock {\em Quarterly Journal of the Royal Meteorological Society},
  47(200):287--292.

\bibitem[Fujiwhara, 1923]{Fujiwhara23}
Fujiwhara, S. (1923).
\newblock On the growth and decay of vortical systems.
\newblock {\em Quarterly Journal of the Royal Meteorological Society},
  49(206):75--104.

\bibitem[Fujiwhara, 1931]{Fujiwhara31}
Fujiwhara, S. (1931).
\newblock Short note on the behavior of two vortices.
\newblock {\em Proc. Phys. Math. Soc. Japan.}, 13:106~110.

\bibitem[Knoll and Keyes, 2004]{Knoll2004}
Knoll, D.~A. and Keyes, D.~E. (2004).
\newblock Jacobian-free newton-krylov methods: a survey of approaches and
  applications.
\newblock {\em J. Comput. Phys.}, 193(2):357--397.

\bibitem[Kossin and Schubert, 2001]{Kossin2001}
Kossin, J.~P. and Schubert, W.~H. (2001).
\newblock Mesovortices, polygonal flow patterns, and rapid pressure falls in
  hurricane-like vortices.
\newblock {\em Journal of the Atmospheric Sciences}, 58:2196--2209.

\bibitem[Kundu and Ira, 2010]{Kundu}
Kundu, P.~K. and Ira, M.~C. (2010).
\newblock {\em Fluid Mechanics}.
\newblock Elsevier, fourth edition edition.

\bibitem[Kuo et~al., 2004]{Kuo2004}
Kuo, H.-C., Lin, L.-Y., Chang, C.-P., and Williams, R.~T. (2004).
\newblock The formation of concentric vorticity structures in typhoons.
\newblock {\em J. Atmos. Sci.}, 61:2722~2734.

\bibitem[Lander and Holland, 1993]{Lander93}
Lander, M. and Holland, G.~J. (1993).
\newblock On the interaction of tropical-cyclone-scale vortices. i:
  Observations.
\newblock {\em Quart. J. Roy. Meteor. Soc.}, 119:1347~1361.

\bibitem[Mallat, 1989]{Mallat89}
Mallat, S. (1989).
\newblock Multiresolution approximations and wavelet orthonormal bases of
  {L}$^2${R}.
\newblock {\em Trans. Amer. Math. Soc.}, 315:69--87.

\bibitem[Mallat, 2009]{Mallat2009}
Mallat, S. (2009).
\newblock {\em A wavelet tour of signal processing}.
\newblock Academic press.

\bibitem[Moon et~al., 2010]{Moon2010}
Moon, Y., Nolan, D.~S., and Iskandarni, M. (2010).
\newblock On the use of two-dimensional incompressible flow to study secondary
  eyewall formation in tropical cyclones.
\newblock {\em Journal of the Atmospheric Sciences}, 67:3765--3773.

\bibitem[Pielke, 2002]{Pielke2002}
Pielke, R.~A. (2002).
\newblock {\em Mesoscale Meteorological Modeling}.
\newblock Academic press, second edition.

\bibitem[Prieto et~al., 2003]{Prieto2003}
Prieto, R., Mcnoldy, B.~D., Fulton, S.~R., and Schubert, W.~H. (2003).
\newblock A classification of binary tropical cyclone~like vortex interactions.
\newblock {\em Monthly Weather Review}, 131:2656--2666.

\bibitem[Schneider and Vasilyev, 2010]{Vasilyev2010}
Schneider, K. and Vasilyev, O.~V. (2010).
\newblock Wavelet methods in computational fluid dynamics*.
\newblock {\em Annual Review of Fluid Mechanics}, 42(1):473--503.

\bibitem[Skamarock and Klemp, 2008]{Skamarock2008}
Skamarock, W.~C. and Klemp, J.~B. (2008).
\newblock A time-split nonhydrostatic atmospheric model for weather research
  and forecasting applications.
\newblock {\em J. Comput. Phys.}, 227(7):3465--3485.

\bibitem[Smagorinsky, 1963]{smagorinsky}
Smagorinsky, J. (1963).
\newblock General circulation experiments with the primitive equations.
\newblock {\em Monthly Weather Review}, 91:99.

\bibitem[Smolarkiewicz et~al., 2014]{Piotr2014}
Smolarkiewicz, P.~K., K\"{u}hnlein, C., and Wedi, N.~P. (2014).
\newblock A consistent framework for discrete integrations of soundproof and
  compressible pdes of atmospheric dynamics.
\newblock {\em J. Comput. Phys.}, 263:185--205.

\bibitem[Stevens et~al., 2002]{Stevens2002}
Stevens, B., Duan, J.~J., McWilliams, J.~C., Munnich, M., and Neelin, J.~D.
  (2002).
\newblock {Entrainment, Rayleigh friction, and boundary layer winds over the
  tropical Pacific}.
\newblock {\em {Journal of Climate}}, 15:30--44.

\bibitem[Tannehill et~al., 1997]{Tannehill97}
Tannehill, J.~C., Anderson, D.~A., and Pletcher, R.~H. (1997).
\newblock {\em Computational Fluid Mechanics Heat Transfer}.
\newblock Taylor and Francis.

\bibitem[Vasilyev and Bowman, 2000]{Oleg2000}
Vasilyev, O.~V. and Bowman, C. (2000).
\newblock Second-generation wavelet collocation method for the solution of
  partial differential equations.
\newblock {\em J. Comput. Phys.}, 165:660--693.

\bibitem[Vasilyev and Kevlahan, 2005]{Oleg2005}
Vasilyev, O.~V. and Kevlahan, N.-R. (2005).
\newblock An adaptive multilevel wavelet collocation method for elliptic
  problems.
\newblock {\em J. Comput. Phys.}, 206:412--431.

\end{thebibliography}

\end{document}